\documentclass[aps,
prx,
twocolumn,
amsmath,
amssymb,
floatfix]{revtex4-1}

\usepackage{graphicx}
\usepackage{dcolumn}
\usepackage{bm}
\usepackage{dsfont}
\usepackage{physics}
\usepackage{amsmath,amsthm,amsfonts,amssymb,amscd}
\usepackage[mathlines]{lineno}
\usepackage[usenames, dvipsnames]{color}
\usepackage{mathtools}
\usepackage{textcomp}
\usepackage{hyperref}
\hypersetup{
    colorlinks=true,
    linkcolor=black,
    filecolor=black,      
    urlcolor=black,
    citecolor = black
}

\usepackage{bbold}
\usepackage{titlesec}
\titlespacing*{\section} {0pt}{3.5ex plus 2ex minus .2ex}{2.3ex plus .2ex}

\begin{document}
\preprint{}
\title{A 10-qubit solid-state spin register with quantum memory up to one minute}

\author{C. E. Bradley$^{1,2}$}\thanks{These authors contributed equally to this work.}
\author{J. Randall$^{1,2}$}\thanks{These authors contributed equally to this work.}
\author{M. H. Abobeih$^{1,2}$}
\author{R. C. Berrevoets$^{1,2}$}
\author{M. J. Degen$^{1,2}$}
\author{\\ M. A. Bakker$^{1,2}$}
\author{M. Markham$^3$} 
\author{D. J. Twitchen$^3$}
\author{T. H. Taminiau$^{1,2}$}
 \email{T.H.Taminiau@TUDelft.nl}

\date{\today}

\affiliation{$^{1}$QuTech, Delft University of Technology, PO Box 5046, 2600 GA Delft, The Netherlands}
\affiliation{$^{2}$Kavli Institute of Nanoscience Delft, Delft University of Technology, PO Box 5046, 2600 GA Delft, The Netherlands}
\affiliation{$^{3}$Element Six, Fermi Avenue, Harwell Oxford, Didcot, \\ Oxfordshire, OX11 0QR, United Kingdom}

\date{\today}

\begin{abstract}

 Spins associated to single defects in solids provide promising qubits for quantum information processing and quantum networks. Recent experiments have demonstrated long coherence times, high-fidelity operations and long-range entanglement. However, control has so far been limited to a few qubits, with entangled states of three spins demonstrated. Realizing larger multi-qubit registers is challenging due to the need for quantum gates that avoid crosstalk and protect the coherence of the complete register. In this paper, we present novel decoherence-protected gates that combine dynamical decoupling of an electron spin with selective phase-controlled driving of nuclear spins. We use these gates to realize a 10-qubit quantum register consisting of the electron spin of a nitrogen-vacancy center and 9 nuclear spins in diamond. We show that the register is fully connected by generating entanglement between all 45 possible qubit pairs, and realize genuine multipartite entangled states with up to 7 qubits. Finally, we investigate the register as a multi-qubit memory. We show coherence times up to 63(2) seconds - the longest reported for a single solid-state qubit - and demonstrate that two-qubit entangled states can be stored for over 10 seconds. Our results enable the control of large quantum registers with long coherence times and therefore open the door to advanced quantum algorithms and quantum networks with solid-state spin qubits.
 
\end{abstract}

\maketitle

\section{INTRODUCTION}

Electron and nuclear spins associated with single defects in solids provide a promising platform for quantum networks and quantum computations \cite{awschalom2018quantum, Zwanenburg_RMP2013}. In these hybrid registers, different types of spins fulfill different roles. Electron spins offer fast control \cite{de2010universal,Fuchs_science2009,christle2015isolated,seo2016quantum,sukachev2017silicon,iwasaki2015germanium,siyushev2017optical,becker2018all,pingault2017coherent,trusheim2018transform,rugar2018characterization} and high fidelity readout \cite{robledo2011high,Pla_nature2013,sukachev2017silicon}, and can be used to control and connect nuclear spins \cite{Cramer_NatureComm2016,waldherr2014quantum,Pla_nature2013,wolfowicz201629si,tosi2017silicon,metsch2019initialization,hensen2019silicon}. Furthermore, electron-electron couplings enable on-chip connectivity between defects \cite{dolde2013room,yamamoto2013strongly,tosi2017silicon}, whilst coupling to photons \cite{yang2016high,togan2010quantum,christle2017isolated,sipahigil2016integrated,trusheim2018transform,evans2018photon} allows for the realization of long-range entanglement links \cite{bernien2013heralded, hensen2015loophole, humphreys2018deterministic}. Nuclear spins provide additional qubits with long coherence times that can be used to store and process quantum states \cite{maurer2012room,muhonen2014storing, waldherr2014quantum,Cramer_NatureComm2016,dehollain2016bell,yang2016high,hensen2019silicon}. 

Recent experiments have demonstrated various schemes for high-fidelity two-qubit gates \cite{dehollain2016bell,van2012decoherence,taminiau2014universal,rong2015experimental, zaiser2016enhancing, unden2018revealing, huang2019experimental}, as well as basic quantum algorithms \cite{van2012decoherence, kalb2017entanglement} and error correction codes \cite{waldherr2014quantum, Cramer_NatureComm2016}. However, to date, these systems have been restricted to few-qubit registers: the largest reported entangled state contains 3 qubits \cite{waldherr2014quantum,Cramer_NatureComm2016,van2019multipartite}. Larger quantum registers are desired for investigating advanced algorithms and quantum networks \cite{terhal2015quantum, nickerson2013topological, nickerson2014scalable}. Such multi-qubit registers are challenging to realize due to the required gates that selectively control the qubits and at the same time decouple unwanted interactions in order to protect coherence in the complete register. 

\begin{figure}
    \centering
    \includegraphics[]{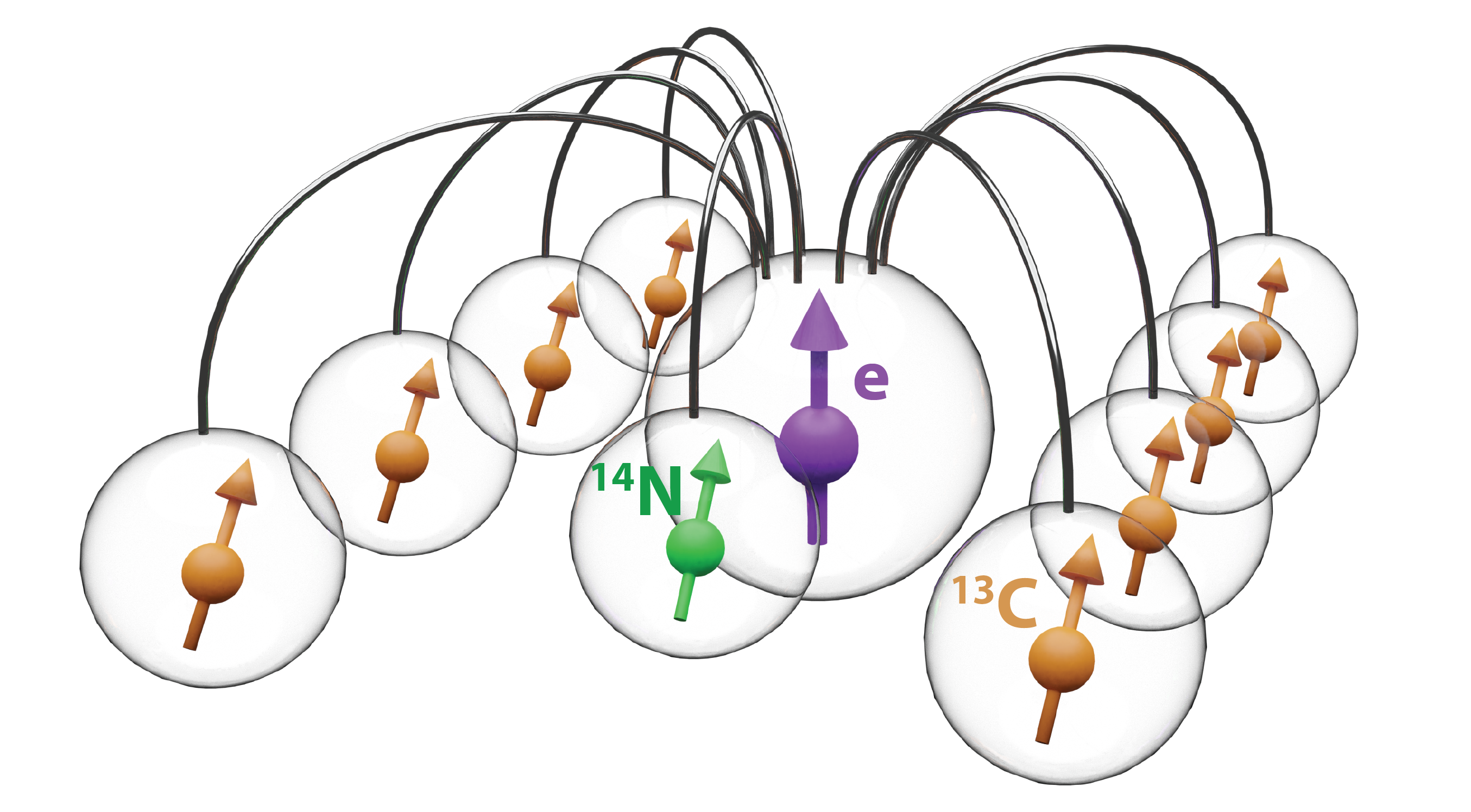}
    \caption{Illustration of the 10-qubit register developed in this work. The electron spin of a single NV center in diamond acts as a central qubit and is connected by two-qubit gates to the intrinsic $^{14}$N nuclear spin, and a further 8 $^{13}$C nuclear spins surrounding the NV center.}
    \label{fig:register}
\end{figure}

In this work, we develop a novel gate scheme based upon selective phase-controlled driving of nuclear spins interleaved with decoupling sequences on an electron spin. These gates enable high-fidelity control of hitherto inaccessible nuclear spin qubits. We combine these gates with previously developed control techniques \cite{robledo2011high,Cramer_NatureComm2016,abobeih2018one} to realize a 10-qubit register composed of a diamond nitrogen-vacancy (NV) center, its $^{14}$N nuclear spin and 8 $^{13}$C spins (Fig. \ref{fig:register}). We show that the register is fully connected by preparing entangled states for all possible pairs of qubits. Furthermore, by also decoupling nuclear-nuclear interactions through echo sequences, we generate $N$-qubit Greenberger-Horne-Zeilinger (GHZ) states, and witness genuine multipartite entanglement for up to 7 spins. Finally, we investigate the coherence properties of the register. We show that single qubit states can be stored for up to 63(2) seconds and two-qubit entangled states can be stored for over 10 seconds. 

\section{TWO-QUBIT GATES: THEORY}\label{sec:ddrf}

We consider an NV center in diamond and surrounding $^{13}$C nuclear spins. To realize a multi-qubit register, we design single-qubit gates and electron-nuclear two-qubit gates to control the NV $^{14}$N spin and several individual $^{13}$C spins. Key challenges in these hybrid systems of multiple coupled spins are to maintain coherence on the electron spin qubit and to avoid unwanted crosstalk. In particular, the electron spin continuously couples to all $^{13}$C spins through the hyperfine interaction, and the dynamics of the electron spin and nuclear spins typically occur on very different timescales \cite{van2012decoherence}. To address these issues, a variety of decoherence-protected gates, in which decoupling sequences on the electron spin are combined with nuclear spin control, have been investigated \cite{van2012decoherence, Kolkowitz_PRL2012, Taminiau_PRL2012, Zhao_NatureNano2012, taminiau2014universal, zaiser2016enhancing, wang2017delayed, pfender2017nonvolatile, unden2018revealing}. Here we develop and demonstrate a novel electron-nuclear two-qubit gate based upon phase-controlled radio-frequency (RF) driving of nuclear spins, interleaved with dynamical decoupling (DD) of the electron spin. We will refer to this scheme as a DDRF gate. Our scheme enables the control of additional $^{13}$C spins while offering improved flexibility in dynamical decoupling to optimize the electron spin coherence and avoid unwanted crosstalk. 

To design a selective two-qubit gate, we utilize the hyperfine interaction which couples each nuclear spin to the electron spin.  As this interaction depends on the relative position of the spin to the NV, different nuclear spins can be distinguished by their precession frequencies \cite{Kolkowitz_PRL2012, Taminiau_PRL2012, Zhao_NatureNano2012}. In the interaction picture with respect to the electron energy splitting, and neglecting non-secular terms, the Hamiltonian describing the electron and a single $^{13}$C nuclear spin is given by \cite{Kolkowitz_PRL2012, Taminiau_PRL2012, Zhao_NatureNano2012}

\begin{equation}\label{eq:HeC}
    H = \omega_{L} I_{z} + A_{\parallel} S_{z} I_{z} + A_{\bot} S_{z} I_{x}, \end{equation}
where $\omega_L = \gamma B_z$ is the nuclear Larmor frequency set by the external magnetic field $B_z$ along the NV axis, $\gamma$ is the $^{13}$C gyromagnetic ratio, $S_{\alpha}$ and $I_{\alpha}$ are the spin-1 and spin-1/2 operators of the electron and nuclear spins respectively, and $A_{\parallel}$ and $A_{\bot}$ are the parallel and perpendicular hyperfine components.

\begin{figure}
    \centering
    \includegraphics[width=0.5\textwidth]{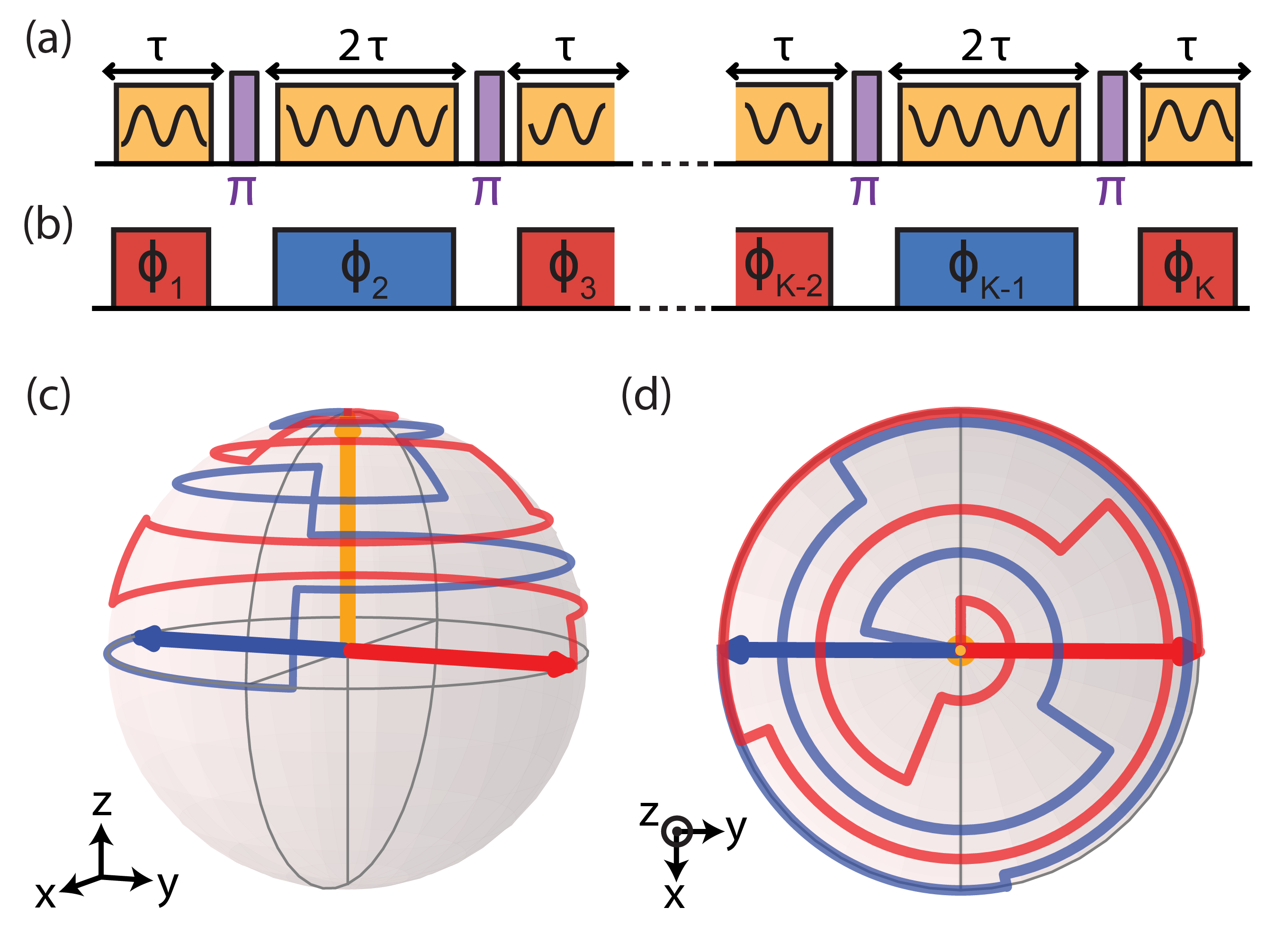}
    \caption{(a) Illustration of the pulse sequence employed to realize a DDRF gate. Dynamical decoupling pulses on the electron spin (purple) are interleaved with RF pulses (yellow) which selectively drive a single nuclear spin. (b) Illustration denoting the RF pulses which are resonant with the nuclear spin given that the electron is in the state $\ket{0}$ ($m_{s} = 0$, blue) or $\ket{1}$ ($m_{s} = -1$, red) at the start of the two-qubit gate. The phase of each RF pulse is adapted to create the desired nuclear spin trajectory accounting for periods of free precession, according to Eq. \ref{eq:rfphases}. (c) Nuclear spin evolution on the Bloch sphere for an example case with $N = 8$ electron decoupling pulses. Starting from the initial state $\ket{\uparrow}$ (yellow), the blue (red) path shows the nuclear spin evolution for the case where the electron starts in the state $\ket{0}$ ($\ket{1}$). The final state vectors are anti-parallel along the equator: therefore, the gate is a maximally entangling two-qubit gate. (d) Top-down view of (c).}
    \label{fig:DDRF}
\end{figure}

To control the nuclear spin, we apply RF pulses of Rabi frequency $\Omega$, phase $\phi$ and frequency $\omega$. To target a specific nuclear spin, we set $\omega$ = $\omega_{1}$, where $\omega_{1} = \sqrt{(\omega_L - A_\parallel)^2 + A_\perp^2}$ is the nuclear spin precession frequency when the electron is in the $m_s = -1$ spin projection. In the following we assume $(\omega_{L}-\omega_1)\gg\Omega$, such that driving of the nuclear spin is negligible while the electron is in the $m_s = 0$ spin projection, and set $A_\perp = 0$ for simplicity (see the Supplemental Material \cite{supp} for the general case). Considering only the $m_s = \{0,-1\}$ subspace, with the addition of RF driving and in a rotating frame at the RF frequency, the Hamiltonian of Eq. \ref{eq:HeC} becomes \cite{van2012decoherence, supp}

\begin{equation}\label{eq:Hamiltonian}\begin{split}
    H = &\ket{0}\bra{0} \otimes (\omega_L - \omega_1)I_z \\ 
    & + \ket{1}\bra{1}\otimes \Omega(\cos(\phi)I_x + \sin(\phi)I_y),
\end{split}\end{equation}
where $\ket{0}$ ($\ket{1}$) indicates the electron $m_s = 0$ ($m_s = -1$) spin projection. In this picture, for the electron in state $\ket{0}$, the nuclear spin undergoes precession around the $\hat{z}$-axis at frequency $(\omega_L - \omega_{1}) = A_\parallel$. Conversely, while the electron is in the state $\ket{1}$, the nuclear spin is driven around a rotation axis in the $\hat{x}$-$\hat{y}$ plane defined by the phase of the RF field $\phi$. 

To simultaneously decouple the electron spin from the environment, we interleave the RF pulses in a sequence of the form ($\tau$ - $\pi$ - $2\tau$ - $\pi$ - $\tau$)$^{N/2}$, where $\pi$ is a $\pi$-pulse on the electron spin, $2\tau$ is the interpulse delay, and $N$ is the total number of electron decoupling pulses (Fig. \ref{fig:DDRF}(a)) \cite{Kolkowitz_PRL2012, Taminiau_PRL2012, Zhao_NatureNano2012}. We consider the evolution of the nuclear spin during this sequence separately for the two initial electron eigenstates: $\ket{0}$ and $\ket{1}$ \cite{Kolkowitz_PRL2012, Taminiau_PRL2012, Zhao_NatureNano2012}. We label each successive RF pulse by integer $k = 1,...,K$, where $K = N + 1$ is the total number of RF pulses. If the inital electron spin state is $\ket{0}$, only the even $k$ RF pulses will be resonant and drive the nuclear spin (Fig. \ref{fig:DDRF}(b)). Conversely, for initial state $\ket{1}$, the odd $k$ pulses are resonant. The desired nuclear spin evolution can now be created by setting the phases $\phi_k$ of the RF pulses.

We construct both an unconditional rotation (single-qubit gate) and a conditional rotation (two-qubit gate). To ensure that the sequential RF rotations build up constructively, the phases of each RF pulse should be set to account for the periods of nuclear spin precession between them, which build up in integer multiples of $\phi_\tau = (\omega_L-\omega_1)\tau$. For the case where the electron starts in the state $\ket{0}$ (even $k$), the required sequence of phases is $\phi_\tau, 3\phi_\tau, 5\phi_\tau,\dots$, while for the case where the electron starts in the state $\ket{1}$ (odd $k$) we require the sequence $0, 2\phi_\tau, 4\phi_\tau,\dots$. The required phases are therefore given by \cite{supp}

\begin{equation}\label{eq:rfphases}
    \phi_k' = \left\{ 
    \begin{array}{ll} 
    (k-1)\phi_\tau + \pi & \quad k\text{ odd} \\
    (k-1)\phi_\tau & \quad k\text{ even},
    \end{array}
    \right.
\end{equation}
where the (optional) $\pi$ phase shift for the odd $k$ sequence converts the unconditional rotation into a conditional rotation. By adding a further phase $\varphi$ to all pulses, we can also set the rotation axis of the gate. The RF pulse phases are thus summarized by $\phi_k = \varphi + \phi_k'$.

With this choice of phases, the total evolution of the two-qubit system is given by $V = V_z \cdot V_\text{CROT}$. Here, $V_z$ is an unconditional rotation of the nuclear spin around $z$ \cite{supp} and $V_\text{CROT}$ is a conditional rotation of the nuclear spin depending on the electron state, given by

\begin{equation}\label{eq:VCROT}\begin{split}
V_\text{CROT} = &\ket{0}\bra{0}\otimes R_\varphi(N\Omega\tau) \\
&+ \ket{1}\bra{1}\otimes R_\varphi(-N\Omega\tau),
\end{split}\end{equation}
where $R_\varphi(\theta) = e^{-i\theta (\cos(\varphi)I_x + \sin(\varphi)I_y)/\hbar}$. $V_\text{CROT}$ describes a controlled rotation of the nuclear spin with tuneable rotation angle (set by $N$, $\Omega$ and $\tau$) and rotation axis (set by $\varphi$). Setting $N\Omega \tau = \pi/2$, a maximally entangling two-qubit operation is achieved, equivalent to a controlled-not (CNOT) gate up to local rotations. Example dynamics for a nuclear spin evolving under such a sequence are shown in Figs. \ref{fig:DDRF}(c) and (d). 

\begin{figure}
    \centering
    \includegraphics[width=0.5\textwidth]{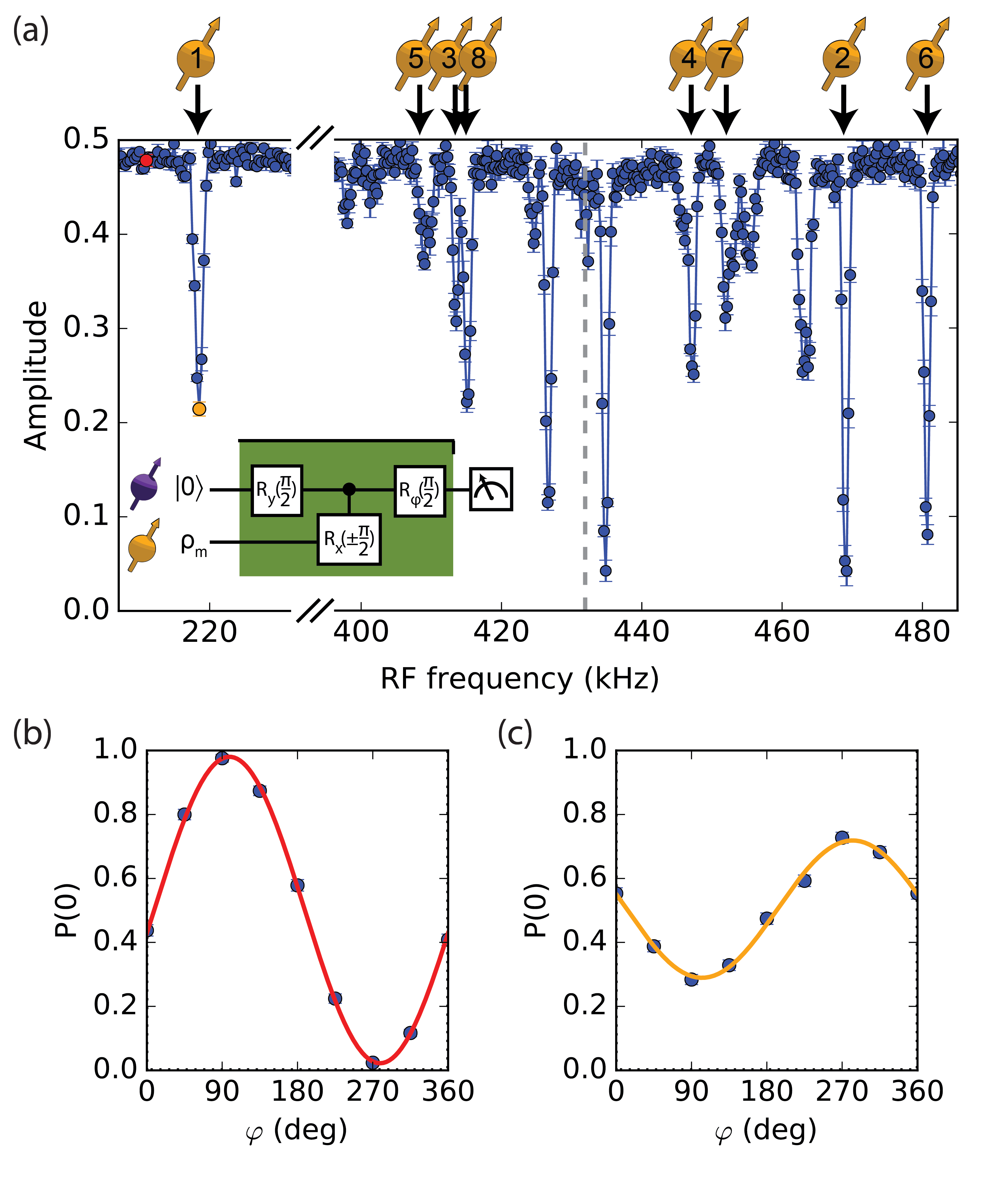}
    \caption{(a) Nuclear spin spectroscopy. After preparing the electron in a superposition state, the DDRF gate (controlled $\pm \pi/2$ rotation, see Eq. \ref{eq:VCROT}) is applied for different RF frequencies $\omega$. The electron spin is then measured along a basis in the equatorial plane defined by angle $\varphi$ (see inset). Each data point in (a) corresponds to the fitted amplitude A of the function $f(\varphi) = a + A \cos(\varphi + \varphi_{0})$, where $\varphi$ is swept from $0$ to $360$ deg and $\varphi_{0}$ accounts for deterministic phase shifts induced on the electron by the RF field. By fitting the amplitude, we distinguish such deterministic phase shifts from loss of coherence due to entangling interactions. The signals due to interaction with the 8 $^{13}$C spins used in this work are labelled. The dashed gray line indicates the $^{13}$C Larmor frequency $\omega_L$. A detailed analysis of the spectrum is given in the Supplemental Material \cite{supp}. (b,c) Example phase sweeps for two data points highlighted in red (b) and orange (c) in (a). Solid lines are fits to $f(\varphi)$. The DDRF gate parameters are $N = 48$ and $\tau = 8\tau_L$, where $\tau_L = 2\pi/\omega_L$ ($\approx2.3\,$\textmu s).}
    \label{fig:DDRFspectroscopy}
\end{figure}

\begin{figure*}
    \centering
    \includegraphics[width=0.8\textwidth]{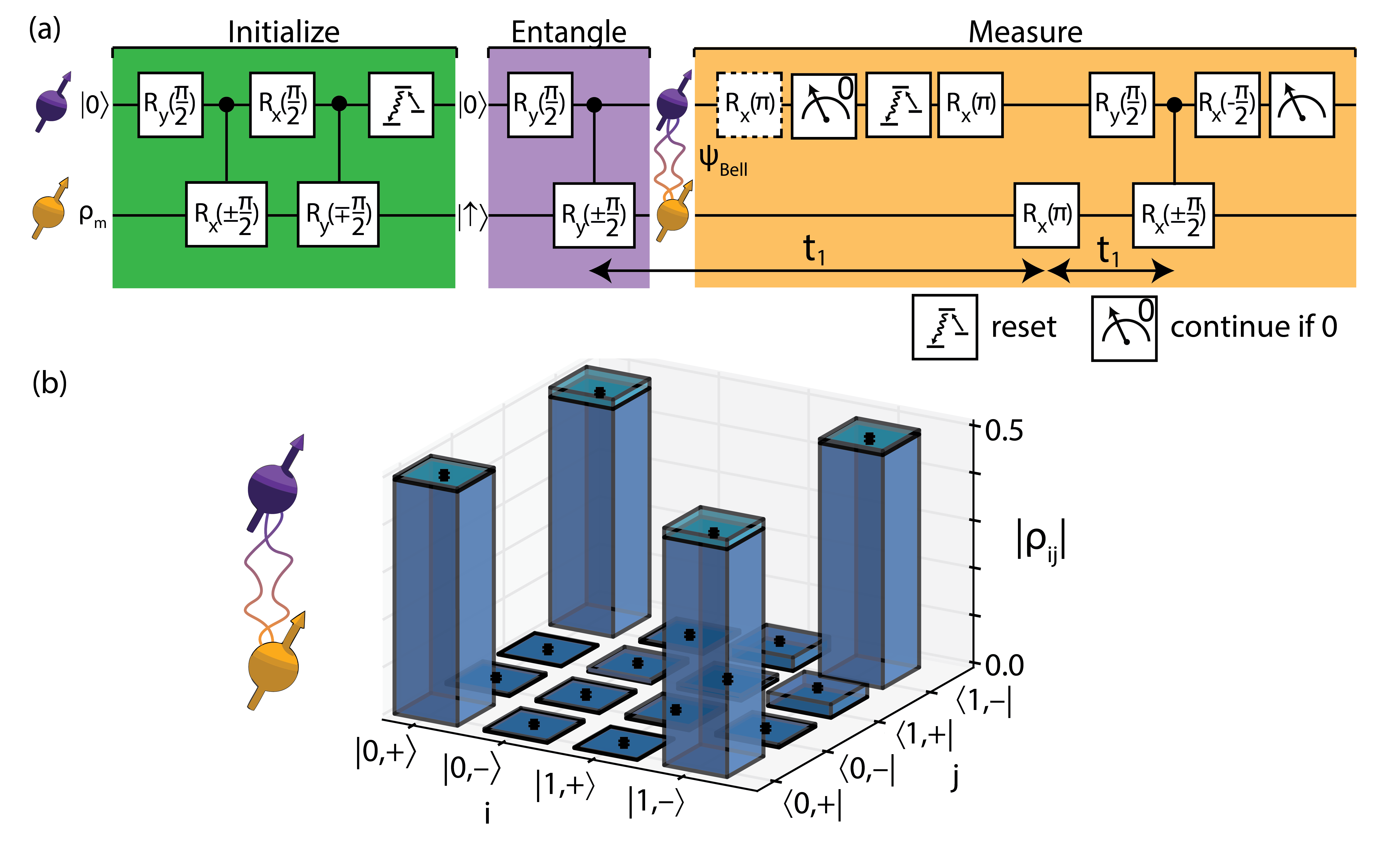}
    \caption{(a) Experimental sequence for preparation of an electron-nuclear Bell state and measurement of the expectation value of the two-qubit operator $ZX$. A series of single and two-qubit gates are used to initialize the nuclear spin \cite{taminiau2014universal,Cramer_NatureComm2016}. A subsequent $\pi/2$ rotation and two-qubit gate generate the Bell state $\ket{\psi_\text{Bell}} = (\ket{0+} + \ket{1-})/\sqrt{2}$. A measurement of the electron spin in the $Z$-basis is followed by an $X$-basis measurement of the nuclear spin through the electron spin. These measurements are separated by a nuclear spin echo, which is implemented to mitigate dephasing of the nuclear spin. The entire sequence is applied with and without an additional electron $\pi$-pulse (dashed box) before the first electron readout, in order to reconstruct the electron state while ensuring that the measurement does not disturb the nuclear spin state \cite{Cramer_NatureComm2016, kalb2017entanglement}. (b) Density matrix of the electron-nuclear state after applying the sequence shown in (a) to qubit C1, reconstructed with state tomography. The DDRF gate parameters are $N = 8$, $\tau = 17\tau_L \approx 39.4\,$\textmu s, $\Omega/2\pi = 1.09(3)\,$kHz, and the total gate duration is $629\,$\textmu s, compared with the nuclear spin $T_{2}^*= 12.0(6)\,$ms. We use erf pulse envelopes with a $7.5\,$\textmu s rise / fall time for each RF pulse to mitigate pulse distortions induced by the RF electronics \cite{supp}. The fidelity with the target Bell state is measured to be $\mathcal{F}_\text{Bell} = 0.972(8)$. Lighter blue shading indicates the density matrix for the ideal state $\ket{\psi_\text{Bell}}$.}
    \label{fig:eC}
\end{figure*}

Our design has several advantages. First, the gate allows nuclear spins with small or negligible $A_\perp$ to be controlled, thereby increasing the number of accessible nuclear spin qubits. Second, because the targeted dynamics are achieved by setting the RF phases and amplitudes, the interpulse delay $\tau$ of the decoupling sequence can be freely optimized to protect the electron coherence. This is in contrast to the gates described in van der Sar et al. \cite{van2012decoherence}, for which $\tau$ is restricted to a specific resonance condition for each spin, making multi-qubit control challenging. Third, because our method does not rely on an average frequency shift over the two electron spin states \cite{taminiau2014universal}, our gates can also be used for selective control of nuclear spins coupled to spin-1/2 defects (such as the negatively-charged group-IV color centers \cite{metsch2019initialization,muller2014optical,sukachev2017silicon,evans2018photon,siyushev2017optical,rugar2018characterization,thiering2018ab}), and via a contact hyperfine coupling, such as for donor spins in silicon \cite{dehollain2016bell} and SiMOS quantum dots \cite{hensen2019silicon}. Finally, because control is achieved through the RF field, a multitude of avenues for future investigation are opened up, such as parallelizing gates by frequency multiplexing and using shaped and composite pulses to mitigate dephasing and crosstalk \cite{vandersypen2005nmr, dolde2014high, rong2015experimental}. 

\section{TWO-QUBIT GATES: EXPERIMENT}

Our experiments are performed at $3.7\,$K using a single NV center in diamond with natural abundance of carbon isotopes ($1.1\%$ $^{13}$C). Further details of the sample and experimental setup can be found in the Supplemental Material \cite{supp}. As a starting point, we use the DDRF gate to identify and characterize $^{13}$C nuclear spin qubits surrounding the NV center. If the electron spin is prepared in a superposition state and the RF frequency is resonant with a nuclear spin in the environment, the entangling interaction (Eq. \ref{eq:VCROT}) decoheres the electron spin. Therefore, varying the RF frequency ($\omega$) performs spectroscopy of the nuclear spin environment. Fig. \ref{fig:DDRFspectroscopy} shows that multiple dips in the electron coherence can be observed, indicating selective interactions with several individual nuclear spins. Importantly, like other RF-based approaches \cite{zaiser2016enhancing, pfender2017nonvolatile}, the DDRF sequence is sensitive to nuclear spins with small or negligible $A_\perp$. Besides extending the number of qubits that can be controlled with a single NV center, this also enables the detection of additional spins when using the NV as a quantum sensor, which we exploit in parallel work to realize 3D imaging of large spin clusters \cite{Abobeih_arXiv2019}.   

\begin{figure*}
    \centering
    \includegraphics[width=0.8\textwidth]{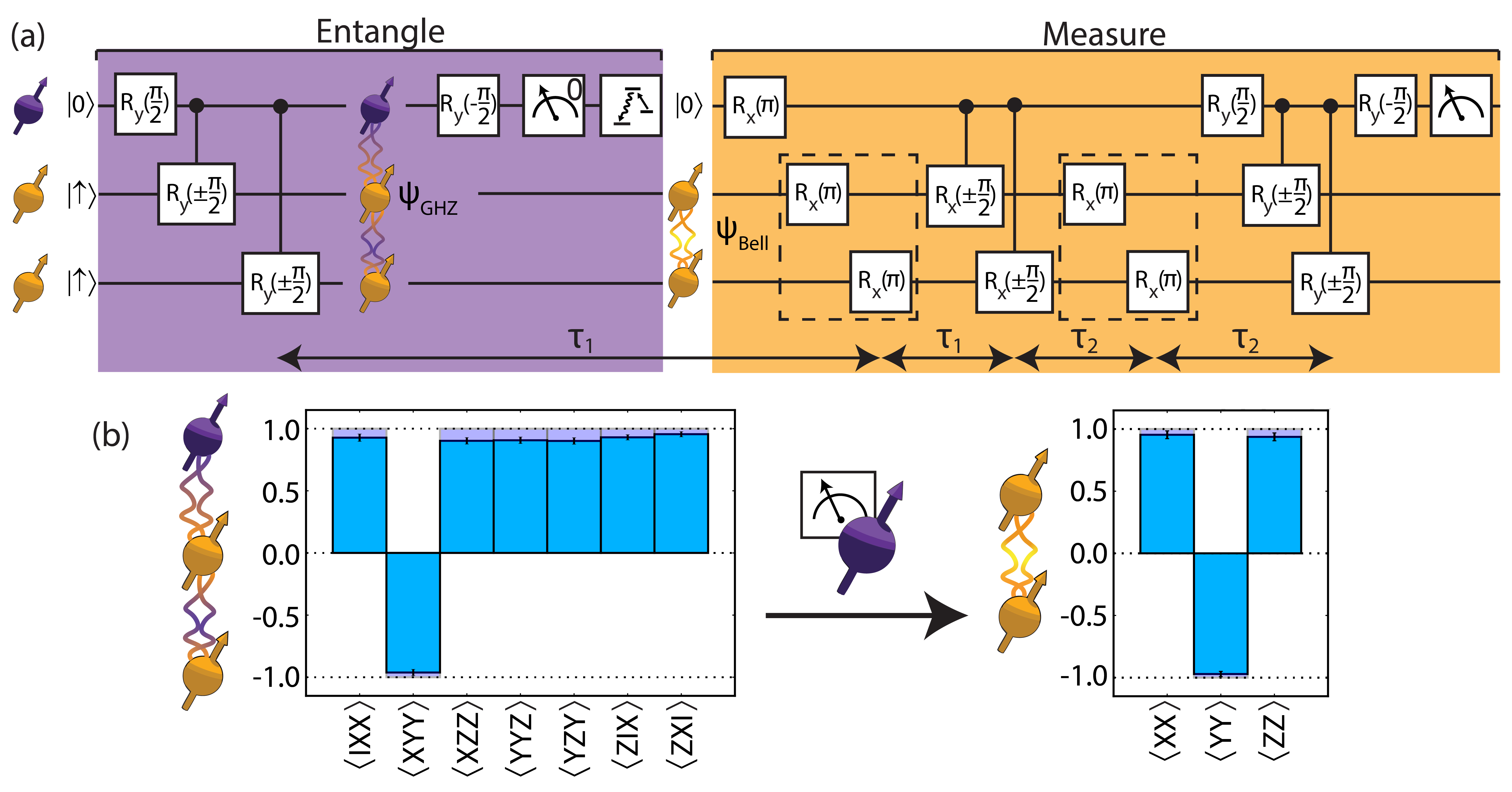}
    \caption{(a) Experimental sequence for the preparation of a nuclear-nuclear Bell state and measurement of the two-qubit operator $ZZ$. After preparation of the electron-nuclear-nuclear GHZ state $\ket{\text{GHZ}_3} = (\ket{0++} + \ket{1--})/\sqrt{2}$, an $X$-basis measurement on the ancilla (electron spin) projects the nuclear spins into the Bell state $\ket{\Phi^+} = (\ket{++} + \ket{--})/\sqrt{2}$.  Measurement of the two-qubit correlations between the nuclear spins is then performed by parity measurements through the electron spin. Spin echoes (dashed boxes) built into the measurement sequence protect the nuclear spins from dephasing errors. (b) Measured expectation values (non-zero terms of the ideal state only) for the electron-nuclear-nuclear state $\ket{\text{GHZ}_3}$, and for the nuclear-nuclear state $\ket{\Phi^+}$. Blue (purple) bars show the experimental (ideal) expectation values for each operator. The nuclear-nuclear correlations are well preserved after a nondestructive measurement of the electron spin in the $X$-basis.} 
    \label{fig:CC}
\end{figure*}

To verify the control offered by the DDRF two-qubit gate, we first demonstrate high fidelity ancilla-based initialization and readout by preparation and tomography of a maximally entangled electron-nuclear state. To test the gate, we select a $^{13}$C spin (spin C1, Fig. \ref{fig:DDRFspectroscopy}) with a strong parallel hyperfine component of $A_\parallel/2\pi = 213.154(1)\,$kHz, but a weak perpendicular hyperfine component $A_\perp/2\pi = 3.0(4)\,$kHz \cite{supp}. We exploit the freedom in choosing the interpulse delay by setting $\tau$ to an integer multiple of the $^{13}$C Larmor period, $\tau_L = 2\pi/\omega_L$, so that unwanted interactions between the electron spin and other $^{13}$C spins in the environment are effectively decoupled \cite{childress2006coherent,abobeih2018one}. 
  
The sequence to perform the state preparation and tomography experiment is shown in Fig. \ref{fig:eC}(a) \cite{taminiau2014universal,Cramer_NatureComm2016}. We first initialize the electron spin in the state $\ket{0}$ by resonant optical excitation \cite{robledo2011high}. We then swap the state of the electron spin onto the $^{13}$C spin and reset the electron spin. Next, we prepare the electron in a superposition state before performing the DDRF controlled-rotation gate, ideally preparing the electron-nuclear Bell state $\ket{\psi_\text{Bell}} = (\ket{0+} + \ket{1-})/\sqrt{2}$, where $\ket{\pm} = (\ket{\downarrow} \pm \ket{\uparrow})/\sqrt{2}$. 

To perform quantum state tomography on the two-qubit state, we first measure the electron spin along a chosen axis by appropriate basis rotations followed by $Z$-basis optical readout \cite{robledo2011high}. To mitigate potential dephasing of the nuclear spin induced by the electron spin measurement, we make the electron spin measurement non-destructive by using a short, weak laser pulse and conditioning progression of the sequence on the outcome $\ket{0}$ (detection of a photon) \cite{Cramer_NatureComm2016, kalb2017entanglement}. Following appropriate basis rotations, we then use the electron spin to measure the nuclear spin in the $X$-basis \cite{Cramer_NatureComm2016}. In this measurement the electron is read out in a single-shot with average fidelity $0.945(2)$ \cite{robledo2011high}. In order to reconstruct the full electron-nuclear state, we perform the sequence with and without an additional electron $\pi$-pulse before the first readout \cite{supp}. 

The reconstructed density matrix from quantum state tomography is shown in Fig. \ref{fig:eC}(b). The prepared state $\rho$ exhibits a fidelity with the target Bell state of $\mathcal{F}_\text{Bell} = \bra{\psi_\text{Bell}}\rho\ket{\psi_\text{Bell}} =  0.972(8)$. Based upon a simple depolarizing noise model, we estimate the two-qubit gate fidelity to be $\mathcal{F}_\text{gate} = 0.991(9)$ \cite{supp}. Additional characterization measurements in combination with numerical simulations indicate that the remaining infidelity can be mostly attributed to electron spin dephasing due to noise from the electronic hardware \cite{supp}.

\section{A 10-QUBIT SOLID-STATE SPIN REGISTER}

We now show how the combination of our DDRF gate with previously developed gates and control techniques \cite{taminiau2014universal, Cramer_NatureComm2016} enables high-fidelity control of a 10-qubit hybrid spin register associated to a single NV-center. Our register is composed of the electron and $^{14}$N spins of the NV-center, along with 8 $^{13}$C nuclear spins (Fig. \ref{fig:register}). 
Our quantum register is connected via the central electron spin. To demonstrate this, we first show that all nuclear spins can be entangled with the electron spin by following the protocol shown in Fig. \ref{fig:eC}(a). For the case of the nitrogen spin, initialization is performed by a measurement-based scheme which heralds the preparation in a particular eigenstate. Compared to previous work \cite{Kalb_NatureComm2016}, we realize an improved initialization fidelity ($\mathcal{F}_\text{init} = 0.997(11)$) by pre-preparing the electron in the $m_{s}=-1$ state instead of a mixed state of $m_{s}=-1$ and $+1$, and by repeating the measurement-based initialization sequence twice \cite{supp}. After initialization, we work in the $m_I = \{0,-1\}$ subspace, and perform operations analogous to those for the $^{13}$C nuclear spins, including the two-qubit gates using the DDRF scheme. Genuine entanglement is probed by measuring the non-zero matrix elements of the target state, and confirmed by negativity of the entanglement witness $\mathcal{W}_\text{Bell} = \mathds{1} - 2\ket{\psi_\text{Bell}}\bra{\psi_\text{Bell}}$ \cite{guhne2009entanglement}.

Next, we show that the register is fully connected by preparing entangled states for all possible pairs of spins. To prepare nuclear-nuclear entanglement, we implement a probabilistic measurement-based scheme \cite{pfaff2013demonstration}, as shown in Fig. \ref{fig:CC}(a). We first prepare a three-qubit GHZ state comprising the electron and two nuclear spins, $\ket{\text{GHZ}_3} = (\ket{0++} + \ket{1--})/\sqrt{2}$, before performing a non-destructive $X$-basis measurement on the electron spin. The measurement ideally prepares the Bell state $\ket{\Phi^+} = (\ket{++} + \ket{--})/\sqrt{2}$ on the targeted pair of nuclear spins. Finally, we measure the necessary expectation values in order to reconstruct the non-zero matrix elements of this state and confirm bipartite entanglement (Fig. \ref{fig:CC}(b)).

\begin{figure}
    \centering
    \includegraphics[width=0.5\textwidth]{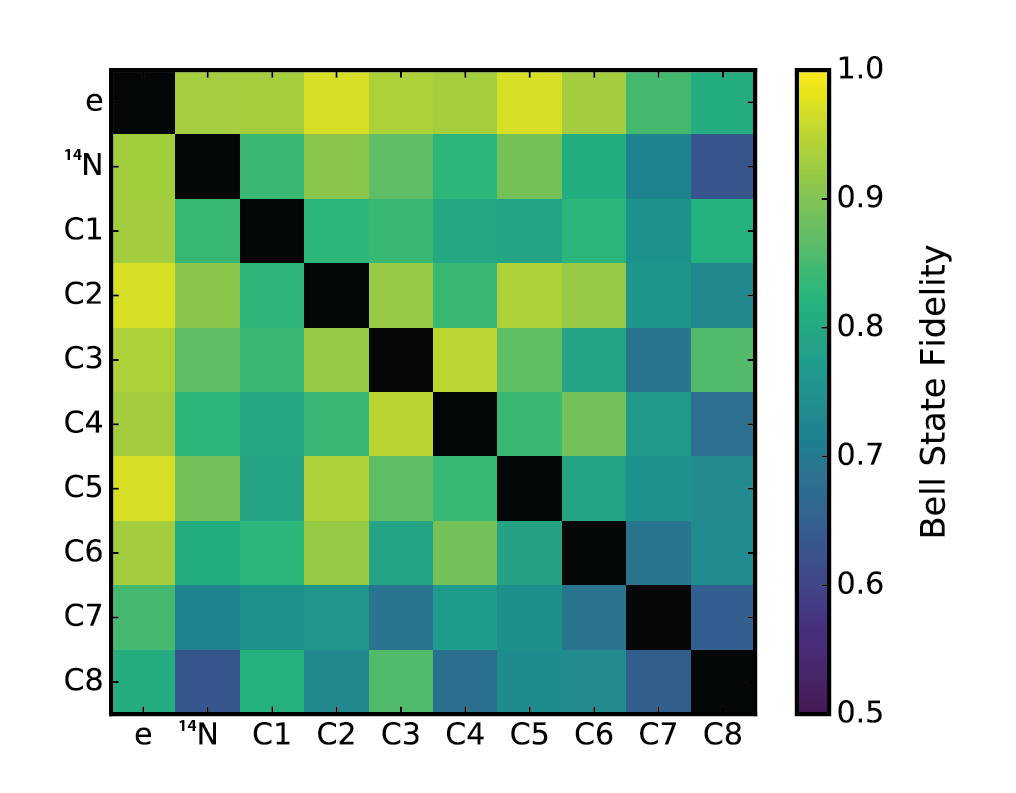}
    \caption{Measured Bell state fidelities for all pairs of qubits in the 10-qubit register. Genuine entanglement is confirmed in all cases, as witnessed by a fidelity exceeding 0.5 with the target state. Qubits C1, C7, C8 and $^{14}$N are controlled using DDRF gates (section \ref{sec:ddrf}). Qubits C2, C3, C4, C5, and C6 are controlled using the gates described in Taminiau et al. \cite{taminiau2014universal}.} 
    \label{fig:grid}
\end{figure}

The measured Bell state fidelities, ranging from 0.63(3) to 0.97(1), are shown in Fig. \ref{fig:grid}. We attribute the variations in the measured values to differences in the two-qubit gate fidelities for each spin. In particular, the lower values measured for $^{13}$C spins C7 and C8 are due to short coherence times in combination with long two-qubit gate durations, necessitated by close spectral proximity to other spins \cite{supp}. All data is measured using a single set of gate parameters, and using a single hardware configuration, rather than separately optimizing for each pair of qubits.

\section{GENERATION OF N-QUBIT GHZ STATES}

\begin{figure*}
    \centering
    \includegraphics[width=0.9\textwidth]{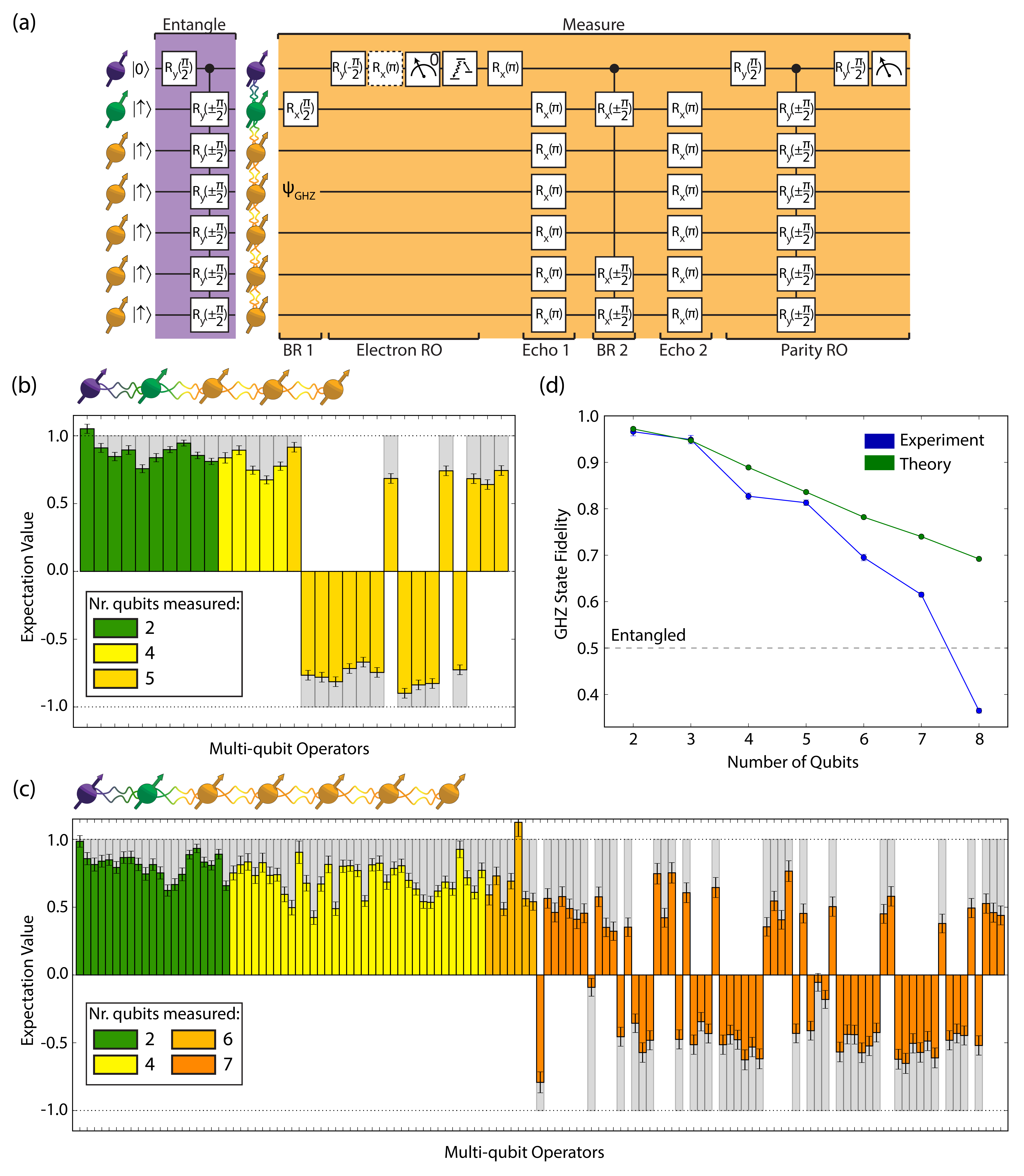}
    \caption{(a) Experimental sequence for the preparation of a 7-qubit GHZ state $\ket{\text{GHZ}_7}$ (purple) and measurement of the 7-qubit operator $XYYYYZZ$ (yellow). The measurement sequence is broken down into basis rotations (BR 1,2), an electron readout (RO), nuclear spin echoes (Echo 1,2), and a parity readout of the nuclear spins. All operations are applied sequentially (in the same way as shown in Fig. \ref{fig:CC}), but some are shown in parallel for clarity. (b-c) Bar plots showing the measured expectation values (non-zero terms of the ideal state only) after preparing the 5-spin (b) and 7-spin (c) GHZ states. The colors indicate the number of qubits for which the measurement basis is not identity, shown in the inset. Gray bars show the ideal expectation values. See the Supplemental Material \cite{supp} for the operator corresponding to each bar. The fidelity with the target state is 0.804(6) (b) and 0.589(5) (c), confirming genuine multipartite entanglement in both cases. (d) Plot of GHZ state fidelity against the number of constituent qubits. A value above 0.5 confirms genuine $N$-qubit entanglement. The blue points are the measured data, while the green points are theoretical predictions assuming a simple depolarizing noise model whose parameters are extracted from single- and two-qubit experiments. Numerical values are given in the Supplemental Material \cite{supp}.} 
    \label{fig:Nqubit}
\end{figure*}

Quantum information processing tasks such as computations and error correction will require the execution of complex algorithms comprising a large number of qubits. An important requirement for a quantum processor is thus the ability to perform operations on many of its constituents within a single algorithm. We test this capability by generating $N$-qubit GHZ type states, defined as 
\begin{equation}\label{eq:GHZN}
    \ket{\text{GHZ}_N} = \frac{1}{\sqrt{2}} \left(\ket{0}\otimes\ket{+}^{\otimes (N-1)} + \ket{1}\otimes\ket{-}^{\otimes (N-1)}\right).
\end{equation}

To generate such states, we follow the sequence shown in Fig. \ref{fig:Nqubit}(a). First, $N-1$ nuclear spins are initialized in the state $\ket{\uparrow}$. Next, we prepare the electron spin in a superposition state, and perform sequential controlled rotation gates between the electron and nuclear spins.

Characterizing the full quantum state for a system of this size is an expensive task due to the dimensionality of the associated Hilbert space. However, we can determine if the state exhibits genuine multipartite entanglement of all $N$ qubits using an entanglement witness with a reduced subset of measurement bases \cite{guhne2009entanglement}. For a GHZ state with system size $N$, there exist 2$^{N}$ operators from which the non-zero elements of the density matrix can be reconstructed by linear inversion, and from which a fidelity with the target state can be calculated. Negativity of the entanglement witness $\mathcal{W}_\text{GHZ} = \mathds{1}/2 - \ket{\text{GHZ}_{N}}\bra{\text{GHZ}_{N}}$ heralds genuine multipartite entanglement \cite{guhne2009entanglement}. We measure the required expectation values of products of Pauli operators on the register via multi-qubit parity measurements. In these experiments, the readout sequence is modified slightly. Prior to the readout of the electron spin state, we rotate the nitrogen such that the desired measurement basis is mapped to the $Z$-basis. This ensures that the population in the measurement basis is protected from dephasing caused by optically reading out the electron spin \cite{blok2014manipulating, supp}.

As the number of qubits is increased, a new challenge arises: the total sequence time becomes comparable to, or even exceeds the natural dephasing times ($T_2^*$) of the nuclear spins. In order to preserve the nuclear spin coherence, we insert spin (Hahn) echo pulses (RF $\pi$-pulses) into the sequence to refocus each spin at the point of the next operation performed upon it. In the Supplemental Material \cite{supp}, we derive a general solution that can be used to algorithmically construct echo sequences that avoid any overlap in gates and that minimize idle time with the electron spin in a superposition.

In Figs. \ref{fig:Nqubit}(b,c), we show measurements for $N=5$ and $N=7$ qubits. In Fig. \ref{fig:Nqubit}(d), we present the measured fidelities with the target GHZ states for 2 to 8 qubits, along with theoretical values as predicted by a depolarizing-noise model based on the individual two-qubit gate fidelities \cite{supp}. The growing discrepancy between the measured and predicted values for larger $N$ suggests residual crosstalk between the qubits, which is not taken into account in the model. For registers comprising up to 7 spins we observe negativity of the witness $\mathcal{W}_\text{GHZ}$, revealing genuine $N$-qubit entanglement of up to 7 qubits with high statistical significance. 

\section{A LONG LIVED QUANTUM MEMORY}

The nuclear spin qubits surrounding the NV center are promising candidates for quantum memories with long coherence times \cite{maurer2012room,KalbQmem}. Here we investigate the coherence properties of single-qubit states and two-qubit entangled states under dynamical decoupling. We show that single qubit states can be stored for up to 63(2) seconds and two-qubit entangled states can be stored for over 10 seconds, thereby demonstrating a long-lived quantum memory within our 10-qubit register.

We first investigate the coherence of individual nuclear spin qubits under dynamical decoupling. After initializing the nuclear spin in the state $\ket{+}$, we prepare the electron in the state $\ket{1}$ (electron $T_{1}=3.6(3) \ \times \ 10^{3}$ s \cite{abobeih2018one}). This has two effects. Firstly, it allows us to perform selective RF $\pi$-pulses on the target nuclear spin. Secondly, the magnetic field gradient imposed by the electron-nuclear hyperfine interaction induces a frozen core, which suppresses flip-flop interactions between nuclear spins \cite{khutsishvili1962spin,guichard2015decoherence} and thereby reduces the noise the spins are exposed to. 

The observed spin-echo coherence times $T_{2}^{\alpha=1}$, with $\alpha$ the number of RF pulses, vary between $0.26(3)$ s to $0.77(4)$ s for the 8 $^{13}$C spins. For the $^{14}$N spin we find $2.3(2)$ s, consistent with the smaller gyromagnetic ratio by factor $3.4$. The range of coherence times observed for the $^{13}$C spins is likely caused by differences in the microscopic environment of each spin. In particular, $^{13}$C spins close to the NV center are in the heart of the frozen core, and, generally tend to couple predominantly to the part of the spin environment for which the dynamics are also suppressed most strongly. Spins farther from the NV tend to couple more strongly to the spin environment outside the frozen core. This explanation is consistent with the observation that the spin with the longest $T_{2}^{\alpha=1}$ of $0.77(4)$ s is located closest (C1, $r= 0.53(5)$ nm \cite{Abobeih_arXiv2019}) to the vacancy lattice site, while the shortest $T_{2}^{\alpha=1}$ of $0.26(3)$ s is found for a spin at a larger distance (C8, $r= 1.04(4)$ nm \cite{Abobeih_arXiv2019}).             

\begin{figure*}
    \centering
    \includegraphics[width=0.8\textwidth]{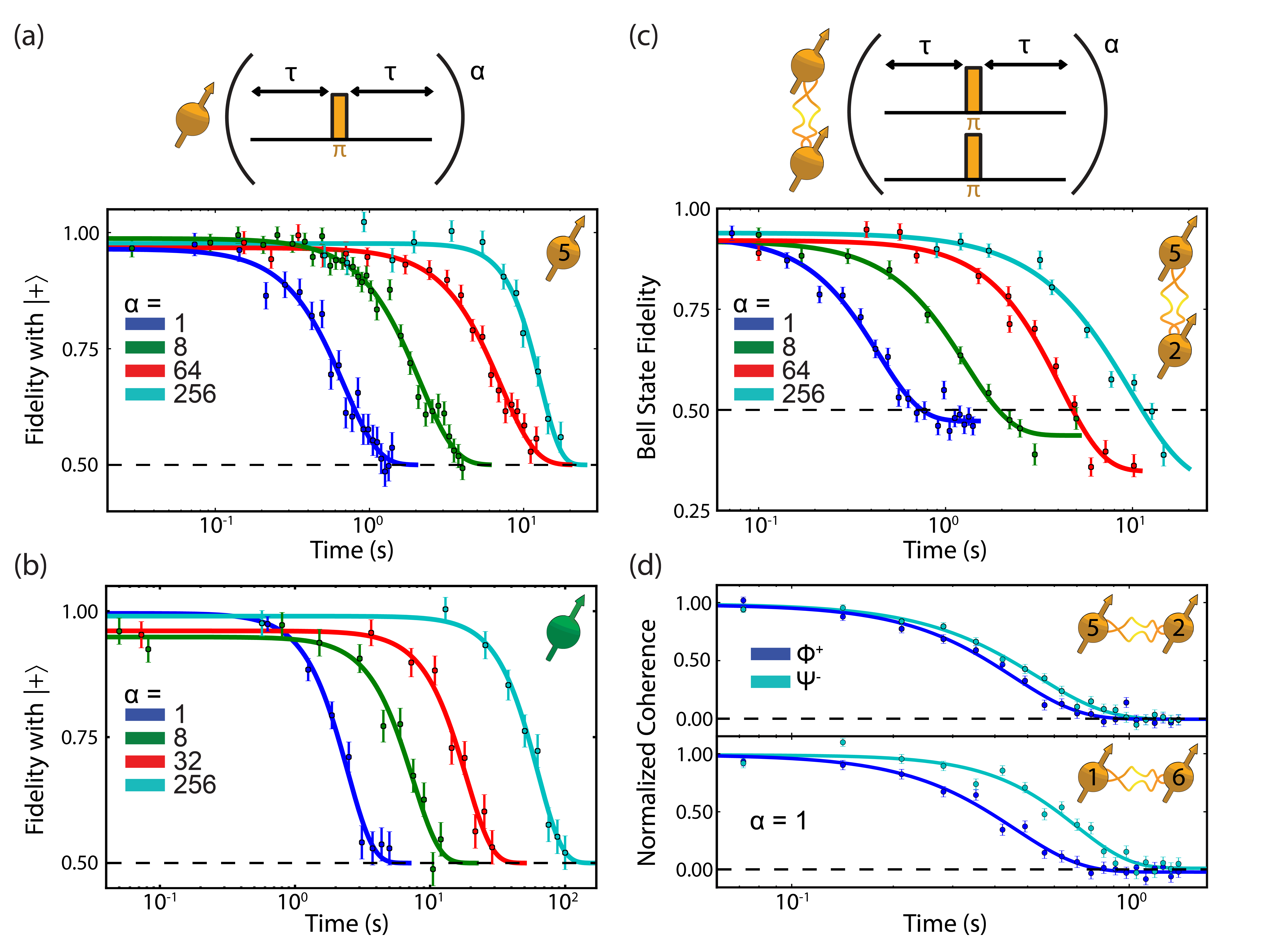}
    \caption{(a) Dynamical decoupling of a single $^{13}$C spin (C5) with $\alpha$ = 1, 8, 64 and 256 pulses. The nuclear spin is initialized and read out in the $X$-basis. We fit the data to the function $f(t) = A + B e^{-(t/T)^{n}}$ (solid lines), where $A=0.5$, and $B$, $T$ and $n$ are fit parameters which account for the decay of the fidelity due to interactions with the nuclear spin bath, external noise and pulse errors. With 256 decoupling pulses, the fitted coherence time is $T_2^{\alpha=256} = 12.9(4)\,$s. (b) Dynamical decoupling of the $^{14}$N spin with $\alpha$ = 1, 8, 32 and 256 pulses. The nuclear spin is initialized and read out in the $X$-basis. Solid lines are again fit to $f(t)$ with $A=0.5$. With 256 decoupling pulses, the fitted coherence time is $T_2^{\alpha=256} = 63(2)\,$s. (c) Dynamical decoupling of a pair of $^{13}$C spins prepared in the Bell state $\ket{\Phi^+}$. Solid lines are fits to $f(t)$, but with $A$ as a free parameter. With 256 decoupling pulses, genuine two-qubit entanglement is witnessed at times up to 10.2 s, where we observe a fidelity of 0.57(2) with the target Bell state.  (d) Normalized coherence $(\langle XX \rangle \pm \langle YY \rangle)/2\mathcal{N}$, where $\mathcal{N}$ is a normalization factor, for two pairs of $^{13}$C spins prepared in both the even and odd parity Bell states $\ket{\Phi^+} = (\ket{\downarrow\downarrow} + \ket{\uparrow\uparrow})/\sqrt{2}$ and $\ket{\Psi^-} = (\ket{\downarrow\uparrow} - \ket{\downarrow\uparrow})/\sqrt{2}$. Solid lines are fits to $f(t)$ with $A=0$ and $B=1$. For pair 1, the fitted decay times are 0.45(2) s and 0.54(1) s for the states $\ket{\Phi^+}$ and $\ket{\Psi^-}$ respectively. For pair 2, the equivalent values are 0.46(2) s and 0.70(3) s.} 
    \label{fig:Coherence}
\end{figure*}

As expected, increasing the number of decoupling pulses leads to an increase in the measured coherence times. Examples of the measured decay curves are shown in Fig. \ref{fig:Coherence}(a) (C5) and (b) ($^{14}$N), where $\alpha$ is varied from 1 to 256. For $\alpha=256$ pulses, the decay time of C5 reaches $T_2^{\alpha=256} = 12.9(4)\,$s, while for the $^{14}$N spin, we measure $T_2^{\alpha=256} = 63(2)\,$s. For the other $^{13}$C spins for which we measure $T_2^{\alpha=256}$, we find a range of values from $4(1)$ to $25(4)$ seconds \cite{supp}. These coherence times are the longest reported for individual qubits in the solid state and exceed values for isolated nuclear spin qubits in isotopically purified materials \cite{maurer2012room,muhonen2014storing,yang2016high}. More importantly, however, in our register we realize these long coherence times while maintaining access to 10 coupled spin qubits.

We exploit the multi-qubit nature of the register to investigate the storage of entangled states of two $^{13}$C spin qubits. After preparing the Bell state $\ket{\Phi^+} = (\ket{++} + \ket{--})/\sqrt{2}$ following the sequence shown in Fig. \ref{fig:CC}(a), we again prepare the electron in the state $\ket{1}$. We then measure the Bell state fidelity as a function of total evolution time for $\alpha=1$ to $\alpha=256$ pulses. Note that since $\ket{\Phi^+}$ is an eigenstate of $ZZ$, its evolution is not affected by the coupling between the two qubits ($1.32(4)$ Hz \cite{Abobeih_arXiv2019}). The measured fidelities are plotted in Fig. \ref{fig:Coherence}(c). For $\alpha=256$ decoupling pulses, we confirm the preservation of entanglement for $>10\,$s, as quantified by a fidelity exceeding $0.5$ with the desired Bell state.

With the capability to store multi-qubit quantum states, it becomes important to consider additional effects that may affect their coherence, such as the presence of correlated noise. As a first experimental step towards understanding such effects, we use entangled states of nuclear spins to explore spatial correlations within the noise environment. We perform experiments on two pairs of $^{13}$C spins. We prepare two Bell states for each pair, one exhibiting even $ZZ$ parity, which, written in the $Z$-basis, is given by $\ket{\Phi^+} = (\ket{\downarrow\downarrow} + \ket{\uparrow\uparrow})/\sqrt{2}$, and another exhibiting odd $ZZ$ parity, $\ket{\Psi^-} = (\ket{\downarrow\uparrow}-\ket{\uparrow\downarrow})/\sqrt{2}$. The difference in the coherence times of these two states gives an indication of the amount of correlation in the noise experienced by the two spins \cite{kwiatkowski2018decoherence}. In the case of perfectly correlated noise, one would expect the state $\ket{\Phi^+}$ to decay at four times the single qubit decay rate (superdecoherence), while the state $\ket{\Psi^-}$ would form a decoherence-free subspace \cite{lidar1998decoherence,reiserer2016robust}. In contrast, for completely uncorrelated noise, the coherence times for the two states would be identical.

We measure the coherence times for the two Bell states, varying the total evolution time for the case of a single spin-echo pulse ($\alpha=1$) with the electron spin prepared in the state $\ket{1}$. In Fig. \ref{fig:Coherence}(d), we plot the normalized coherence signal for both Bell states and for both pairs of qubits. A statistically significant difference between the decay curves for the two Bell states is found for both pairs, where the odd-parity state $\ket{\Psi^-}$ decays more slowly than the even-parity state $\ket{\Phi^+}$, indicating partly correlated noise in the system. We can relate the size of the effect to the distance between the spins in the pairs, which has been characterized in separate work \cite{Abobeih_arXiv2019}. This reveals that the pair with a smaller separation (C1 and C6, distance $0.96(3)\,$nm) shows more correlation than the pair with a larger separation (C5 and C2, $1.38(7)\,$nm). This observation is consistent with the idea that proximity generally promotes correlated noise, although large deviations from this rule are expected to be possible for specific cases \cite{kwiatkowski2018decoherence}. Characterizing such correlated noise provides new opportunities to investigate the physics of decoherence in spin baths \cite{kwiatkowski2018decoherence}, and to develop and test quantum error correction schemes that are tailored for specific correlated noise in such systems \cite{monz201114, layden2019efficient}. 

\section{CONCLUSION}
In conclusion, we have developed a novel electron-nuclear two-qubit gate and applied these gates to realize a 10-qubit solid-state spin register that can store quantum states for up to one minute. The techniques developed in this work can be readily implemented for multi-qubit control in a variety of other donor and defect platforms, including spin-1/2 \cite{metsch2019initialization,muller2014optical,sukachev2017silicon,evans2018photon,siyushev2017optical,rugar2018characterization,thiering2018ab} and contact hyperfine \cite{dehollain2016bell,hensen2019silicon} systems, for which many previous gate designs are challenging to apply \cite{Kolkowitz_PRL2012, Taminiau_PRL2012, Zhao_NatureNano2012, taminiau2014universal}. Further improvements in selectivity and fidelity of the gates are anticipated to be possible by (optimal) shaping of the RF pulses  \cite{vandersypen2005nmr, dolde2014high, rong2015experimental} and by reducing electronic noise. Additionally, the use of direct RF driving opens the possibility to perform gates in parallel on multiple qubits. Combined with already demonstrated long-range optical entanglement \cite{bernien2013heralded,hensen2015loophole,humphreys2018deterministic}, our multi-qubit register paves the way for the realization of rudimentary few-node quantum networks comprising tens of qubits. This will enable the investigation of basic error correction codes and algorithms over quantum networks \cite{terhal2015quantum,nickerson2013topological,nickerson2014scalable}. Finally, looking beyond quantum information, the gate sequences developed here also enable new quantum sensing methods \cite{Abobeih_arXiv2019}.

\section*{ACKNOWLEDGEMENTS}
We thank T. van der Sar for valuable discussions and preliminary experiments. We thank V. V. Dobrovitski and R. Hanson for valuable discussions, R. F. L. Vermeulen and R. N. Schouten for assistance with the RF electronics, and M. Eschen for assistance with the experimental setup. We acknowledge support from the Netherlands Organization for Scientific Research (NWO) through a Vidi grant.  

\newpage

\bibliography{bib}

\begin{thebibliography}{70}%
\makeatletter
\providecommand \@ifxundefined [1]{%
 \@ifx{#1\undefined}
}%
\providecommand \@ifnum [1]{%
 \ifnum #1\expandafter \@firstoftwo
 \else \expandafter \@secondoftwo
 \fi
}%
\providecommand \@ifx [1]{%
 \ifx #1\expandafter \@firstoftwo
 \else \expandafter \@secondoftwo
 \fi
}%
\providecommand \natexlab [1]{#1}%
\providecommand \enquote  [1]{``#1''}%
\providecommand \bibnamefont  [1]{#1}%
\providecommand \bibfnamefont [1]{#1}%
\providecommand \citenamefont [1]{#1}%
\providecommand \href@noop [0]{\@secondoftwo}%
\providecommand \href [0]{\begingroup \@sanitize@url \@href}%
\providecommand \@href[1]{\@@startlink{#1}\@@href}%
\providecommand \@@href[1]{\endgroup#1\@@endlink}%
\providecommand \@sanitize@url [0]{\catcode `\\12\catcode `\$12\catcode
  `\&12\catcode `\#12\catcode `\^12\catcode `\_12\catcode `\%12\relax}%
\providecommand \@@startlink[1]{}%
\providecommand \@@endlink[0]{}%
\providecommand \url  [0]{\begingroup\@sanitize@url \@url }%
\providecommand \@url [1]{\endgroup\@href {#1}{\urlprefix }}%
\providecommand \urlprefix  [0]{URL }%
\providecommand \Eprint [0]{\href }%
\providecommand \doibase [0]{http://dx.doi.org/}%
\providecommand \selectlanguage [0]{\@gobble}%
\providecommand \bibinfo  [0]{\@secondoftwo}%
\providecommand \bibfield  [0]{\@secondoftwo}%
\providecommand \translation [1]{[#1]}%
\providecommand \BibitemOpen [0]{}%
\providecommand \bibitemStop [0]{}%
\providecommand \bibitemNoStop [0]{.\EOS\space}%
\providecommand \EOS [0]{\spacefactor3000\relax}%
\providecommand \BibitemShut  [1]{\csname bibitem#1\endcsname}%
\let\auto@bib@innerbib\@empty
\bibitem [{\citenamefont {Awschalom}\ \emph {et~al.}(2018)\citenamefont
  {Awschalom}, \citenamefont {Hanson}, \citenamefont {Wrachtrup},\ and\
  \citenamefont {Zhou}}]{awschalom2018quantum}%
  \BibitemOpen
  \bibfield  {author} {\bibinfo {author} {\bibfnamefont {D.~D.}\ \bibnamefont
  {Awschalom}}, \bibinfo {author} {\bibfnamefont {R.}~\bibnamefont {Hanson}},
  \bibinfo {author} {\bibfnamefont {J.}~\bibnamefont {Wrachtrup}}, \ and\
  \bibinfo {author} {\bibfnamefont {B.~B.}\ \bibnamefont {Zhou}},\ }\href
  {\doibase 10.1038/s41566-018-0232-2} {\bibfield  {journal} {\bibinfo
  {journal} {Nat. Photonics}\ }\textbf {\bibinfo {volume} {12}},\ \bibinfo
  {pages} {516} (\bibinfo {year} {2018})}\BibitemShut {NoStop}%
\bibitem [{\citenamefont {Zwanenburg}\ \emph {et~al.}(2013)\citenamefont
  {Zwanenburg}, \citenamefont {Dzurak}, \citenamefont {Morello}, \citenamefont
  {Simmons}, \citenamefont {Hollenberg}, \citenamefont {Klimeck}, \citenamefont
  {Rogge}, \citenamefont {Coppersmith},\ and\ \citenamefont
  {Eriksson}}]{Zwanenburg_RMP2013}%
  \BibitemOpen
  \bibfield  {author} {\bibinfo {author} {\bibfnamefont {F.~A.}\ \bibnamefont
  {Zwanenburg}}, \bibinfo {author} {\bibfnamefont {A.~S.}\ \bibnamefont
  {Dzurak}}, \bibinfo {author} {\bibfnamefont {A.}~\bibnamefont {Morello}},
  \bibinfo {author} {\bibfnamefont {M.~Y.}\ \bibnamefont {Simmons}}, \bibinfo
  {author} {\bibfnamefont {L.~C.~L.}\ \bibnamefont {Hollenberg}}, \bibinfo
  {author} {\bibfnamefont {G.}~\bibnamefont {Klimeck}}, \bibinfo {author}
  {\bibfnamefont {S.}~\bibnamefont {Rogge}}, \bibinfo {author} {\bibfnamefont
  {S.~N.}\ \bibnamefont {Coppersmith}}, \ and\ \bibinfo {author} {\bibfnamefont
  {M.~A.}\ \bibnamefont {Eriksson}},\ }\href {\doibase
  10.1103/RevModPhys.85.961} {\bibfield  {journal} {\bibinfo  {journal} {Rev.
  Mod. Phys.}\ }\textbf {\bibinfo {volume} {85}},\ \bibinfo {pages} {961}
  (\bibinfo {year} {2013})}\BibitemShut {NoStop}%
\bibitem [{\citenamefont {De~Lange}\ \emph {et~al.}(2010)\citenamefont
  {De~Lange}, \citenamefont {Wang}, \citenamefont {Riste}, \citenamefont
  {Dobrovitski},\ and\ \citenamefont {Hanson}}]{de2010universal}%
  \BibitemOpen
  \bibfield  {author} {\bibinfo {author} {\bibfnamefont {G.}~\bibnamefont
  {De~Lange}}, \bibinfo {author} {\bibfnamefont {Z.}~\bibnamefont {Wang}},
  \bibinfo {author} {\bibfnamefont {D.}~\bibnamefont {Riste}}, \bibinfo
  {author} {\bibfnamefont {V.}~\bibnamefont {Dobrovitski}}, \ and\ \bibinfo
  {author} {\bibfnamefont {R.}~\bibnamefont {Hanson}},\ }\href {\doibase
  10.1126/science.1192739} {\bibfield  {journal} {\bibinfo  {journal}
  {Science}\ }\textbf {\bibinfo {volume} {330}},\ \bibinfo {pages} {60}
  (\bibinfo {year} {2010})}\BibitemShut {NoStop}%
\bibitem [{\citenamefont {Fuchs}\ \emph {et~al.}(2009)\citenamefont {Fuchs},
  \citenamefont {Dobrovitski}, \citenamefont {Toyli}, \citenamefont
  {Heremans},\ and\ \citenamefont {Awschalom}}]{Fuchs_science2009}%
  \BibitemOpen
  \bibfield  {author} {\bibinfo {author} {\bibfnamefont {G.~D.}\ \bibnamefont
  {Fuchs}}, \bibinfo {author} {\bibfnamefont {V.~V.}\ \bibnamefont
  {Dobrovitski}}, \bibinfo {author} {\bibfnamefont {D.~M.}\ \bibnamefont
  {Toyli}}, \bibinfo {author} {\bibfnamefont {F.~J.}\ \bibnamefont {Heremans}},
  \ and\ \bibinfo {author} {\bibfnamefont {D.~D.}\ \bibnamefont {Awschalom}},\
  }\href {\doibase 10.1126/science.1181193} {\bibfield  {journal} {\bibinfo
  {journal} {Science}\ }\textbf {\bibinfo {volume} {326}},\ \bibinfo {pages}
  {1520} (\bibinfo {year} {2009})}\BibitemShut {NoStop}%
\bibitem [{\citenamefont {Christle}\ \emph {et~al.}(2015)\citenamefont
  {Christle}, \citenamefont {Falk}, \citenamefont {Andrich}, \citenamefont
  {Klimov}, \citenamefont {Hassan}, \citenamefont {Son}, \citenamefont
  {Janz{\'e}n}, \citenamefont {Ohshima},\ and\ \citenamefont
  {Awschalom}}]{christle2015isolated}%
  \BibitemOpen
  \bibfield  {author} {\bibinfo {author} {\bibfnamefont {D.~J.}\ \bibnamefont
  {Christle}}, \bibinfo {author} {\bibfnamefont {A.~L.}\ \bibnamefont {Falk}},
  \bibinfo {author} {\bibfnamefont {P.}~\bibnamefont {Andrich}}, \bibinfo
  {author} {\bibfnamefont {P.~V.}\ \bibnamefont {Klimov}}, \bibinfo {author}
  {\bibfnamefont {J.~U.}\ \bibnamefont {Hassan}}, \bibinfo {author}
  {\bibfnamefont {N.~T.}\ \bibnamefont {Son}}, \bibinfo {author} {\bibfnamefont
  {E.}~\bibnamefont {Janz{\'e}n}}, \bibinfo {author} {\bibfnamefont
  {T.}~\bibnamefont {Ohshima}}, \ and\ \bibinfo {author} {\bibfnamefont
  {D.~D.}\ \bibnamefont {Awschalom}},\ }\href {\doibase 10.1038/nmat4144}
  {\bibfield  {journal} {\bibinfo  {journal} {Nat. Mater.}\ }\textbf {\bibinfo
  {volume} {14}},\ \bibinfo {pages} {160} (\bibinfo {year} {2015})}\BibitemShut
  {NoStop}%
\bibitem [{\citenamefont {Seo}\ \emph {et~al.}(2016)\citenamefont {Seo},
  \citenamefont {Falk}, \citenamefont {Klimov}, \citenamefont {Miao},
  \citenamefont {Galli},\ and\ \citenamefont {Awschalom}}]{seo2016quantum}%
  \BibitemOpen
  \bibfield  {author} {\bibinfo {author} {\bibfnamefont {H.}~\bibnamefont
  {Seo}}, \bibinfo {author} {\bibfnamefont {A.~L.}\ \bibnamefont {Falk}},
  \bibinfo {author} {\bibfnamefont {P.~V.}\ \bibnamefont {Klimov}}, \bibinfo
  {author} {\bibfnamefont {K.~C.}\ \bibnamefont {Miao}}, \bibinfo {author}
  {\bibfnamefont {G.}~\bibnamefont {Galli}}, \ and\ \bibinfo {author}
  {\bibfnamefont {D.~D.}\ \bibnamefont {Awschalom}},\ }\href {\doibase
  10.1038/ncomms12935} {\bibfield  {journal} {\bibinfo  {journal} {Nat.
  Commun.}\ }\textbf {\bibinfo {volume} {7}},\ \bibinfo {pages} {12935}
  (\bibinfo {year} {2016})}\BibitemShut {NoStop}%
\bibitem [{\citenamefont {Sukachev}\ \emph {et~al.}(2017)\citenamefont
  {Sukachev}, \citenamefont {Sipahigil}, \citenamefont {Nguyen}, \citenamefont
  {Bhaskar}, \citenamefont {Evans}, \citenamefont {Jelezko},\ and\
  \citenamefont {Lukin}}]{sukachev2017silicon}%
  \BibitemOpen
  \bibfield  {author} {\bibinfo {author} {\bibfnamefont {D.~D.}\ \bibnamefont
  {Sukachev}}, \bibinfo {author} {\bibfnamefont {A.}~\bibnamefont {Sipahigil}},
  \bibinfo {author} {\bibfnamefont {C.~T.}\ \bibnamefont {Nguyen}}, \bibinfo
  {author} {\bibfnamefont {M.~K.}\ \bibnamefont {Bhaskar}}, \bibinfo {author}
  {\bibfnamefont {R.~E.}\ \bibnamefont {Evans}}, \bibinfo {author}
  {\bibfnamefont {F.}~\bibnamefont {Jelezko}}, \ and\ \bibinfo {author}
  {\bibfnamefont {M.~D.}\ \bibnamefont {Lukin}},\ }\href {\doibase
  10.1103/PhysRevLett.119.223602} {\bibfield  {journal} {\bibinfo  {journal}
  {Phys. Rev. Lett.}\ }\textbf {\bibinfo {volume} {119}},\ \bibinfo {pages}
  {223602} (\bibinfo {year} {2017})}\BibitemShut {NoStop}%
\bibitem [{\citenamefont {Iwasaki}\ \emph {et~al.}(2015)\citenamefont
  {Iwasaki}, \citenamefont {Ishibashi}, \citenamefont {Miyamoto}, \citenamefont
  {Doi}, \citenamefont {Kobayashi}, \citenamefont {Miyazaki}, \citenamefont
  {Tahara}, \citenamefont {Jahnke}, \citenamefont {Rogers}, \citenamefont
  {Naydenov} \emph {et~al.}}]{iwasaki2015germanium}%
  \BibitemOpen
  \bibfield  {author} {\bibinfo {author} {\bibfnamefont {T.}~\bibnamefont
  {Iwasaki}}, \bibinfo {author} {\bibfnamefont {F.}~\bibnamefont {Ishibashi}},
  \bibinfo {author} {\bibfnamefont {Y.}~\bibnamefont {Miyamoto}}, \bibinfo
  {author} {\bibfnamefont {Y.}~\bibnamefont {Doi}}, \bibinfo {author}
  {\bibfnamefont {S.}~\bibnamefont {Kobayashi}}, \bibinfo {author}
  {\bibfnamefont {T.}~\bibnamefont {Miyazaki}}, \bibinfo {author}
  {\bibfnamefont {K.}~\bibnamefont {Tahara}}, \bibinfo {author} {\bibfnamefont
  {K.~D.}\ \bibnamefont {Jahnke}}, \bibinfo {author} {\bibfnamefont {L.~J.}\
  \bibnamefont {Rogers}}, \bibinfo {author} {\bibfnamefont {B.}~\bibnamefont
  {Naydenov}},  \emph {et~al.},\ }\href {\doibase 10.1038/srep12882} {\bibfield
   {journal} {\bibinfo  {journal} {Sci. Rep.}\ }\textbf {\bibinfo {volume}
  {5}},\ \bibinfo {pages} {12882} (\bibinfo {year} {2015})}\BibitemShut
  {NoStop}%
\bibitem [{\citenamefont {Siyushev}\ \emph {et~al.}(2017)\citenamefont
  {Siyushev}, \citenamefont {Metsch}, \citenamefont {Ijaz}, \citenamefont
  {Binder}, \citenamefont {Bhaskar}, \citenamefont {Sukachev}, \citenamefont
  {Sipahigil}, \citenamefont {Evans}, \citenamefont {Nguyen}, \citenamefont
  {Lukin} \emph {et~al.}}]{siyushev2017optical}%
  \BibitemOpen
  \bibfield  {author} {\bibinfo {author} {\bibfnamefont {P.}~\bibnamefont
  {Siyushev}}, \bibinfo {author} {\bibfnamefont {M.~H.}\ \bibnamefont
  {Metsch}}, \bibinfo {author} {\bibfnamefont {A.}~\bibnamefont {Ijaz}},
  \bibinfo {author} {\bibfnamefont {J.~M.}\ \bibnamefont {Binder}}, \bibinfo
  {author} {\bibfnamefont {M.~K.}\ \bibnamefont {Bhaskar}}, \bibinfo {author}
  {\bibfnamefont {D.~D.}\ \bibnamefont {Sukachev}}, \bibinfo {author}
  {\bibfnamefont {A.}~\bibnamefont {Sipahigil}}, \bibinfo {author}
  {\bibfnamefont {R.~E.}\ \bibnamefont {Evans}}, \bibinfo {author}
  {\bibfnamefont {C.~T.}\ \bibnamefont {Nguyen}}, \bibinfo {author}
  {\bibfnamefont {M.~D.}\ \bibnamefont {Lukin}},  \emph {et~al.},\ }\href
  {\doibase 10.1103/PhysRevB.96.081201} {\bibfield  {journal} {\bibinfo
  {journal} {Phys. Rev. B}\ }\textbf {\bibinfo {volume} {96}},\ \bibinfo
  {pages} {081201(R)} (\bibinfo {year} {2017})}\BibitemShut {NoStop}%
\bibitem [{\citenamefont {Becker}\ \emph {et~al.}(2018)\citenamefont {Becker},
  \citenamefont {Pingault}, \citenamefont {Gro{\ss}}, \citenamefont
  {G{\"u}ndo{\u{g}}an}, \citenamefont {Kukharchyk}, \citenamefont {Markham},
  \citenamefont {Edmonds}, \citenamefont {Atat{\"u}re}, \citenamefont
  {Bushev},\ and\ \citenamefont {Becher}}]{becker2018all}%
  \BibitemOpen
  \bibfield  {author} {\bibinfo {author} {\bibfnamefont {J.~N.}\ \bibnamefont
  {Becker}}, \bibinfo {author} {\bibfnamefont {B.}~\bibnamefont {Pingault}},
  \bibinfo {author} {\bibfnamefont {D.}~\bibnamefont {Gro{\ss}}}, \bibinfo
  {author} {\bibfnamefont {M.}~\bibnamefont {G{\"u}ndo{\u{g}}an}}, \bibinfo
  {author} {\bibfnamefont {N.}~\bibnamefont {Kukharchyk}}, \bibinfo {author}
  {\bibfnamefont {M.}~\bibnamefont {Markham}}, \bibinfo {author} {\bibfnamefont
  {A.}~\bibnamefont {Edmonds}}, \bibinfo {author} {\bibfnamefont
  {M.}~\bibnamefont {Atat{\"u}re}}, \bibinfo {author} {\bibfnamefont
  {P.}~\bibnamefont {Bushev}}, \ and\ \bibinfo {author} {\bibfnamefont
  {C.}~\bibnamefont {Becher}},\ }\href {\doibase
  10.1103/PhysRevLett.120.053603} {\bibfield  {journal} {\bibinfo  {journal}
  {Phys. Rev. Lett.}\ }\textbf {\bibinfo {volume} {120}},\ \bibinfo {pages}
  {053603} (\bibinfo {year} {2018})}\BibitemShut {NoStop}%
\bibitem [{\citenamefont {Pingault}\ \emph {et~al.}(2017)\citenamefont
  {Pingault}, \citenamefont {Jarausch}, \citenamefont {Hepp}, \citenamefont
  {Klintberg}, \citenamefont {Becker}, \citenamefont {Markham}, \citenamefont
  {Becher},\ and\ \citenamefont {Atat{\"u}re}}]{pingault2017coherent}%
  \BibitemOpen
  \bibfield  {author} {\bibinfo {author} {\bibfnamefont {B.}~\bibnamefont
  {Pingault}}, \bibinfo {author} {\bibfnamefont {D.-D.}\ \bibnamefont
  {Jarausch}}, \bibinfo {author} {\bibfnamefont {C.}~\bibnamefont {Hepp}},
  \bibinfo {author} {\bibfnamefont {L.}~\bibnamefont {Klintberg}}, \bibinfo
  {author} {\bibfnamefont {J.~N.}\ \bibnamefont {Becker}}, \bibinfo {author}
  {\bibfnamefont {M.}~\bibnamefont {Markham}}, \bibinfo {author} {\bibfnamefont
  {C.}~\bibnamefont {Becher}}, \ and\ \bibinfo {author} {\bibfnamefont
  {M.}~\bibnamefont {Atat{\"u}re}},\ }\href {\doibase 10.1038/ncomms15579}
  {\bibfield  {journal} {\bibinfo  {journal} {Nat. Commun.}\ }\textbf {\bibinfo
  {volume} {8}},\ \bibinfo {pages} {15579} (\bibinfo {year}
  {2017})}\BibitemShut {NoStop}%
\bibitem [{\citenamefont {Trusheim}\ \emph {et~al.}(2018)\citenamefont
  {Trusheim}, \citenamefont {Pingault}, \citenamefont {Wan}, \citenamefont
  {De~Santis}, \citenamefont {Chen}, \citenamefont {Walsh}, \citenamefont
  {Rose}, \citenamefont {Becker}, \citenamefont {Bersin}, \citenamefont
  {Malladi} \emph {et~al.}}]{trusheim2018transform}%
  \BibitemOpen
  \bibfield  {author} {\bibinfo {author} {\bibfnamefont {M.~E.}\ \bibnamefont
  {Trusheim}}, \bibinfo {author} {\bibfnamefont {B.}~\bibnamefont {Pingault}},
  \bibinfo {author} {\bibfnamefont {N.~H.}\ \bibnamefont {Wan}}, \bibinfo
  {author} {\bibfnamefont {L.}~\bibnamefont {De~Santis}}, \bibinfo {author}
  {\bibfnamefont {K.~C.}\ \bibnamefont {Chen}}, \bibinfo {author}
  {\bibfnamefont {M.}~\bibnamefont {Walsh}}, \bibinfo {author} {\bibfnamefont
  {J.~J.}\ \bibnamefont {Rose}}, \bibinfo {author} {\bibfnamefont {J.~N.}\
  \bibnamefont {Becker}}, \bibinfo {author} {\bibfnamefont {E.}~\bibnamefont
  {Bersin}}, \bibinfo {author} {\bibfnamefont {G.}~\bibnamefont {Malladi}},
  \emph {et~al.},\ }\href {https://arxiv.org/abs/1811.07777} {\bibfield
  {journal} {\bibinfo  {journal} {arXiv preprint arXiv:1811.07777}\ } (\bibinfo
  {year} {2018})}\BibitemShut {NoStop}%
\bibitem [{\citenamefont {Rugar}\ \emph {et~al.}(2018)\citenamefont {Rugar},
  \citenamefont {Dory}, \citenamefont {Sun},\ and\ \citenamefont
  {Vu{\v{c}}kovi{\'c}}}]{rugar2018characterization}%
  \BibitemOpen
  \bibfield  {author} {\bibinfo {author} {\bibfnamefont {A.~E.}\ \bibnamefont
  {Rugar}}, \bibinfo {author} {\bibfnamefont {C.}~\bibnamefont {Dory}},
  \bibinfo {author} {\bibfnamefont {S.}~\bibnamefont {Sun}}, \ and\ \bibinfo
  {author} {\bibfnamefont {J.}~\bibnamefont {Vu{\v{c}}kovi{\'c}}},\ }\href
  {https://arxiv.org/abs/1811.09941} {\bibfield  {journal} {\bibinfo  {journal}
  {arXiv preprint arXiv:1811.09941}\ } (\bibinfo {year} {2018})}\BibitemShut
  {NoStop}%
\bibitem [{\citenamefont {Robledo}\ \emph {et~al.}(2011)\citenamefont
  {Robledo}, \citenamefont {Childress}, \citenamefont {Bernien}, \citenamefont
  {Hensen}, \citenamefont {Alkemade},\ and\ \citenamefont
  {Hanson}}]{robledo2011high}%
  \BibitemOpen
  \bibfield  {author} {\bibinfo {author} {\bibfnamefont {L.}~\bibnamefont
  {Robledo}}, \bibinfo {author} {\bibfnamefont {L.}~\bibnamefont {Childress}},
  \bibinfo {author} {\bibfnamefont {H.}~\bibnamefont {Bernien}}, \bibinfo
  {author} {\bibfnamefont {B.}~\bibnamefont {Hensen}}, \bibinfo {author}
  {\bibfnamefont {P.~F.}\ \bibnamefont {Alkemade}}, \ and\ \bibinfo {author}
  {\bibfnamefont {R.}~\bibnamefont {Hanson}},\ }\href {\doibase
  10.1038/nature10401} {\bibfield  {journal} {\bibinfo  {journal} {Nature}\
  }\textbf {\bibinfo {volume} {477}},\ \bibinfo {pages} {574} (\bibinfo {year}
  {2011})}\BibitemShut {NoStop}%
\bibitem [{\citenamefont {Pla}\ \emph {et~al.}(2013)\citenamefont {Pla},
  \citenamefont {Tan}, \citenamefont {Dehollain}, \citenamefont {Lim},
  \citenamefont {Morton}, \citenamefont {Zwanenburg}, \citenamefont {Jamieson},
  \citenamefont {Dzurak},\ and\ \citenamefont {Morello}}]{Pla_nature2013}%
  \BibitemOpen
  \bibfield  {author} {\bibinfo {author} {\bibfnamefont {J.~J.}\ \bibnamefont
  {Pla}}, \bibinfo {author} {\bibfnamefont {K.~Y.}\ \bibnamefont {Tan}},
  \bibinfo {author} {\bibfnamefont {J.~P.}\ \bibnamefont {Dehollain}}, \bibinfo
  {author} {\bibfnamefont {W.~H.}\ \bibnamefont {Lim}}, \bibinfo {author}
  {\bibfnamefont {J.~J.~L.}\ \bibnamefont {Morton}}, \bibinfo {author}
  {\bibfnamefont {F.~A.}\ \bibnamefont {Zwanenburg}}, \bibinfo {author}
  {\bibfnamefont {D.~N.}\ \bibnamefont {Jamieson}}, \bibinfo {author}
  {\bibfnamefont {A.~S.}\ \bibnamefont {Dzurak}}, \ and\ \bibinfo {author}
  {\bibfnamefont {A.}~\bibnamefont {Morello}},\ }\href {\doibase
  10.1038/nature12011} {\bibfield  {journal} {\bibinfo  {journal} {Nature}\
  }\textbf {\bibinfo {volume} {496}},\ \bibinfo {pages} {334} (\bibinfo {year}
  {2013})}\BibitemShut {NoStop}%
\bibitem [{\citenamefont {Cramer}\ \emph {et~al.}(2016)\citenamefont {Cramer},
  \citenamefont {Kalb}, \citenamefont {Rol}, \citenamefont {Hensen},
  \citenamefont {Blok}, \citenamefont {Markham}, \citenamefont {Twitchen},
  \citenamefont {Hanson},\ and\ \citenamefont
  {Taminiau}}]{Cramer_NatureComm2016}%
  \BibitemOpen
  \bibfield  {author} {\bibinfo {author} {\bibfnamefont {J.}~\bibnamefont
  {Cramer}}, \bibinfo {author} {\bibfnamefont {N.}~\bibnamefont {Kalb}},
  \bibinfo {author} {\bibfnamefont {M.~A.}\ \bibnamefont {Rol}}, \bibinfo
  {author} {\bibfnamefont {B.}~\bibnamefont {Hensen}}, \bibinfo {author}
  {\bibfnamefont {M.~S.}\ \bibnamefont {Blok}}, \bibinfo {author}
  {\bibfnamefont {M.}~\bibnamefont {Markham}}, \bibinfo {author} {\bibfnamefont
  {D.~J.}\ \bibnamefont {Twitchen}}, \bibinfo {author} {\bibfnamefont
  {R.}~\bibnamefont {Hanson}}, \ and\ \bibinfo {author} {\bibfnamefont {T.~H.}\
  \bibnamefont {Taminiau}},\ }\href {\doibase 10.1038/ncomms11526} {\bibfield
  {journal} {\bibinfo  {journal} {Nat. Commun.}\ }\textbf {\bibinfo {volume}
  {7}},\ \bibinfo {pages} {11526} (\bibinfo {year} {2016})}\BibitemShut
  {NoStop}%
\bibitem [{\citenamefont {Waldherr}\ \emph {et~al.}(2014)\citenamefont
  {Waldherr}, \citenamefont {Wang}, \citenamefont {Zaiser}, \citenamefont
  {Jamali}, \citenamefont {Schulte-Herbruggen}, \citenamefont {Abe},
  \citenamefont {Ohshima}, \citenamefont {Isoya}, \citenamefont {Du},
  \citenamefont {Neumann} \emph {et~al.}}]{waldherr2014quantum}%
  \BibitemOpen
  \bibfield  {author} {\bibinfo {author} {\bibfnamefont {G.}~\bibnamefont
  {Waldherr}}, \bibinfo {author} {\bibfnamefont {Y.}~\bibnamefont {Wang}},
  \bibinfo {author} {\bibfnamefont {S.}~\bibnamefont {Zaiser}}, \bibinfo
  {author} {\bibfnamefont {M.}~\bibnamefont {Jamali}}, \bibinfo {author}
  {\bibfnamefont {T.}~\bibnamefont {Schulte-Herbruggen}}, \bibinfo {author}
  {\bibfnamefont {H.}~\bibnamefont {Abe}}, \bibinfo {author} {\bibfnamefont
  {T.}~\bibnamefont {Ohshima}}, \bibinfo {author} {\bibfnamefont
  {J.}~\bibnamefont {Isoya}}, \bibinfo {author} {\bibfnamefont
  {J.}~\bibnamefont {Du}}, \bibinfo {author} {\bibfnamefont {P.}~\bibnamefont
  {Neumann}},  \emph {et~al.},\ }\href {\doibase 10.1038/nature12919}
  {\bibfield  {journal} {\bibinfo  {journal} {Nature}\ }\textbf {\bibinfo
  {volume} {506}},\ \bibinfo {pages} {204} (\bibinfo {year}
  {2014})}\BibitemShut {NoStop}%
\bibitem [{\citenamefont {Wolfowicz}\ \emph {et~al.}(2016)\citenamefont
  {Wolfowicz}, \citenamefont {Mortemousque}, \citenamefont {Guichard},
  \citenamefont {Simmons}, \citenamefont {Thewalt}, \citenamefont {Itoh},\ and\
  \citenamefont {Morton}}]{wolfowicz201629si}%
  \BibitemOpen
  \bibfield  {author} {\bibinfo {author} {\bibfnamefont {G.}~\bibnamefont
  {Wolfowicz}}, \bibinfo {author} {\bibfnamefont {P.-A.}\ \bibnamefont
  {Mortemousque}}, \bibinfo {author} {\bibfnamefont {R.}~\bibnamefont
  {Guichard}}, \bibinfo {author} {\bibfnamefont {S.}~\bibnamefont {Simmons}},
  \bibinfo {author} {\bibfnamefont {M.~L.}\ \bibnamefont {Thewalt}}, \bibinfo
  {author} {\bibfnamefont {K.~M.}\ \bibnamefont {Itoh}}, \ and\ \bibinfo
  {author} {\bibfnamefont {J.~J.}\ \bibnamefont {Morton}},\ }\href {\doibase
  10.1088/1367-2630/18/2/023021} {\bibfield  {journal} {\bibinfo  {journal}
  {New J. Phys.}\ }\textbf {\bibinfo {volume} {18}},\ \bibinfo {pages} {023021}
  (\bibinfo {year} {2016})}\BibitemShut {NoStop}%
\bibitem [{\citenamefont {Tosi}\ \emph {et~al.}(2017)\citenamefont {Tosi},
  \citenamefont {Mohiyaddin}, \citenamefont {Schmitt}, \citenamefont {Tenberg},
  \citenamefont {Rahman}, \citenamefont {Klimeck},\ and\ \citenamefont
  {Morello}}]{tosi2017silicon}%
  \BibitemOpen
  \bibfield  {author} {\bibinfo {author} {\bibfnamefont {G.}~\bibnamefont
  {Tosi}}, \bibinfo {author} {\bibfnamefont {F.~A.}\ \bibnamefont
  {Mohiyaddin}}, \bibinfo {author} {\bibfnamefont {V.}~\bibnamefont {Schmitt}},
  \bibinfo {author} {\bibfnamefont {S.}~\bibnamefont {Tenberg}}, \bibinfo
  {author} {\bibfnamefont {R.}~\bibnamefont {Rahman}}, \bibinfo {author}
  {\bibfnamefont {G.}~\bibnamefont {Klimeck}}, \ and\ \bibinfo {author}
  {\bibfnamefont {A.}~\bibnamefont {Morello}},\ }\href {\doibase
  10.1038/s41467-017-00378-x} {\bibfield  {journal} {\bibinfo  {journal} {Nat.
  Commun.}\ }\textbf {\bibinfo {volume} {8}},\ \bibinfo {pages} {450} (\bibinfo
  {year} {2017})}\BibitemShut {NoStop}%
\bibitem [{\citenamefont {Metsch}\ \emph {et~al.}(2019)\citenamefont {Metsch},
  \citenamefont {Senkalla}, \citenamefont {Tratzmiller}, \citenamefont
  {Scheuer}, \citenamefont {Kern}, \citenamefont {Achard}, \citenamefont
  {Tallaire}, \citenamefont {Plenio}, \citenamefont {Siyushev},\ and\
  \citenamefont {Jelezko}}]{metsch2019initialization}%
  \BibitemOpen
  \bibfield  {author} {\bibinfo {author} {\bibfnamefont {M.~H.}\ \bibnamefont
  {Metsch}}, \bibinfo {author} {\bibfnamefont {K.}~\bibnamefont {Senkalla}},
  \bibinfo {author} {\bibfnamefont {B.}~\bibnamefont {Tratzmiller}}, \bibinfo
  {author} {\bibfnamefont {J.}~\bibnamefont {Scheuer}}, \bibinfo {author}
  {\bibfnamefont {M.}~\bibnamefont {Kern}}, \bibinfo {author} {\bibfnamefont
  {J.}~\bibnamefont {Achard}}, \bibinfo {author} {\bibfnamefont
  {A.}~\bibnamefont {Tallaire}}, \bibinfo {author} {\bibfnamefont {M.~B.}\
  \bibnamefont {Plenio}}, \bibinfo {author} {\bibfnamefont {P.}~\bibnamefont
  {Siyushev}}, \ and\ \bibinfo {author} {\bibfnamefont {F.}~\bibnamefont
  {Jelezko}},\ }\href {https://arxiv.org/abs/1902.02965} {\bibfield  {journal}
  {\bibinfo  {journal} {arXiv preprint arXiv:1902.02965}\ } (\bibinfo {year}
  {2019})}\BibitemShut {NoStop}%
\bibitem [{\citenamefont {Hensen}\ \emph {et~al.}(2019)\citenamefont {Hensen},
  \citenamefont {Huang}, \citenamefont {Yang}, \citenamefont {Chan},
  \citenamefont {Yoneda}, \citenamefont {Tanttu}, \citenamefont {Hudson},
  \citenamefont {Laucht}, \citenamefont {Itoh}, \citenamefont {Morello},\ and\
  \citenamefont {Dzurak}}]{hensen2019silicon}%
  \BibitemOpen
  \bibfield  {author} {\bibinfo {author} {\bibfnamefont {B.}~\bibnamefont
  {Hensen}}, \bibinfo {author} {\bibfnamefont {W.}~\bibnamefont {Huang}},
  \bibinfo {author} {\bibfnamefont {C.-H.}\ \bibnamefont {Yang}}, \bibinfo
  {author} {\bibfnamefont {K.~W.}\ \bibnamefont {Chan}}, \bibinfo {author}
  {\bibfnamefont {J.}~\bibnamefont {Yoneda}}, \bibinfo {author} {\bibfnamefont
  {T.}~\bibnamefont {Tanttu}}, \bibinfo {author} {\bibfnamefont {F.~E.}\
  \bibnamefont {Hudson}}, \bibinfo {author} {\bibfnamefont {A.}~\bibnamefont
  {Laucht}}, \bibinfo {author} {\bibfnamefont {K.~M.}\ \bibnamefont {Itoh}},
  \bibinfo {author} {\bibfnamefont {A.}~\bibnamefont {Morello}}, \ and\
  \bibinfo {author} {\bibfnamefont {A.~S.}\ \bibnamefont {Dzurak}},\ }\href
  {https://arxiv.org/abs/1904.08260} {\bibfield  {journal} {\bibinfo  {journal}
  {arXiv preprint arXiv:1904.08260}\ } (\bibinfo {year} {2019})}\BibitemShut
  {NoStop}%
\bibitem [{\citenamefont {Dolde}\ \emph {et~al.}(2013)\citenamefont {Dolde},
  \citenamefont {Jakobi}, \citenamefont {Naydenov}, \citenamefont {Zhao},
  \citenamefont {Pezzagna}, \citenamefont {Trautmann}, \citenamefont {Meijer},
  \citenamefont {Neumann}, \citenamefont {Jelezko},\ and\ \citenamefont
  {Wrachtrup}}]{dolde2013room}%
  \BibitemOpen
  \bibfield  {author} {\bibinfo {author} {\bibfnamefont {F.}~\bibnamefont
  {Dolde}}, \bibinfo {author} {\bibfnamefont {I.}~\bibnamefont {Jakobi}},
  \bibinfo {author} {\bibfnamefont {B.}~\bibnamefont {Naydenov}}, \bibinfo
  {author} {\bibfnamefont {N.}~\bibnamefont {Zhao}}, \bibinfo {author}
  {\bibfnamefont {S.}~\bibnamefont {Pezzagna}}, \bibinfo {author}
  {\bibfnamefont {C.}~\bibnamefont {Trautmann}}, \bibinfo {author}
  {\bibfnamefont {J.}~\bibnamefont {Meijer}}, \bibinfo {author} {\bibfnamefont
  {P.}~\bibnamefont {Neumann}}, \bibinfo {author} {\bibfnamefont
  {F.}~\bibnamefont {Jelezko}}, \ and\ \bibinfo {author} {\bibfnamefont
  {J.}~\bibnamefont {Wrachtrup}},\ }\href {\doibase 10.1038/nphys2545}
  {\bibfield  {journal} {\bibinfo  {journal} {Nat. Phys.}\ }\textbf {\bibinfo
  {volume} {9}},\ \bibinfo {pages} {139} (\bibinfo {year} {2013})}\BibitemShut
  {NoStop}%
\bibitem [{\citenamefont {Yamamoto}\ \emph {et~al.}(2013)\citenamefont
  {Yamamoto}, \citenamefont {M{\"u}ller}, \citenamefont {McGuinness},
  \citenamefont {Teraji}, \citenamefont {Naydenov}, \citenamefont {Onoda},
  \citenamefont {Ohshima}, \citenamefont {Wrachtrup}, \citenamefont {Jelezko},\
  and\ \citenamefont {Isoya}}]{yamamoto2013strongly}%
  \BibitemOpen
  \bibfield  {author} {\bibinfo {author} {\bibfnamefont {T.}~\bibnamefont
  {Yamamoto}}, \bibinfo {author} {\bibfnamefont {C.}~\bibnamefont
  {M{\"u}ller}}, \bibinfo {author} {\bibfnamefont {L.~P.}\ \bibnamefont
  {McGuinness}}, \bibinfo {author} {\bibfnamefont {T.}~\bibnamefont {Teraji}},
  \bibinfo {author} {\bibfnamefont {B.}~\bibnamefont {Naydenov}}, \bibinfo
  {author} {\bibfnamefont {S.}~\bibnamefont {Onoda}}, \bibinfo {author}
  {\bibfnamefont {T.}~\bibnamefont {Ohshima}}, \bibinfo {author} {\bibfnamefont
  {J.}~\bibnamefont {Wrachtrup}}, \bibinfo {author} {\bibfnamefont
  {F.}~\bibnamefont {Jelezko}}, \ and\ \bibinfo {author} {\bibfnamefont
  {J.}~\bibnamefont {Isoya}},\ }\href {\doibase 10.1103/PhysRevB.88.201201}
  {\bibfield  {journal} {\bibinfo  {journal} {Phys. Rev. B}\ }\textbf {\bibinfo
  {volume} {88}},\ \bibinfo {pages} {201201(R)} (\bibinfo {year}
  {2013})}\BibitemShut {NoStop}%
\bibitem [{\citenamefont {Yang}\ \emph {et~al.}(2016)\citenamefont {Yang},
  \citenamefont {Wang}, \citenamefont {Rao}, \citenamefont {Tran},
  \citenamefont {Momenzadeh}, \citenamefont {Markham}, \citenamefont
  {Twitchen}, \citenamefont {Wang}, \citenamefont {Yang}, \citenamefont
  {St{\"o}hr} \emph {et~al.}}]{yang2016high}%
  \BibitemOpen
  \bibfield  {author} {\bibinfo {author} {\bibfnamefont {S.}~\bibnamefont
  {Yang}}, \bibinfo {author} {\bibfnamefont {Y.}~\bibnamefont {Wang}}, \bibinfo
  {author} {\bibfnamefont {D.~B.}\ \bibnamefont {Rao}}, \bibinfo {author}
  {\bibfnamefont {T.~H.}\ \bibnamefont {Tran}}, \bibinfo {author}
  {\bibfnamefont {A.~S.}\ \bibnamefont {Momenzadeh}}, \bibinfo {author}
  {\bibfnamefont {M.}~\bibnamefont {Markham}}, \bibinfo {author} {\bibfnamefont
  {D.}~\bibnamefont {Twitchen}}, \bibinfo {author} {\bibfnamefont
  {P.}~\bibnamefont {Wang}}, \bibinfo {author} {\bibfnamefont {W.}~\bibnamefont
  {Yang}}, \bibinfo {author} {\bibfnamefont {R.}~\bibnamefont {St{\"o}hr}},
  \emph {et~al.},\ }\href {\doibase 10.1038/nphoton.2016.103} {\bibfield
  {journal} {\bibinfo  {journal} {Nat. Photonics}\ }\textbf {\bibinfo {volume}
  {10}},\ \bibinfo {pages} {507} (\bibinfo {year} {2016})}\BibitemShut
  {NoStop}%
\bibitem [{\citenamefont {Togan}\ \emph {et~al.}(2010)\citenamefont {Togan},
  \citenamefont {Chu}, \citenamefont {Trifonov}, \citenamefont {Jiang},
  \citenamefont {Maze}, \citenamefont {Childress}, \citenamefont {Dutt},
  \citenamefont {S{\o}rensen}, \citenamefont {Hemmer}, \citenamefont {Zibrov}
  \emph {et~al.}}]{togan2010quantum}%
  \BibitemOpen
  \bibfield  {author} {\bibinfo {author} {\bibfnamefont {E.}~\bibnamefont
  {Togan}}, \bibinfo {author} {\bibfnamefont {Y.}~\bibnamefont {Chu}}, \bibinfo
  {author} {\bibfnamefont {A.}~\bibnamefont {Trifonov}}, \bibinfo {author}
  {\bibfnamefont {L.}~\bibnamefont {Jiang}}, \bibinfo {author} {\bibfnamefont
  {J.}~\bibnamefont {Maze}}, \bibinfo {author} {\bibfnamefont {L.}~\bibnamefont
  {Childress}}, \bibinfo {author} {\bibfnamefont {M.~G.}\ \bibnamefont {Dutt}},
  \bibinfo {author} {\bibfnamefont {A.~S.}\ \bibnamefont {S{\o}rensen}},
  \bibinfo {author} {\bibfnamefont {P.}~\bibnamefont {Hemmer}}, \bibinfo
  {author} {\bibfnamefont {A.~S.}\ \bibnamefont {Zibrov}},  \emph {et~al.},\
  }\href {\doibase 10.1038/nature09256} {\bibfield  {journal} {\bibinfo
  {journal} {Nature}\ }\textbf {\bibinfo {volume} {466}},\ \bibinfo {pages}
  {730} (\bibinfo {year} {2010})}\BibitemShut {NoStop}%
\bibitem [{\citenamefont {Christle}\ \emph {et~al.}(2017)\citenamefont
  {Christle}, \citenamefont {Klimov}, \citenamefont {delasCasas}, \citenamefont
  {Sz{\'a}sz}, \citenamefont {Iv{\'a}dy}, \citenamefont {Jokubavicius},
  \citenamefont {Hassan}, \citenamefont {Syv{\"a}j{\"a}rvi}, \citenamefont
  {Koehl}, \citenamefont {Ohshima} \emph {et~al.}}]{christle2017isolated}%
  \BibitemOpen
  \bibfield  {author} {\bibinfo {author} {\bibfnamefont {D.~J.}\ \bibnamefont
  {Christle}}, \bibinfo {author} {\bibfnamefont {P.~V.}\ \bibnamefont
  {Klimov}}, \bibinfo {author} {\bibfnamefont {C.~F.}\ \bibnamefont
  {delasCasas}}, \bibinfo {author} {\bibfnamefont {K.}~\bibnamefont
  {Sz{\'a}sz}}, \bibinfo {author} {\bibfnamefont {V.}~\bibnamefont
  {Iv{\'a}dy}}, \bibinfo {author} {\bibfnamefont {V.}~\bibnamefont
  {Jokubavicius}}, \bibinfo {author} {\bibfnamefont {J.~U.}\ \bibnamefont
  {Hassan}}, \bibinfo {author} {\bibfnamefont {M.}~\bibnamefont
  {Syv{\"a}j{\"a}rvi}}, \bibinfo {author} {\bibfnamefont {W.~F.}\ \bibnamefont
  {Koehl}}, \bibinfo {author} {\bibfnamefont {T.}~\bibnamefont {Ohshima}},
  \emph {et~al.},\ }\href {\doibase 10.1103/PhysRevX.7.021046} {\bibfield
  {journal} {\bibinfo  {journal} {Phys. Rev. X}\ }\textbf {\bibinfo {volume}
  {7}},\ \bibinfo {pages} {021046} (\bibinfo {year} {2017})}\BibitemShut
  {NoStop}%
\bibitem [{\citenamefont {Sipahigil}\ \emph {et~al.}(2016)\citenamefont
  {Sipahigil}, \citenamefont {Evans}, \citenamefont {Sukachev}, \citenamefont
  {Burek}, \citenamefont {Borregaard}, \citenamefont {Bhaskar}, \citenamefont
  {Nguyen}, \citenamefont {Pacheco}, \citenamefont {Atikian}, \citenamefont
  {Meuwly} \emph {et~al.}}]{sipahigil2016integrated}%
  \BibitemOpen
  \bibfield  {author} {\bibinfo {author} {\bibfnamefont {A.}~\bibnamefont
  {Sipahigil}}, \bibinfo {author} {\bibfnamefont {R.~E.}\ \bibnamefont
  {Evans}}, \bibinfo {author} {\bibfnamefont {D.~D.}\ \bibnamefont {Sukachev}},
  \bibinfo {author} {\bibfnamefont {M.~J.}\ \bibnamefont {Burek}}, \bibinfo
  {author} {\bibfnamefont {J.}~\bibnamefont {Borregaard}}, \bibinfo {author}
  {\bibfnamefont {M.~K.}\ \bibnamefont {Bhaskar}}, \bibinfo {author}
  {\bibfnamefont {C.~T.}\ \bibnamefont {Nguyen}}, \bibinfo {author}
  {\bibfnamefont {J.~L.}\ \bibnamefont {Pacheco}}, \bibinfo {author}
  {\bibfnamefont {H.~A.}\ \bibnamefont {Atikian}}, \bibinfo {author}
  {\bibfnamefont {C.}~\bibnamefont {Meuwly}},  \emph {et~al.},\ }\href
  {\doibase 10.1126/science.aah6875} {\bibfield  {journal} {\bibinfo  {journal}
  {Science}\ }\textbf {\bibinfo {volume} {354}},\ \bibinfo {pages} {847}
  (\bibinfo {year} {2016})}\BibitemShut {NoStop}%
\bibitem [{\citenamefont {Evans}\ \emph {et~al.}(2018)\citenamefont {Evans},
  \citenamefont {Bhaskar}, \citenamefont {Sukachev}, \citenamefont {Nguyen},
  \citenamefont {Sipahigil}, \citenamefont {Burek}, \citenamefont {Machielse},
  \citenamefont {Zhang}, \citenamefont {Zibrov}, \citenamefont {Bielejec} \emph
  {et~al.}}]{evans2018photon}%
  \BibitemOpen
  \bibfield  {author} {\bibinfo {author} {\bibfnamefont {R.~E.}\ \bibnamefont
  {Evans}}, \bibinfo {author} {\bibfnamefont {M.~K.}\ \bibnamefont {Bhaskar}},
  \bibinfo {author} {\bibfnamefont {D.~D.}\ \bibnamefont {Sukachev}}, \bibinfo
  {author} {\bibfnamefont {C.~T.}\ \bibnamefont {Nguyen}}, \bibinfo {author}
  {\bibfnamefont {A.}~\bibnamefont {Sipahigil}}, \bibinfo {author}
  {\bibfnamefont {M.~J.}\ \bibnamefont {Burek}}, \bibinfo {author}
  {\bibfnamefont {B.}~\bibnamefont {Machielse}}, \bibinfo {author}
  {\bibfnamefont {G.~H.}\ \bibnamefont {Zhang}}, \bibinfo {author}
  {\bibfnamefont {A.~S.}\ \bibnamefont {Zibrov}}, \bibinfo {author}
  {\bibfnamefont {E.}~\bibnamefont {Bielejec}},  \emph {et~al.},\ }\href
  {\doibase 10.1126/science.aau4691} {\bibfield  {journal} {\bibinfo  {journal}
  {Science}\ }\textbf {\bibinfo {volume} {362}},\ \bibinfo {pages} {662}
  (\bibinfo {year} {2018})}\BibitemShut {NoStop}%
\bibitem [{\citenamefont {Bernien}\ \emph {et~al.}(2013)\citenamefont
  {Bernien}, \citenamefont {Hensen}, \citenamefont {Pfaff}, \citenamefont
  {Koolstra}, \citenamefont {Blok}, \citenamefont {Robledo}, \citenamefont
  {Taminiau}, \citenamefont {Markham}, \citenamefont {Twitchen}, \citenamefont
  {Childress} \emph {et~al.}}]{bernien2013heralded}%
  \BibitemOpen
  \bibfield  {author} {\bibinfo {author} {\bibfnamefont {H.}~\bibnamefont
  {Bernien}}, \bibinfo {author} {\bibfnamefont {B.}~\bibnamefont {Hensen}},
  \bibinfo {author} {\bibfnamefont {W.}~\bibnamefont {Pfaff}}, \bibinfo
  {author} {\bibfnamefont {G.}~\bibnamefont {Koolstra}}, \bibinfo {author}
  {\bibfnamefont {M.}~\bibnamefont {Blok}}, \bibinfo {author} {\bibfnamefont
  {L.}~\bibnamefont {Robledo}}, \bibinfo {author} {\bibfnamefont
  {T.}~\bibnamefont {Taminiau}}, \bibinfo {author} {\bibfnamefont
  {M.}~\bibnamefont {Markham}}, \bibinfo {author} {\bibfnamefont
  {D.}~\bibnamefont {Twitchen}}, \bibinfo {author} {\bibfnamefont
  {L.}~\bibnamefont {Childress}},  \emph {et~al.},\ }\href {\doibase
  10.1038/nature12016} {\bibfield  {journal} {\bibinfo  {journal} {Nature}\
  }\textbf {\bibinfo {volume} {497}},\ \bibinfo {pages} {86} (\bibinfo {year}
  {2013})}\BibitemShut {NoStop}%
\bibitem [{\citenamefont {Hensen}\ \emph {et~al.}(2015)\citenamefont {Hensen},
  \citenamefont {Bernien}, \citenamefont {Dr{\'e}au}, \citenamefont {Reiserer},
  \citenamefont {Kalb}, \citenamefont {Blok}, \citenamefont {Ruitenberg},
  \citenamefont {Vermeulen}, \citenamefont {Schouten}, \citenamefont
  {Abell{\'a}n} \emph {et~al.}}]{hensen2015loophole}%
  \BibitemOpen
  \bibfield  {author} {\bibinfo {author} {\bibfnamefont {B.}~\bibnamefont
  {Hensen}}, \bibinfo {author} {\bibfnamefont {H.}~\bibnamefont {Bernien}},
  \bibinfo {author} {\bibfnamefont {A.~E.}\ \bibnamefont {Dr{\'e}au}}, \bibinfo
  {author} {\bibfnamefont {A.}~\bibnamefont {Reiserer}}, \bibinfo {author}
  {\bibfnamefont {N.}~\bibnamefont {Kalb}}, \bibinfo {author} {\bibfnamefont
  {M.~S.}\ \bibnamefont {Blok}}, \bibinfo {author} {\bibfnamefont
  {J.}~\bibnamefont {Ruitenberg}}, \bibinfo {author} {\bibfnamefont {R.~F.}\
  \bibnamefont {Vermeulen}}, \bibinfo {author} {\bibfnamefont {R.~N.}\
  \bibnamefont {Schouten}}, \bibinfo {author} {\bibfnamefont {C.}~\bibnamefont
  {Abell{\'a}n}},  \emph {et~al.},\ }\href {\doibase 10.1038/nature15759}
  {\bibfield  {journal} {\bibinfo  {journal} {Nature}\ }\textbf {\bibinfo
  {volume} {526}},\ \bibinfo {pages} {682} (\bibinfo {year}
  {2015})}\BibitemShut {NoStop}%
\bibitem [{\citenamefont {Humphreys}\ \emph {et~al.}(2018)\citenamefont
  {Humphreys}, \citenamefont {Kalb}, \citenamefont {Morits}, \citenamefont
  {Schouten}, \citenamefont {Vermeulen}, \citenamefont {Twitchen},
  \citenamefont {Markham},\ and\ \citenamefont
  {Hanson}}]{humphreys2018deterministic}%
  \BibitemOpen
  \bibfield  {author} {\bibinfo {author} {\bibfnamefont {P.~C.}\ \bibnamefont
  {Humphreys}}, \bibinfo {author} {\bibfnamefont {N.}~\bibnamefont {Kalb}},
  \bibinfo {author} {\bibfnamefont {J.~P.}\ \bibnamefont {Morits}}, \bibinfo
  {author} {\bibfnamefont {R.~N.}\ \bibnamefont {Schouten}}, \bibinfo {author}
  {\bibfnamefont {R.~F.}\ \bibnamefont {Vermeulen}}, \bibinfo {author}
  {\bibfnamefont {D.~J.}\ \bibnamefont {Twitchen}}, \bibinfo {author}
  {\bibfnamefont {M.}~\bibnamefont {Markham}}, \ and\ \bibinfo {author}
  {\bibfnamefont {R.}~\bibnamefont {Hanson}},\ }\href {\doibase
  10.1038/s41586-018-0200-5} {\bibfield  {journal} {\bibinfo  {journal}
  {Nature}\ }\textbf {\bibinfo {volume} {558}},\ \bibinfo {pages} {268}
  (\bibinfo {year} {2018})}\BibitemShut {NoStop}%
\bibitem [{\citenamefont {Maurer}\ \emph {et~al.}(2012)\citenamefont {Maurer},
  \citenamefont {Kucsko}, \citenamefont {Latta}, \citenamefont {Jiang},
  \citenamefont {Yao}, \citenamefont {Bennett}, \citenamefont {Pastawski},
  \citenamefont {Hunger}, \citenamefont {Chisholm}, \citenamefont {Markham}
  \emph {et~al.}}]{maurer2012room}%
  \BibitemOpen
  \bibfield  {author} {\bibinfo {author} {\bibfnamefont {P.~C.}\ \bibnamefont
  {Maurer}}, \bibinfo {author} {\bibfnamefont {G.}~\bibnamefont {Kucsko}},
  \bibinfo {author} {\bibfnamefont {C.}~\bibnamefont {Latta}}, \bibinfo
  {author} {\bibfnamefont {L.}~\bibnamefont {Jiang}}, \bibinfo {author}
  {\bibfnamefont {N.~Y.}\ \bibnamefont {Yao}}, \bibinfo {author} {\bibfnamefont
  {S.~D.}\ \bibnamefont {Bennett}}, \bibinfo {author} {\bibfnamefont
  {F.}~\bibnamefont {Pastawski}}, \bibinfo {author} {\bibfnamefont
  {D.}~\bibnamefont {Hunger}}, \bibinfo {author} {\bibfnamefont
  {N.}~\bibnamefont {Chisholm}}, \bibinfo {author} {\bibfnamefont
  {M.}~\bibnamefont {Markham}},  \emph {et~al.},\ }\href {\doibase
  10.1126/science.1220513} {\bibfield  {journal} {\bibinfo  {journal}
  {Science}\ }\textbf {\bibinfo {volume} {336}},\ \bibinfo {pages} {1283}
  (\bibinfo {year} {2012})}\BibitemShut {NoStop}%
\bibitem [{\citenamefont {Muhonen}\ \emph {et~al.}(2014)\citenamefont
  {Muhonen}, \citenamefont {Dehollain}, \citenamefont {Laucht}, \citenamefont
  {Hudson}, \citenamefont {Kalra}, \citenamefont {Sekiguchi}, \citenamefont
  {Itoh}, \citenamefont {Jamieson}, \citenamefont {McCallum}, \citenamefont
  {Dzurak} \emph {et~al.}}]{muhonen2014storing}%
  \BibitemOpen
  \bibfield  {author} {\bibinfo {author} {\bibfnamefont {J.~T.}\ \bibnamefont
  {Muhonen}}, \bibinfo {author} {\bibfnamefont {J.~P.}\ \bibnamefont
  {Dehollain}}, \bibinfo {author} {\bibfnamefont {A.}~\bibnamefont {Laucht}},
  \bibinfo {author} {\bibfnamefont {F.~E.}\ \bibnamefont {Hudson}}, \bibinfo
  {author} {\bibfnamefont {R.}~\bibnamefont {Kalra}}, \bibinfo {author}
  {\bibfnamefont {T.}~\bibnamefont {Sekiguchi}}, \bibinfo {author}
  {\bibfnamefont {K.~M.}\ \bibnamefont {Itoh}}, \bibinfo {author}
  {\bibfnamefont {D.~N.}\ \bibnamefont {Jamieson}}, \bibinfo {author}
  {\bibfnamefont {J.~C.}\ \bibnamefont {McCallum}}, \bibinfo {author}
  {\bibfnamefont {A.~S.}\ \bibnamefont {Dzurak}},  \emph {et~al.},\ }\href
  {\doibase 10.1038/nnano.2014.211} {\bibfield  {journal} {\bibinfo  {journal}
  {Nat. Nanotechnol.}\ }\textbf {\bibinfo {volume} {9}},\ \bibinfo {pages}
  {986} (\bibinfo {year} {2014})}\BibitemShut {NoStop}%
\bibitem [{\citenamefont {Dehollain}\ \emph {et~al.}(2016)\citenamefont
  {Dehollain}, \citenamefont {Simmons}, \citenamefont {Muhonen}, \citenamefont
  {Kalra}, \citenamefont {Laucht}, \citenamefont {Hudson}, \citenamefont
  {Itoh}, \citenamefont {Jamieson}, \citenamefont {McCallum}, \citenamefont
  {Dzurak} \emph {et~al.}}]{dehollain2016bell}%
  \BibitemOpen
  \bibfield  {author} {\bibinfo {author} {\bibfnamefont {J.~P.}\ \bibnamefont
  {Dehollain}}, \bibinfo {author} {\bibfnamefont {S.}~\bibnamefont {Simmons}},
  \bibinfo {author} {\bibfnamefont {J.~T.}\ \bibnamefont {Muhonen}}, \bibinfo
  {author} {\bibfnamefont {R.}~\bibnamefont {Kalra}}, \bibinfo {author}
  {\bibfnamefont {A.}~\bibnamefont {Laucht}}, \bibinfo {author} {\bibfnamefont
  {F.}~\bibnamefont {Hudson}}, \bibinfo {author} {\bibfnamefont {K.~M.}\
  \bibnamefont {Itoh}}, \bibinfo {author} {\bibfnamefont {D.~N.}\ \bibnamefont
  {Jamieson}}, \bibinfo {author} {\bibfnamefont {J.~C.}\ \bibnamefont
  {McCallum}}, \bibinfo {author} {\bibfnamefont {A.~S.}\ \bibnamefont
  {Dzurak}},  \emph {et~al.},\ }\href {\doibase 10.1038/nnano.2015.262}
  {\bibfield  {journal} {\bibinfo  {journal} {Nat. Nanotechnol.}\ }\textbf
  {\bibinfo {volume} {11}},\ \bibinfo {pages} {242} (\bibinfo {year}
  {2016})}\BibitemShut {NoStop}%
\bibitem [{\citenamefont {Van~der Sar}\ \emph {et~al.}(2012)\citenamefont
  {Van~der Sar}, \citenamefont {Wang}, \citenamefont {Blok}, \citenamefont
  {Bernien}, \citenamefont {Taminiau}, \citenamefont {Toyli}, \citenamefont
  {Lidar}, \citenamefont {Awschalom}, \citenamefont {Hanson},\ and\
  \citenamefont {Dobrovitski}}]{van2012decoherence}%
  \BibitemOpen
  \bibfield  {author} {\bibinfo {author} {\bibfnamefont {T.}~\bibnamefont
  {Van~der Sar}}, \bibinfo {author} {\bibfnamefont {Z.}~\bibnamefont {Wang}},
  \bibinfo {author} {\bibfnamefont {M.}~\bibnamefont {Blok}}, \bibinfo {author}
  {\bibfnamefont {H.}~\bibnamefont {Bernien}}, \bibinfo {author} {\bibfnamefont
  {T.}~\bibnamefont {Taminiau}}, \bibinfo {author} {\bibfnamefont
  {D.}~\bibnamefont {Toyli}}, \bibinfo {author} {\bibfnamefont
  {D.}~\bibnamefont {Lidar}}, \bibinfo {author} {\bibfnamefont
  {D.}~\bibnamefont {Awschalom}}, \bibinfo {author} {\bibfnamefont
  {R.}~\bibnamefont {Hanson}}, \ and\ \bibinfo {author} {\bibfnamefont
  {V.}~\bibnamefont {Dobrovitski}},\ }\href {\doibase 10.1038/nature10900}
  {\bibfield  {journal} {\bibinfo  {journal} {Nature}\ }\textbf {\bibinfo
  {volume} {484}},\ \bibinfo {pages} {82} (\bibinfo {year} {2012})}\BibitemShut
  {NoStop}%
\bibitem [{\citenamefont {Taminiau}\ \emph {et~al.}(2014)\citenamefont
  {Taminiau}, \citenamefont {Cramer}, \citenamefont {van~der Sar},
  \citenamefont {Dobrovitski},\ and\ \citenamefont
  {Hanson}}]{taminiau2014universal}%
  \BibitemOpen
  \bibfield  {author} {\bibinfo {author} {\bibfnamefont {T.~H.}\ \bibnamefont
  {Taminiau}}, \bibinfo {author} {\bibfnamefont {J.}~\bibnamefont {Cramer}},
  \bibinfo {author} {\bibfnamefont {T.}~\bibnamefont {van~der Sar}}, \bibinfo
  {author} {\bibfnamefont {V.~V.}\ \bibnamefont {Dobrovitski}}, \ and\ \bibinfo
  {author} {\bibfnamefont {R.}~\bibnamefont {Hanson}},\ }\href {\doibase
  10.1038/nnano.2014.2} {\bibfield  {journal} {\bibinfo  {journal} {Nat.
  Nanotechnol.}\ }\textbf {\bibinfo {volume} {9}},\ \bibinfo {pages} {171}
  (\bibinfo {year} {2014})}\BibitemShut {NoStop}%
\bibitem [{\citenamefont {Rong}\ \emph {et~al.}(2015)\citenamefont {Rong},
  \citenamefont {Geng}, \citenamefont {Shi}, \citenamefont {Liu}, \citenamefont
  {Xu}, \citenamefont {Ma}, \citenamefont {Kong}, \citenamefont {Jiang},
  \citenamefont {Wu},\ and\ \citenamefont {Du}}]{rong2015experimental}%
  \BibitemOpen
  \bibfield  {author} {\bibinfo {author} {\bibfnamefont {X.}~\bibnamefont
  {Rong}}, \bibinfo {author} {\bibfnamefont {J.}~\bibnamefont {Geng}}, \bibinfo
  {author} {\bibfnamefont {F.}~\bibnamefont {Shi}}, \bibinfo {author}
  {\bibfnamefont {Y.}~\bibnamefont {Liu}}, \bibinfo {author} {\bibfnamefont
  {K.}~\bibnamefont {Xu}}, \bibinfo {author} {\bibfnamefont {W.}~\bibnamefont
  {Ma}}, \bibinfo {author} {\bibfnamefont {F.}~\bibnamefont {Kong}}, \bibinfo
  {author} {\bibfnamefont {Z.}~\bibnamefont {Jiang}}, \bibinfo {author}
  {\bibfnamefont {Y.}~\bibnamefont {Wu}}, \ and\ \bibinfo {author}
  {\bibfnamefont {J.}~\bibnamefont {Du}},\ }\href {\doibase 10.1038/ncomms9748}
  {\bibfield  {journal} {\bibinfo  {journal} {Nat. Commun.}\ }\textbf {\bibinfo
  {volume} {6}},\ \bibinfo {pages} {8748} (\bibinfo {year} {2015})}\BibitemShut
  {NoStop}%
\bibitem [{\citenamefont {Zaiser}\ \emph {et~al.}(2016)\citenamefont {Zaiser},
  \citenamefont {Rendler}, \citenamefont {Jakobi}, \citenamefont {Wolf},
  \citenamefont {Lee}, \citenamefont {Wagner}, \citenamefont {Bergholm},
  \citenamefont {Schulte-Herbr{\"u}ggen}, \citenamefont {Neumann},\ and\
  \citenamefont {Wrachtrup}}]{zaiser2016enhancing}%
  \BibitemOpen
  \bibfield  {author} {\bibinfo {author} {\bibfnamefont {S.}~\bibnamefont
  {Zaiser}}, \bibinfo {author} {\bibfnamefont {T.}~\bibnamefont {Rendler}},
  \bibinfo {author} {\bibfnamefont {I.}~\bibnamefont {Jakobi}}, \bibinfo
  {author} {\bibfnamefont {T.}~\bibnamefont {Wolf}}, \bibinfo {author}
  {\bibfnamefont {S.-Y.}\ \bibnamefont {Lee}}, \bibinfo {author} {\bibfnamefont
  {S.}~\bibnamefont {Wagner}}, \bibinfo {author} {\bibfnamefont
  {V.}~\bibnamefont {Bergholm}}, \bibinfo {author} {\bibfnamefont
  {T.}~\bibnamefont {Schulte-Herbr{\"u}ggen}}, \bibinfo {author} {\bibfnamefont
  {P.}~\bibnamefont {Neumann}}, \ and\ \bibinfo {author} {\bibfnamefont
  {J.}~\bibnamefont {Wrachtrup}},\ }\href {\doibase 10.1038/ncomms12279}
  {\bibfield  {journal} {\bibinfo  {journal} {Nat. Commun.}\ }\textbf {\bibinfo
  {volume} {7}},\ \bibinfo {pages} {12279} (\bibinfo {year}
  {2016})}\BibitemShut {NoStop}%
\bibitem [{\citenamefont {Unden}\ \emph {et~al.}(2018)\citenamefont {Unden},
  \citenamefont {Louzon}, \citenamefont {Zwolak}, \citenamefont {Zurek},\ and\
  \citenamefont {Jelezko}}]{unden2018revealing}%
  \BibitemOpen
  \bibfield  {author} {\bibinfo {author} {\bibfnamefont {T.}~\bibnamefont
  {Unden}}, \bibinfo {author} {\bibfnamefont {D.}~\bibnamefont {Louzon}},
  \bibinfo {author} {\bibfnamefont {M.}~\bibnamefont {Zwolak}}, \bibinfo
  {author} {\bibfnamefont {W.}~\bibnamefont {Zurek}}, \ and\ \bibinfo {author}
  {\bibfnamefont {F.}~\bibnamefont {Jelezko}},\ }\href
  {https://arxiv.org/abs/1809.10456} {\bibfield  {journal} {\bibinfo  {journal}
  {arXiv preprint arXiv:1809.10456}\ } (\bibinfo {year} {2018})}\BibitemShut
  {NoStop}%
\bibitem [{\citenamefont {Huang}\ \emph {et~al.}(2019)\citenamefont {Huang},
  \citenamefont {Wu}, \citenamefont {Wang}, \citenamefont {Hou}, \citenamefont
  {Wang}, \citenamefont {Zhang}, \citenamefont {Lian}, \citenamefont {Liu},
  \citenamefont {Wang}, \citenamefont {Zhang} \emph
  {et~al.}}]{huang2019experimental}%
  \BibitemOpen
  \bibfield  {author} {\bibinfo {author} {\bibfnamefont {Y.-Y.}\ \bibnamefont
  {Huang}}, \bibinfo {author} {\bibfnamefont {Y.-K.}\ \bibnamefont {Wu}},
  \bibinfo {author} {\bibfnamefont {F.}~\bibnamefont {Wang}}, \bibinfo {author}
  {\bibfnamefont {P.-Y.}\ \bibnamefont {Hou}}, \bibinfo {author} {\bibfnamefont
  {W.-B.}\ \bibnamefont {Wang}}, \bibinfo {author} {\bibfnamefont {W.-G.}\
  \bibnamefont {Zhang}}, \bibinfo {author} {\bibfnamefont {W.-Q.}\ \bibnamefont
  {Lian}}, \bibinfo {author} {\bibfnamefont {Y.-Q.}\ \bibnamefont {Liu}},
  \bibinfo {author} {\bibfnamefont {H.-Y.}\ \bibnamefont {Wang}}, \bibinfo
  {author} {\bibfnamefont {H.-Y.}\ \bibnamefont {Zhang}},  \emph {et~al.},\
  }\href {\doibase 10.1103/PhysRevLett.122.010503} {\bibfield  {journal}
  {\bibinfo  {journal} {Phys. Rev. Lett.}\ }\textbf {\bibinfo {volume} {122}},\
  \bibinfo {pages} {010503} (\bibinfo {year} {2019})}\BibitemShut {NoStop}%
\bibitem [{\citenamefont {Kalb}\ \emph {et~al.}(2017)\citenamefont {Kalb},
  \citenamefont {Reiserer}, \citenamefont {Humphreys}, \citenamefont
  {Bakermans}, \citenamefont {Kamerling}, \citenamefont {Nickerson},
  \citenamefont {Benjamin}, \citenamefont {Twitchen}, \citenamefont {Markham},\
  and\ \citenamefont {Hanson}}]{kalb2017entanglement}%
  \BibitemOpen
  \bibfield  {author} {\bibinfo {author} {\bibfnamefont {N.}~\bibnamefont
  {Kalb}}, \bibinfo {author} {\bibfnamefont {A.~A.}\ \bibnamefont {Reiserer}},
  \bibinfo {author} {\bibfnamefont {P.~C.}\ \bibnamefont {Humphreys}}, \bibinfo
  {author} {\bibfnamefont {J.~J.}\ \bibnamefont {Bakermans}}, \bibinfo {author}
  {\bibfnamefont {S.~J.}\ \bibnamefont {Kamerling}}, \bibinfo {author}
  {\bibfnamefont {N.~H.}\ \bibnamefont {Nickerson}}, \bibinfo {author}
  {\bibfnamefont {S.~C.}\ \bibnamefont {Benjamin}}, \bibinfo {author}
  {\bibfnamefont {D.~J.}\ \bibnamefont {Twitchen}}, \bibinfo {author}
  {\bibfnamefont {M.}~\bibnamefont {Markham}}, \ and\ \bibinfo {author}
  {\bibfnamefont {R.}~\bibnamefont {Hanson}},\ }\href {\doibase
  10.1126/science.aan0070} {\bibfield  {journal} {\bibinfo  {journal}
  {Science}\ }\textbf {\bibinfo {volume} {356}},\ \bibinfo {pages} {928}
  (\bibinfo {year} {2017})}\BibitemShut {NoStop}%
\bibitem [{\citenamefont {van Dam}\ \emph {et~al.}(2019)\citenamefont {van
  Dam}, \citenamefont {Cramer}, \citenamefont {Taminiau},\ and\ \citenamefont
  {Hanson}}]{van2019multipartite}%
  \BibitemOpen
  \bibfield  {author} {\bibinfo {author} {\bibfnamefont {S.~B.}\ \bibnamefont
  {van Dam}}, \bibinfo {author} {\bibfnamefont {J.}~\bibnamefont {Cramer}},
  \bibinfo {author} {\bibfnamefont {T.~H.}\ \bibnamefont {Taminiau}}, \ and\
  \bibinfo {author} {\bibfnamefont {R.}~\bibnamefont {Hanson}},\ }\href
  {https://arxiv.org/abs/1902.08842} {\bibfield  {journal} {\bibinfo  {journal}
  {arXiv preprint arXiv:1902.08842}\ } (\bibinfo {year} {2019})}\BibitemShut
  {NoStop}%
\bibitem [{\citenamefont {Terhal}(2015)}]{terhal2015quantum}%
  \BibitemOpen
  \bibfield  {author} {\bibinfo {author} {\bibfnamefont {B.~M.}\ \bibnamefont
  {Terhal}},\ }\href {\doibase 10.1103/RevModPhys.87.307} {\bibfield  {journal}
  {\bibinfo  {journal} {Rev. Mod. Phys.}\ }\textbf {\bibinfo {volume} {87}},\
  \bibinfo {pages} {307} (\bibinfo {year} {2015})}\BibitemShut {NoStop}%
\bibitem [{\citenamefont {Nickerson}\ \emph {et~al.}(2013)\citenamefont
  {Nickerson}, \citenamefont {Li},\ and\ \citenamefont
  {Benjamin}}]{nickerson2013topological}%
  \BibitemOpen
  \bibfield  {author} {\bibinfo {author} {\bibfnamefont {N.~H.}\ \bibnamefont
  {Nickerson}}, \bibinfo {author} {\bibfnamefont {Y.}~\bibnamefont {Li}}, \
  and\ \bibinfo {author} {\bibfnamefont {S.~C.}\ \bibnamefont {Benjamin}},\
  }\href {\doibase 10.1038/ncomms2773} {\bibfield  {journal} {\bibinfo
  {journal} {Nat. Commun.}\ }\textbf {\bibinfo {volume} {4}},\ \bibinfo {pages}
  {1756} (\bibinfo {year} {2013})}\BibitemShut {NoStop}%
\bibitem [{\citenamefont {Nickerson}\ \emph {et~al.}(2014)\citenamefont
  {Nickerson}, \citenamefont {Fitzsimons},\ and\ \citenamefont
  {Benjamin}}]{nickerson2014scalable}%
  \BibitemOpen
  \bibfield  {author} {\bibinfo {author} {\bibfnamefont {N.~H.}\ \bibnamefont
  {Nickerson}}, \bibinfo {author} {\bibfnamefont {J.~F.}\ \bibnamefont
  {Fitzsimons}}, \ and\ \bibinfo {author} {\bibfnamefont {S.~C.}\ \bibnamefont
  {Benjamin}},\ }\href {\doibase 10.1103/PhysRevX.4.041041} {\bibfield
  {journal} {\bibinfo  {journal} {Phys. Rev. X}\ }\textbf {\bibinfo {volume}
  {4}},\ \bibinfo {pages} {041041} (\bibinfo {year} {2014})}\BibitemShut
  {NoStop}%
\bibitem [{\citenamefont {Abobeih}\ \emph {et~al.}(2018)\citenamefont
  {Abobeih}, \citenamefont {Cramer}, \citenamefont {Bakker}, \citenamefont
  {Kalb}, \citenamefont {Markham}, \citenamefont {Twitchen},\ and\
  \citenamefont {Taminiau}}]{abobeih2018one}%
  \BibitemOpen
  \bibfield  {author} {\bibinfo {author} {\bibfnamefont {M.~H.}\ \bibnamefont
  {Abobeih}}, \bibinfo {author} {\bibfnamefont {J.}~\bibnamefont {Cramer}},
  \bibinfo {author} {\bibfnamefont {M.~A.}\ \bibnamefont {Bakker}}, \bibinfo
  {author} {\bibfnamefont {N.}~\bibnamefont {Kalb}}, \bibinfo {author}
  {\bibfnamefont {M.}~\bibnamefont {Markham}}, \bibinfo {author} {\bibfnamefont
  {D.}~\bibnamefont {Twitchen}}, \ and\ \bibinfo {author} {\bibfnamefont
  {T.~H.}\ \bibnamefont {Taminiau}},\ }\href {\doibase
  10.1038/s41467-018-04916-z} {\bibfield  {journal} {\bibinfo  {journal} {Nat.
  Commun.}\ }\textbf {\bibinfo {volume} {9}},\ \bibinfo {pages} {2552}
  (\bibinfo {year} {2018})}\BibitemShut {NoStop}%
\bibitem [{\citenamefont {Kolkowitz}\ \emph {et~al.}(2012)\citenamefont
  {Kolkowitz}, \citenamefont {Unterreithmeier}, \citenamefont {Bennett},\ and\
  \citenamefont {Lukin}}]{Kolkowitz_PRL2012}%
  \BibitemOpen
  \bibfield  {author} {\bibinfo {author} {\bibfnamefont {S.}~\bibnamefont
  {Kolkowitz}}, \bibinfo {author} {\bibfnamefont {Q.~P.}\ \bibnamefont
  {Unterreithmeier}}, \bibinfo {author} {\bibfnamefont {S.~D.}\ \bibnamefont
  {Bennett}}, \ and\ \bibinfo {author} {\bibfnamefont {M.~D.}\ \bibnamefont
  {Lukin}},\ }\href {\doibase 10.1103/PhysRevLett.109.137601} {\bibfield
  {journal} {\bibinfo  {journal} {Phys. Rev. Lett.}\ }\textbf {\bibinfo
  {volume} {109}},\ \bibinfo {pages} {137601} (\bibinfo {year}
  {2012})}\BibitemShut {NoStop}%
\bibitem [{\citenamefont {Taminiau}\ \emph {et~al.}(2012)\citenamefont
  {Taminiau}, \citenamefont {Wagenaar}, \citenamefont {van~der Sar},
  \citenamefont {Jelezko}, \citenamefont {Dobrovitski},\ and\ \citenamefont
  {Hanson}}]{Taminiau_PRL2012}%
  \BibitemOpen
  \bibfield  {author} {\bibinfo {author} {\bibfnamefont {T.~H.}\ \bibnamefont
  {Taminiau}}, \bibinfo {author} {\bibfnamefont {J.~J.~T.}\ \bibnamefont
  {Wagenaar}}, \bibinfo {author} {\bibfnamefont {T.}~\bibnamefont {van~der
  Sar}}, \bibinfo {author} {\bibfnamefont {F.}~\bibnamefont {Jelezko}},
  \bibinfo {author} {\bibfnamefont {V.~V.}\ \bibnamefont {Dobrovitski}}, \ and\
  \bibinfo {author} {\bibfnamefont {R.}~\bibnamefont {Hanson}},\ }\href
  {\doibase 10.1103/PhysRevLett.109.137602} {\bibfield  {journal} {\bibinfo
  {journal} {Phys. Rev. Lett.}\ }\textbf {\bibinfo {volume} {109}},\ \bibinfo
  {pages} {137602} (\bibinfo {year} {2012})}\BibitemShut {NoStop}%
\bibitem [{\citenamefont {Zhao}\ \emph {et~al.}(2012)\citenamefont {Zhao},
  \citenamefont {Honert}, \citenamefont {Schmid}, \citenamefont {Klas},
  \citenamefont {Isoya}, \citenamefont {Markham}, \citenamefont {Twitchen},
  \citenamefont {Jelezko}, \citenamefont {Liu}, \citenamefont {Fedder},\ and\
  \citenamefont {Wrachtrup}}]{Zhao_NatureNano2012}%
  \BibitemOpen
  \bibfield  {author} {\bibinfo {author} {\bibfnamefont {N.}~\bibnamefont
  {Zhao}}, \bibinfo {author} {\bibfnamefont {J.}~\bibnamefont {Honert}},
  \bibinfo {author} {\bibfnamefont {B.}~\bibnamefont {Schmid}}, \bibinfo
  {author} {\bibfnamefont {M.}~\bibnamefont {Klas}}, \bibinfo {author}
  {\bibfnamefont {J.}~\bibnamefont {Isoya}}, \bibinfo {author} {\bibfnamefont
  {M.}~\bibnamefont {Markham}}, \bibinfo {author} {\bibfnamefont
  {D.}~\bibnamefont {Twitchen}}, \bibinfo {author} {\bibfnamefont
  {F.}~\bibnamefont {Jelezko}}, \bibinfo {author} {\bibfnamefont {R.-B.}\
  \bibnamefont {Liu}}, \bibinfo {author} {\bibfnamefont {H.}~\bibnamefont
  {Fedder}}, \ and\ \bibinfo {author} {\bibfnamefont {J.}~\bibnamefont
  {Wrachtrup}},\ }\href {\doibase 10.1038/nnano.2012.152} {\bibfield  {journal}
  {\bibinfo  {journal} {Nat. Nanotechnol.}\ }\textbf {\bibinfo {volume} {7}},\
  \bibinfo {pages} {657} (\bibinfo {year} {2012})}\BibitemShut {NoStop}%
\bibitem [{\citenamefont {Wang}\ \emph {et~al.}(2017)\citenamefont {Wang},
  \citenamefont {Casanova},\ and\ \citenamefont {Plenio}}]{wang2017delayed}%
  \BibitemOpen
  \bibfield  {author} {\bibinfo {author} {\bibfnamefont {Z.-Y.}\ \bibnamefont
  {Wang}}, \bibinfo {author} {\bibfnamefont {J.}~\bibnamefont {Casanova}}, \
  and\ \bibinfo {author} {\bibfnamefont {M.~B.}\ \bibnamefont {Plenio}},\
  }\href {\doibase 10.1038/ncomms14660} {\bibfield  {journal} {\bibinfo
  {journal} {Nat. Commun.}\ }\textbf {\bibinfo {volume} {8}},\ \bibinfo {pages}
  {14660} (\bibinfo {year} {2017})}\BibitemShut {NoStop}%
\bibitem [{\citenamefont {Pfender}\ \emph {et~al.}(2017)\citenamefont
  {Pfender}, \citenamefont {Aslam}, \citenamefont {Sumiya}, \citenamefont
  {Onoda}, \citenamefont {Neumann}, \citenamefont {Isoya}, \citenamefont
  {Meriles},\ and\ \citenamefont {Wrachtrup}}]{pfender2017nonvolatile}%
  \BibitemOpen
  \bibfield  {author} {\bibinfo {author} {\bibfnamefont {M.}~\bibnamefont
  {Pfender}}, \bibinfo {author} {\bibfnamefont {N.}~\bibnamefont {Aslam}},
  \bibinfo {author} {\bibfnamefont {H.}~\bibnamefont {Sumiya}}, \bibinfo
  {author} {\bibfnamefont {S.}~\bibnamefont {Onoda}}, \bibinfo {author}
  {\bibfnamefont {P.}~\bibnamefont {Neumann}}, \bibinfo {author} {\bibfnamefont
  {J.}~\bibnamefont {Isoya}}, \bibinfo {author} {\bibfnamefont {C.~A.}\
  \bibnamefont {Meriles}}, \ and\ \bibinfo {author} {\bibfnamefont
  {J.}~\bibnamefont {Wrachtrup}},\ }\href {\doibase 10.1038/s41467-017-00964-z}
  {\bibfield  {journal} {\bibinfo  {journal} {Nat. Commun.}\ }\textbf {\bibinfo
  {volume} {8}},\ \bibinfo {pages} {834} (\bibinfo {year} {2017})}\BibitemShut
  {NoStop}%
\bibitem [{sup()}]{supp}%
  \BibitemOpen
  \href@noop {} {\bibinfo  {journal} {The supplemental materials contains
  experimental details and parameters, all measured coherence times and
  fidelities, modelling of the expected fidelities, and the generalised
  theoretical description of the gate dynamics}\ }\BibitemShut {NoStop}%
\bibitem [{\citenamefont {M{\"u}ller}\ \emph {et~al.}(2014)\citenamefont
  {M{\"u}ller}, \citenamefont {Hepp}, \citenamefont {Pingault}, \citenamefont
  {Neu}, \citenamefont {Gsell}, \citenamefont {Schreck}, \citenamefont
  {Sternschulte}, \citenamefont {Steinm{\"u}ller-Nethl}, \citenamefont
  {Becher},\ and\ \citenamefont {Atat{\"u}re}}]{muller2014optical}%
  \BibitemOpen
\bibfield  {journal} {  }\bibfield  {author} {\bibinfo {author} {\bibfnamefont
  {T.}~\bibnamefont {M{\"u}ller}}, \bibinfo {author} {\bibfnamefont
  {C.}~\bibnamefont {Hepp}}, \bibinfo {author} {\bibfnamefont {B.}~\bibnamefont
  {Pingault}}, \bibinfo {author} {\bibfnamefont {E.}~\bibnamefont {Neu}},
  \bibinfo {author} {\bibfnamefont {S.}~\bibnamefont {Gsell}}, \bibinfo
  {author} {\bibfnamefont {M.}~\bibnamefont {Schreck}}, \bibinfo {author}
  {\bibfnamefont {H.}~\bibnamefont {Sternschulte}}, \bibinfo {author}
  {\bibfnamefont {D.}~\bibnamefont {Steinm{\"u}ller-Nethl}}, \bibinfo {author}
  {\bibfnamefont {C.}~\bibnamefont {Becher}}, \ and\ \bibinfo {author}
  {\bibfnamefont {M.}~\bibnamefont {Atat{\"u}re}},\ }\href {\doibase
  10.1038/ncomms4328} {\bibfield  {journal} {\bibinfo  {journal} {Nat.
  Commun.}\ }\textbf {\bibinfo {volume} {5}},\ \bibinfo {pages} {3328}
  (\bibinfo {year} {2014})}\BibitemShut {NoStop}%
\bibitem [{\citenamefont {Thiering}\ and\ \citenamefont
  {Gali}(2018)}]{thiering2018ab}%
  \BibitemOpen
  \bibfield  {author} {\bibinfo {author} {\bibfnamefont {G.}~\bibnamefont
  {Thiering}}\ and\ \bibinfo {author} {\bibfnamefont {A.}~\bibnamefont
  {Gali}},\ }\href {\doibase 10.1103/PhysRevX.8.021063} {\bibfield  {journal}
  {\bibinfo  {journal} {Phys. Rev. X}\ }\textbf {\bibinfo {volume} {8}},\
  \bibinfo {pages} {021063} (\bibinfo {year} {2018})}\BibitemShut {NoStop}%
\bibitem [{\citenamefont {Vandersypen}\ and\ \citenamefont
  {Chuang}(2005)}]{vandersypen2005nmr}%
  \BibitemOpen
  \bibfield  {author} {\bibinfo {author} {\bibfnamefont {L.~M.}\ \bibnamefont
  {Vandersypen}}\ and\ \bibinfo {author} {\bibfnamefont {I.~L.}\ \bibnamefont
  {Chuang}},\ }\href {\doibase 10.1103/RevModPhys.76.1037} {\bibfield
  {journal} {\bibinfo  {journal} {Rev. Mod. Phys.}\ }\textbf {\bibinfo {volume}
  {76}},\ \bibinfo {pages} {1037} (\bibinfo {year} {2005})}\BibitemShut
  {NoStop}%
\bibitem [{\citenamefont {Dolde}\ \emph {et~al.}(2014)\citenamefont {Dolde},
  \citenamefont {Bergholm}, \citenamefont {Wang}, \citenamefont {Jakobi},
  \citenamefont {Naydenov}, \citenamefont {Pezzagna}, \citenamefont {Meijer},
  \citenamefont {Jelezko}, \citenamefont {Neumann}, \citenamefont
  {Schulte-Herbr{\"u}ggen} \emph {et~al.}}]{dolde2014high}%
  \BibitemOpen
  \bibfield  {author} {\bibinfo {author} {\bibfnamefont {F.}~\bibnamefont
  {Dolde}}, \bibinfo {author} {\bibfnamefont {V.}~\bibnamefont {Bergholm}},
  \bibinfo {author} {\bibfnamefont {Y.}~\bibnamefont {Wang}}, \bibinfo {author}
  {\bibfnamefont {I.}~\bibnamefont {Jakobi}}, \bibinfo {author} {\bibfnamefont
  {B.}~\bibnamefont {Naydenov}}, \bibinfo {author} {\bibfnamefont
  {S.}~\bibnamefont {Pezzagna}}, \bibinfo {author} {\bibfnamefont
  {J.}~\bibnamefont {Meijer}}, \bibinfo {author} {\bibfnamefont
  {F.}~\bibnamefont {Jelezko}}, \bibinfo {author} {\bibfnamefont
  {P.}~\bibnamefont {Neumann}}, \bibinfo {author} {\bibfnamefont
  {T.}~\bibnamefont {Schulte-Herbr{\"u}ggen}},  \emph {et~al.},\ }\href
  {\doibase 10.1038/ncomms4371} {\bibfield  {journal} {\bibinfo  {journal}
  {Nat. Commun.}\ }\textbf {\bibinfo {volume} {5}},\ \bibinfo {pages} {3371}
  (\bibinfo {year} {2014})}\BibitemShut {NoStop}%
\bibitem [{\citenamefont {Abobeih}\ \emph {et~al.}(2019)\citenamefont
  {Abobeih}, \citenamefont {Randall}, \citenamefont {Bradley}, \citenamefont
  {Bartling}, \citenamefont {Bakker}, \citenamefont {Degen}, \citenamefont
  {Markham}, \citenamefont {Twitchen},\ and\ \citenamefont
  {Taminiau}}]{Abobeih_arXiv2019}%
  \BibitemOpen
  \bibfield  {author} {\bibinfo {author} {\bibfnamefont {M.~H.}\ \bibnamefont
  {Abobeih}}, \bibinfo {author} {\bibfnamefont {J.}~\bibnamefont {Randall}},
  \bibinfo {author} {\bibfnamefont {C.~E.}\ \bibnamefont {Bradley}}, \bibinfo
  {author} {\bibfnamefont {H.~P.}\ \bibnamefont {Bartling}}, \bibinfo {author}
  {\bibfnamefont {M.~A.}\ \bibnamefont {Bakker}}, \bibinfo {author}
  {\bibfnamefont {M.~J.}\ \bibnamefont {Degen}}, \bibinfo {author}
  {\bibfnamefont {M.}~\bibnamefont {Markham}}, \bibinfo {author} {\bibfnamefont
  {D.~J.}\ \bibnamefont {Twitchen}}, \ and\ \bibinfo {author} {\bibfnamefont
  {T.~H.}\ \bibnamefont {Taminiau}},\ }\href {https://arxiv.org/abs/1905.02095}
  {\bibfield  {journal} {\bibinfo  {journal} {arXiv preprint arXiv:1905.02095}\
  } (\bibinfo {year} {2019})}\BibitemShut {NoStop}%
\bibitem [{\citenamefont {Childress}\ \emph {et~al.}(2006)\citenamefont
  {Childress}, \citenamefont {Dutt}, \citenamefont {Taylor}, \citenamefont
  {Zibrov}, \citenamefont {Jelezko}, \citenamefont {Wrachtrup}, \citenamefont
  {Hemmer},\ and\ \citenamefont {Lukin}}]{childress2006coherent}%
  \BibitemOpen
  \bibfield  {author} {\bibinfo {author} {\bibfnamefont {L.}~\bibnamefont
  {Childress}}, \bibinfo {author} {\bibfnamefont {M.~G.}\ \bibnamefont {Dutt}},
  \bibinfo {author} {\bibfnamefont {J.}~\bibnamefont {Taylor}}, \bibinfo
  {author} {\bibfnamefont {A.}~\bibnamefont {Zibrov}}, \bibinfo {author}
  {\bibfnamefont {F.}~\bibnamefont {Jelezko}}, \bibinfo {author} {\bibfnamefont
  {J.}~\bibnamefont {Wrachtrup}}, \bibinfo {author} {\bibfnamefont
  {P.}~\bibnamefont {Hemmer}}, \ and\ \bibinfo {author} {\bibfnamefont
  {M.}~\bibnamefont {Lukin}},\ }\href {\doibase 10.1126/science.1131871}
  {\bibfield  {journal} {\bibinfo  {journal} {Science}\ }\textbf {\bibinfo
  {volume} {314}},\ \bibinfo {pages} {281} (\bibinfo {year}
  {2006})}\BibitemShut {NoStop}%
\bibitem [{\citenamefont {Kalb}\ \emph {et~al.}(2016)\citenamefont {Kalb},
  \citenamefont {Cramer}, \citenamefont {Twitchen}, \citenamefont {Markham},
  \citenamefont {Hanson},\ and\ \citenamefont
  {Taminiau}}]{Kalb_NatureComm2016}%
  \BibitemOpen
  \bibfield  {author} {\bibinfo {author} {\bibfnamefont {N.}~\bibnamefont
  {Kalb}}, \bibinfo {author} {\bibfnamefont {J.}~\bibnamefont {Cramer}},
  \bibinfo {author} {\bibfnamefont {D.~J.}\ \bibnamefont {Twitchen}}, \bibinfo
  {author} {\bibfnamefont {M.}~\bibnamefont {Markham}}, \bibinfo {author}
  {\bibfnamefont {R.}~\bibnamefont {Hanson}}, \ and\ \bibinfo {author}
  {\bibfnamefont {T.~H.}\ \bibnamefont {Taminiau}},\ }\href {\doibase
  10.1038/ncomms13111} {\bibfield  {journal} {\bibinfo  {journal} {Nat.
  Commun.}\ }\textbf {\bibinfo {volume} {7}},\ \bibinfo {pages} {13111}
  (\bibinfo {year} {2016})}\BibitemShut {NoStop}%
\bibitem [{\citenamefont {G{\"u}hne}\ and\ \citenamefont
  {T{\'o}th}(2009)}]{guhne2009entanglement}%
  \BibitemOpen
  \bibfield  {author} {\bibinfo {author} {\bibfnamefont {O.}~\bibnamefont
  {G{\"u}hne}}\ and\ \bibinfo {author} {\bibfnamefont {G.}~\bibnamefont
  {T{\'o}th}},\ }\href {\doibase 10.1016/j.physrep.2009.02.004} {\bibfield
  {journal} {\bibinfo  {journal} {Phys. Rep.}\ }\textbf {\bibinfo {volume}
  {474}},\ \bibinfo {pages} {1} (\bibinfo {year} {2009})}\BibitemShut {NoStop}%
\bibitem [{\citenamefont {Pfaff}\ \emph {et~al.}(2013)\citenamefont {Pfaff},
  \citenamefont {Taminiau}, \citenamefont {Robledo}, \citenamefont {Bernien},
  \citenamefont {Markham}, \citenamefont {Twitchen},\ and\ \citenamefont
  {Hanson}}]{pfaff2013demonstration}%
  \BibitemOpen
  \bibfield  {author} {\bibinfo {author} {\bibfnamefont {W.}~\bibnamefont
  {Pfaff}}, \bibinfo {author} {\bibfnamefont {T.~H.}\ \bibnamefont {Taminiau}},
  \bibinfo {author} {\bibfnamefont {L.}~\bibnamefont {Robledo}}, \bibinfo
  {author} {\bibfnamefont {H.}~\bibnamefont {Bernien}}, \bibinfo {author}
  {\bibfnamefont {M.}~\bibnamefont {Markham}}, \bibinfo {author} {\bibfnamefont
  {D.~J.}\ \bibnamefont {Twitchen}}, \ and\ \bibinfo {author} {\bibfnamefont
  {R.}~\bibnamefont {Hanson}},\ }\href {\doibase 10.1038/nphys2444} {\bibfield
  {journal} {\bibinfo  {journal} {Nat. Phys.}\ }\textbf {\bibinfo {volume}
  {9}},\ \bibinfo {pages} {29} (\bibinfo {year} {2013})}\BibitemShut {NoStop}%
\bibitem [{\citenamefont {Blok}\ \emph {et~al.}(2014)\citenamefont {Blok},
  \citenamefont {Bonato}, \citenamefont {Markham}, \citenamefont {Twitchen},
  \citenamefont {Dobrovitski},\ and\ \citenamefont
  {Hanson}}]{blok2014manipulating}%
  \BibitemOpen
  \bibfield  {author} {\bibinfo {author} {\bibfnamefont {M.}~\bibnamefont
  {Blok}}, \bibinfo {author} {\bibfnamefont {C.}~\bibnamefont {Bonato}},
  \bibinfo {author} {\bibfnamefont {M.}~\bibnamefont {Markham}}, \bibinfo
  {author} {\bibfnamefont {D.}~\bibnamefont {Twitchen}}, \bibinfo {author}
  {\bibfnamefont {V.}~\bibnamefont {Dobrovitski}}, \ and\ \bibinfo {author}
  {\bibfnamefont {R.}~\bibnamefont {Hanson}},\ }\href {\doibase
  10.1038/nphys2881} {\bibfield  {journal} {\bibinfo  {journal} {Nat. Phys.}\
  }\textbf {\bibinfo {volume} {10}},\ \bibinfo {pages} {189} (\bibinfo {year}
  {2014})}\BibitemShut {NoStop}%
\bibitem [{\citenamefont {Kalb}\ \emph {et~al.}(2018)\citenamefont {Kalb},
  \citenamefont {Humphreys}, \citenamefont {Slim},\ and\ \citenamefont
  {Hanson}}]{KalbQmem}%
  \BibitemOpen
  \bibfield  {author} {\bibinfo {author} {\bibfnamefont {N.}~\bibnamefont
  {Kalb}}, \bibinfo {author} {\bibfnamefont {P.~C.}\ \bibnamefont {Humphreys}},
  \bibinfo {author} {\bibfnamefont {J.~J.}\ \bibnamefont {Slim}}, \ and\
  \bibinfo {author} {\bibfnamefont {R.}~\bibnamefont {Hanson}},\ }\href
  {\doibase 10.1103/PhysRevA.97.062330} {\bibfield  {journal} {\bibinfo
  {journal} {Phys. Rev. A}\ }\textbf {\bibinfo {volume} {97}},\ \bibinfo
  {pages} {062330} (\bibinfo {year} {2018})}\BibitemShut {NoStop}%
\bibitem [{\citenamefont {Khutsishvili}(1962)}]{khutsishvili1962spin}%
  \BibitemOpen
  \bibfield  {author} {\bibinfo {author} {\bibfnamefont {G.}~\bibnamefont
  {Khutsishvili}},\ }\href
  {https://iopscience.iop.org/article/10.1070/PU1969v011n06ABEH003776}
  {\bibfield  {journal} {\bibinfo  {journal} {Sov. Phys.-JETP}\ }\textbf
  {\bibinfo {volume} {15}} (\bibinfo {year} {1962})}\BibitemShut {NoStop}%
\bibitem [{\citenamefont {Guichard}\ \emph {et~al.}(2015)\citenamefont
  {Guichard}, \citenamefont {Balian}, \citenamefont {Wolfowicz}, \citenamefont
  {Mortemousque},\ and\ \citenamefont {Monteiro}}]{guichard2015decoherence}%
  \BibitemOpen
  \bibfield  {author} {\bibinfo {author} {\bibfnamefont {R.}~\bibnamefont
  {Guichard}}, \bibinfo {author} {\bibfnamefont {S.~J.}\ \bibnamefont
  {Balian}}, \bibinfo {author} {\bibfnamefont {G.}~\bibnamefont {Wolfowicz}},
  \bibinfo {author} {\bibfnamefont {P.~A.}\ \bibnamefont {Mortemousque}}, \
  and\ \bibinfo {author} {\bibfnamefont {T.~S.}\ \bibnamefont {Monteiro}},\
  }\href {\doibase 10.1103/PhysRevB.91.214303} {\bibfield  {journal} {\bibinfo
  {journal} {Phys. Rev. B}\ }\textbf {\bibinfo {volume} {91}},\ \bibinfo
  {pages} {214303} (\bibinfo {year} {2015})}\BibitemShut {NoStop}%
\bibitem [{\citenamefont {Kwiatkowski}\ and\ \citenamefont
  {Cywi{\'n}ski}(2018)}]{kwiatkowski2018decoherence}%
  \BibitemOpen
  \bibfield  {author} {\bibinfo {author} {\bibfnamefont {D.}~\bibnamefont
  {Kwiatkowski}}\ and\ \bibinfo {author} {\bibfnamefont {{\L}.}~\bibnamefont
  {Cywi{\'n}ski}},\ }\href {\doibase 10.1103/PhysRevB.98.155202} {\bibfield
  {journal} {\bibinfo  {journal} {Phys. Rev. B}\ }\textbf {\bibinfo {volume}
  {98}},\ \bibinfo {pages} {155202} (\bibinfo {year} {2018})}\BibitemShut
  {NoStop}%
\bibitem [{\citenamefont {Lidar}\ \emph {et~al.}(1998)\citenamefont {Lidar},
  \citenamefont {Chuang},\ and\ \citenamefont {Whaley}}]{lidar1998decoherence}%
  \BibitemOpen
  \bibfield  {author} {\bibinfo {author} {\bibfnamefont {D.~A.}\ \bibnamefont
  {Lidar}}, \bibinfo {author} {\bibfnamefont {I.~L.}\ \bibnamefont {Chuang}}, \
  and\ \bibinfo {author} {\bibfnamefont {K.~B.}\ \bibnamefont {Whaley}},\
  }\href {\doibase 10.1103/PhysRevLett.81.2594} {\bibfield  {journal} {\bibinfo
   {journal} {Phys. Rev. Lett.}\ }\textbf {\bibinfo {volume} {81}},\ \bibinfo
  {pages} {2594} (\bibinfo {year} {1998})}\BibitemShut {NoStop}%
\bibitem [{\citenamefont {Reiserer}\ \emph {et~al.}(2016)\citenamefont
  {Reiserer}, \citenamefont {Kalb}, \citenamefont {Blok}, \citenamefont {van
  Bemmelen}, \citenamefont {Taminiau}, \citenamefont {Hanson}, \citenamefont
  {Twitchen},\ and\ \citenamefont {Markham}}]{reiserer2016robust}%
  \BibitemOpen
  \bibfield  {author} {\bibinfo {author} {\bibfnamefont {A.}~\bibnamefont
  {Reiserer}}, \bibinfo {author} {\bibfnamefont {N.}~\bibnamefont {Kalb}},
  \bibinfo {author} {\bibfnamefont {M.~S.}\ \bibnamefont {Blok}}, \bibinfo
  {author} {\bibfnamefont {K.~J.~M.}\ \bibnamefont {van Bemmelen}}, \bibinfo
  {author} {\bibfnamefont {T.~H.}\ \bibnamefont {Taminiau}}, \bibinfo {author}
  {\bibfnamefont {R.}~\bibnamefont {Hanson}}, \bibinfo {author} {\bibfnamefont
  {D.~J.}\ \bibnamefont {Twitchen}}, \ and\ \bibinfo {author} {\bibfnamefont
  {M.}~\bibnamefont {Markham}},\ }\href {\doibase 10.1103/PhysRevX.6.021040}
  {\bibfield  {journal} {\bibinfo  {journal} {Phys. Rev. X}\ }\textbf {\bibinfo
  {volume} {6}},\ \bibinfo {pages} {021040} (\bibinfo {year}
  {2016})}\BibitemShut {NoStop}%
\bibitem [{\citenamefont {Monz}\ \emph {et~al.}(2011)\citenamefont {Monz},
  \citenamefont {Schindler}, \citenamefont {Barreiro}, \citenamefont {Chwalla},
  \citenamefont {Nigg}, \citenamefont {Coish}, \citenamefont {Harlander},
  \citenamefont {H{\"a}nsel}, \citenamefont {Hennrich},\ and\ \citenamefont
  {Blatt}}]{monz201114}%
  \BibitemOpen
  \bibfield  {author} {\bibinfo {author} {\bibfnamefont {T.}~\bibnamefont
  {Monz}}, \bibinfo {author} {\bibfnamefont {P.}~\bibnamefont {Schindler}},
  \bibinfo {author} {\bibfnamefont {J.~T.}\ \bibnamefont {Barreiro}}, \bibinfo
  {author} {\bibfnamefont {M.}~\bibnamefont {Chwalla}}, \bibinfo {author}
  {\bibfnamefont {D.}~\bibnamefont {Nigg}}, \bibinfo {author} {\bibfnamefont
  {W.~A.}\ \bibnamefont {Coish}}, \bibinfo {author} {\bibfnamefont
  {M.}~\bibnamefont {Harlander}}, \bibinfo {author} {\bibfnamefont
  {W.}~\bibnamefont {H{\"a}nsel}}, \bibinfo {author} {\bibfnamefont
  {M.}~\bibnamefont {Hennrich}}, \ and\ \bibinfo {author} {\bibfnamefont
  {R.}~\bibnamefont {Blatt}},\ }\href {\doibase 10.1103/PhysRevLett.106.130506}
  {\bibfield  {journal} {\bibinfo  {journal} {Phys. Rev. Lett.}\ }\textbf
  {\bibinfo {volume} {106}},\ \bibinfo {pages} {130506} (\bibinfo {year}
  {2011})}\BibitemShut {NoStop}%
\bibitem [{\citenamefont {Layden}\ \emph {et~al.}(2019)\citenamefont {Layden},
  \citenamefont {Chen},\ and\ \citenamefont
  {Cappellaro}}]{layden2019efficient}%
  \BibitemOpen
  \bibfield  {author} {\bibinfo {author} {\bibfnamefont {D.}~\bibnamefont
  {Layden}}, \bibinfo {author} {\bibfnamefont {M.}~\bibnamefont {Chen}}, \ and\
  \bibinfo {author} {\bibfnamefont {P.}~\bibnamefont {Cappellaro}},\ }\href
  {https://arxiv.org/abs/1903.01046} {\bibfield  {journal} {\bibinfo  {journal}
  {arXiv preprint arXiv:1903.01046}\ } (\bibinfo {year} {2019})}\BibitemShut
  {NoStop}%
\end{thebibliography}%


\begin{thebibliography}{28}%
\makeatletter
\providecommand \@ifxundefined [1]{%
 \@ifx{#1\undefined}
}%
\providecommand \@ifnum [1]{%
 \ifnum #1\expandafter \@firstoftwo
 \else \expandafter \@secondoftwo
 \fi
}%
\providecommand \@ifx [1]{%
 \ifx #1\expandafter \@firstoftwo
 \else \expandafter \@secondoftwo
 \fi
}%
\providecommand \natexlab [1]{#1}%
\providecommand \enquote  [1]{``#1''}%
\providecommand \bibnamefont  [1]{#1}%
\providecommand \bibfnamefont [1]{#1}%
\providecommand \citenamefont [1]{#1}%
\providecommand \href@noop [0]{\@secondoftwo}%
\providecommand \href [0]{\begingroup \@sanitize@url \@href}%
\providecommand \@href[1]{\@@startlink{#1}\@@href}%
\providecommand \@@href[1]{\endgroup#1\@@endlink}%
\providecommand \@sanitize@url [0]{\catcode `\\12\catcode `\$12\catcode
  `\&12\catcode `\#12\catcode `\^12\catcode `\_12\catcode `\%12\relax}%
\providecommand \@@startlink[1]{}%
\providecommand \@@endlink[0]{}%
\providecommand \url  [0]{\begingroup\@sanitize@url \@url }%
\providecommand \@url [1]{\endgroup\@href {#1}{\urlprefix }}%
\providecommand \urlprefix  [0]{URL }%
\providecommand \Eprint [0]{\href }%
\providecommand \doibase [0]{http://dx.doi.org/}%
\providecommand \selectlanguage [0]{\@gobble}%
\providecommand \bibinfo  [0]{\@secondoftwo}%
\providecommand \bibfield  [0]{\@secondoftwo}%
\providecommand \translation [1]{[#1]}%
\providecommand \BibitemOpen [0]{}%
\providecommand \bibitemStop [0]{}%
\providecommand \bibitemNoStop [0]{.\EOS\space}%
\providecommand \EOS [0]{\spacefactor3000\relax}%
\providecommand \BibitemShut  [1]{\csname bibitem#1\endcsname}%
\let\auto@bib@innerbib\@empty
\bibitem [{\citenamefont {Hadden}\ \emph {et~al.}(2010)\citenamefont {Hadden},
  \citenamefont {Harrison}, \citenamefont {Stanley-Clarke}, \citenamefont
  {Marseglia}, \citenamefont {Ho}, \citenamefont {Patton}, \citenamefont
  {O’Brien},\ and\ \citenamefont {Rarity}}]{hadden2010strongly}%
  \BibitemOpen
  \bibfield  {author} {\bibinfo {author} {\bibfnamefont {J.}~\bibnamefont
  {Hadden}}, \bibinfo {author} {\bibfnamefont {J.}~\bibnamefont {Harrison}},
  \bibinfo {author} {\bibfnamefont {A.}~\bibnamefont {Stanley-Clarke}},
  \bibinfo {author} {\bibfnamefont {L.}~\bibnamefont {Marseglia}}, \bibinfo
  {author} {\bibfnamefont {Y.-L.}\ \bibnamefont {Ho}}, \bibinfo {author}
  {\bibfnamefont {B.}~\bibnamefont {Patton}}, \bibinfo {author} {\bibfnamefont
  {J.}~\bibnamefont {O’Brien}}, \ and\ \bibinfo {author} {\bibfnamefont
  {J.}~\bibnamefont {Rarity}},\ }\href {\doibase 10.1063/1.3519847} {\bibfield
  {journal} {\bibinfo  {journal} {Appl. Phys. Lett.}\ }\textbf {\bibinfo
  {volume} {97}},\ \bibinfo {pages} {241901} (\bibinfo {year}
  {2010})}\BibitemShut {NoStop}%
\bibitem [{\citenamefont {Robledo}\ \emph {et~al.}(2011)\citenamefont
  {Robledo}, \citenamefont {Childress}, \citenamefont {Bernien}, \citenamefont
  {Hensen}, \citenamefont {Alkemade},\ and\ \citenamefont
  {Hanson}}]{robledo2011high}%
  \BibitemOpen
  \bibfield  {author} {\bibinfo {author} {\bibfnamefont {L.}~\bibnamefont
  {Robledo}}, \bibinfo {author} {\bibfnamefont {L.}~\bibnamefont {Childress}},
  \bibinfo {author} {\bibfnamefont {H.}~\bibnamefont {Bernien}}, \bibinfo
  {author} {\bibfnamefont {B.}~\bibnamefont {Hensen}}, \bibinfo {author}
  {\bibfnamefont {P.~F.}\ \bibnamefont {Alkemade}}, \ and\ \bibinfo {author}
  {\bibfnamefont {R.}~\bibnamefont {Hanson}},\ }\href {\doibase
  10.1038/nature10401} {\bibfield  {journal} {\bibinfo  {journal} {Nature}\
  }\textbf {\bibinfo {volume} {477}},\ \bibinfo {pages} {574} (\bibinfo {year}
  {2011})}\BibitemShut {NoStop}%
\bibitem [{\citenamefont {Pfaff}\ \emph {et~al.}(2013)\citenamefont {Pfaff},
  \citenamefont {Taminiau}, \citenamefont {Robledo}, \citenamefont {Bernien},
  \citenamefont {Markham}, \citenamefont {Twitchen},\ and\ \citenamefont
  {Hanson}}]{pfaff2013demonstration}%
  \BibitemOpen
  \bibfield  {author} {\bibinfo {author} {\bibfnamefont {W.}~\bibnamefont
  {Pfaff}}, \bibinfo {author} {\bibfnamefont {T.~H.}\ \bibnamefont {Taminiau}},
  \bibinfo {author} {\bibfnamefont {L.}~\bibnamefont {Robledo}}, \bibinfo
  {author} {\bibfnamefont {H.}~\bibnamefont {Bernien}}, \bibinfo {author}
  {\bibfnamefont {M.}~\bibnamefont {Markham}}, \bibinfo {author} {\bibfnamefont
  {D.~J.}\ \bibnamefont {Twitchen}}, \ and\ \bibinfo {author} {\bibfnamefont
  {R.}~\bibnamefont {Hanson}},\ }\href {\doibase 10.1038/nphys2444} {\bibfield
  {journal} {\bibinfo  {journal} {Nat. Phys.}\ }\textbf {\bibinfo {volume}
  {9}},\ \bibinfo {pages} {29} (\bibinfo {year} {2013})}\BibitemShut {NoStop}%
\bibitem [{\citenamefont {Yeung}\ \emph {et~al.}(2012)\citenamefont {Yeung},
  \citenamefont {Le~Sage}, \citenamefont {Pham}, \citenamefont {Stanwix},\ and\
  \citenamefont {Walsworth}}]{yeung2012anti}%
  \BibitemOpen
  \bibfield  {author} {\bibinfo {author} {\bibfnamefont {T.}~\bibnamefont
  {Yeung}}, \bibinfo {author} {\bibfnamefont {D.}~\bibnamefont {Le~Sage}},
  \bibinfo {author} {\bibfnamefont {L.~M.}\ \bibnamefont {Pham}}, \bibinfo
  {author} {\bibfnamefont {P.}~\bibnamefont {Stanwix}}, \ and\ \bibinfo
  {author} {\bibfnamefont {R.~L.}\ \bibnamefont {Walsworth}},\ }\href {\doibase
  10.1063/1.4730401} {\bibfield  {journal} {\bibinfo  {journal} {Appl. Phys.
  Lett.}\ }\textbf {\bibinfo {volume} {100}},\ \bibinfo {pages} {251111}
  (\bibinfo {year} {2012})}\BibitemShut {NoStop}%
\bibitem [{\citenamefont {Abobeih}\ \emph {et~al.}(2018)\citenamefont
  {Abobeih}, \citenamefont {Cramer}, \citenamefont {Bakker}, \citenamefont
  {Kalb}, \citenamefont {Markham}, \citenamefont {Twitchen},\ and\
  \citenamefont {Taminiau}}]{abobeih2018one}%
  \BibitemOpen
  \bibfield  {author} {\bibinfo {author} {\bibfnamefont {M.~H.}\ \bibnamefont
  {Abobeih}}, \bibinfo {author} {\bibfnamefont {J.}~\bibnamefont {Cramer}},
  \bibinfo {author} {\bibfnamefont {M.~A.}\ \bibnamefont {Bakker}}, \bibinfo
  {author} {\bibfnamefont {N.}~\bibnamefont {Kalb}}, \bibinfo {author}
  {\bibfnamefont {M.}~\bibnamefont {Markham}}, \bibinfo {author} {\bibfnamefont
  {D.}~\bibnamefont {Twitchen}}, \ and\ \bibinfo {author} {\bibfnamefont
  {T.~H.}\ \bibnamefont {Taminiau}},\ }\href {\doibase
  10.1038/s41467-018-04916-z} {\bibfield  {journal} {\bibinfo  {journal} {Nat.
  Commun.}\ }\textbf {\bibinfo {volume} {9}},\ \bibinfo {pages} {2552}
  (\bibinfo {year} {2018})}\BibitemShut {NoStop}%
\bibitem [{\citenamefont {Abobeih}\ \emph {et~al.}(2019)\citenamefont
  {Abobeih}, \citenamefont {Randall}, \citenamefont {Bradley}, \citenamefont
  {Bartling}, \citenamefont {Bakker}, \citenamefont {Degen}, \citenamefont
  {Markham}, \citenamefont {Twitchen},\ and\ \citenamefont
  {Taminiau}}]{Abobeih_arXiv2019}%
  \BibitemOpen
  \bibfield  {author} {\bibinfo {author} {\bibfnamefont {M.~H.}\ \bibnamefont
  {Abobeih}}, \bibinfo {author} {\bibfnamefont {J.}~\bibnamefont {Randall}},
  \bibinfo {author} {\bibfnamefont {C.~E.}\ \bibnamefont {Bradley}}, \bibinfo
  {author} {\bibfnamefont {H.~P.}\ \bibnamefont {Bartling}}, \bibinfo {author}
  {\bibfnamefont {M.~A.}\ \bibnamefont {Bakker}}, \bibinfo {author}
  {\bibfnamefont {M.~J.}\ \bibnamefont {Degen}}, \bibinfo {author}
  {\bibfnamefont {M.}~\bibnamefont {Markham}}, \bibinfo {author} {\bibfnamefont
  {D.~J.}\ \bibnamefont {Twitchen}}, \ and\ \bibinfo {author} {\bibfnamefont
  {T.~H.}\ \bibnamefont {Taminiau}},\ }\href {https://arxiv.org/abs/1905.02095}
  {\bibfield  {journal} {\bibinfo  {journal} {arXiv preprint arXiv:1905.02095}\
  } (\bibinfo {year} {2019})}\BibitemShut {NoStop}%
\bibitem [{\citenamefont {Vandersypen}\ and\ \citenamefont
  {Chuang}(2005)}]{vandersypen2005nmr}%
  \BibitemOpen
  \bibfield  {author} {\bibinfo {author} {\bibfnamefont {L.~M.}\ \bibnamefont
  {Vandersypen}}\ and\ \bibinfo {author} {\bibfnamefont {I.~L.}\ \bibnamefont
  {Chuang}},\ }\href {\doibase 10.1103/RevModPhys.76.1037} {\bibfield
  {journal} {\bibinfo  {journal} {Rev. Mod. Phys.}\ }\textbf {\bibinfo {volume}
  {76}},\ \bibinfo {pages} {1037} (\bibinfo {year} {2005})}\BibitemShut
  {NoStop}%
\bibitem [{\citenamefont {Warren}(1984)}]{warren1984effects}%
  \BibitemOpen
  \bibfield  {author} {\bibinfo {author} {\bibfnamefont {W.~S.}\ \bibnamefont
  {Warren}},\ }\href {\doibase 10.1063/1.447644} {\bibfield  {journal}
  {\bibinfo  {journal} {J. Chem. Phys.}\ }\textbf {\bibinfo {volume} {81}},\
  \bibinfo {pages} {5437} (\bibinfo {year} {1984})}\BibitemShut {NoStop}%
\bibitem [{\citenamefont {Gullion}\ \emph {et~al.}(1990)\citenamefont
  {Gullion}, \citenamefont {Baker},\ and\ \citenamefont
  {Conradi}}]{gullion1990new}%
  \BibitemOpen
  \bibfield  {author} {\bibinfo {author} {\bibfnamefont {T.}~\bibnamefont
  {Gullion}}, \bibinfo {author} {\bibfnamefont {D.~B.}\ \bibnamefont {Baker}},
  \ and\ \bibinfo {author} {\bibfnamefont {M.~S.}\ \bibnamefont {Conradi}},\
  }\href {\doibase 10.1016/0022-2364(90)90331-3} {\bibfield  {journal}
  {\bibinfo  {journal} {J. Magn. Reson.}\ }\textbf {\bibinfo {volume} {89}},\
  \bibinfo {pages} {479} (\bibinfo {year} {1990})}\BibitemShut {NoStop}%
\bibitem [{\citenamefont {Khutsishvili}(1962)}]{khutsishvili1962spin}%
  \BibitemOpen
  \bibfield  {author} {\bibinfo {author} {\bibfnamefont {G.}~\bibnamefont
  {Khutsishvili}},\ }\href
  {https://iopscience.iop.org/article/10.1070/PU1969v011n06ABEH003776}
  {\bibfield  {journal} {\bibinfo  {journal} {Sov. Phys.-JETP}\ }\textbf
  {\bibinfo {volume} {15}} (\bibinfo {year} {1962})}\BibitemShut {NoStop}%
\bibitem [{\citenamefont {Taminiau}\ \emph {et~al.}(2014)\citenamefont
  {Taminiau}, \citenamefont {Cramer}, \citenamefont {van~der Sar},
  \citenamefont {Dobrovitski},\ and\ \citenamefont
  {Hanson}}]{taminiau2014universal}%
  \BibitemOpen
  \bibfield  {author} {\bibinfo {author} {\bibfnamefont {T.~H.}\ \bibnamefont
  {Taminiau}}, \bibinfo {author} {\bibfnamefont {J.}~\bibnamefont {Cramer}},
  \bibinfo {author} {\bibfnamefont {T.}~\bibnamefont {van~der Sar}}, \bibinfo
  {author} {\bibfnamefont {V.~V.}\ \bibnamefont {Dobrovitski}}, \ and\ \bibinfo
  {author} {\bibfnamefont {R.}~\bibnamefont {Hanson}},\ }\href {\doibase
  10.1038/nnano.2014.2} {\bibfield  {journal} {\bibinfo  {journal} {Nat.
  Nanotechnol.}\ }\textbf {\bibinfo {volume} {9}},\ \bibinfo {pages} {171}
  (\bibinfo {year} {2014})}\BibitemShut {NoStop}%
\bibitem [{\citenamefont {Van~der Sar}\ \emph {et~al.}(2012)\citenamefont
  {Van~der Sar}, \citenamefont {Wang}, \citenamefont {Blok}, \citenamefont
  {Bernien}, \citenamefont {Taminiau}, \citenamefont {Toyli}, \citenamefont
  {Lidar}, \citenamefont {Awschalom}, \citenamefont {Hanson},\ and\
  \citenamefont {Dobrovitski}}]{van2012decoherence}%
  \BibitemOpen
  \bibfield  {author} {\bibinfo {author} {\bibfnamefont {T.}~\bibnamefont
  {Van~der Sar}}, \bibinfo {author} {\bibfnamefont {Z.}~\bibnamefont {Wang}},
  \bibinfo {author} {\bibfnamefont {M.}~\bibnamefont {Blok}}, \bibinfo {author}
  {\bibfnamefont {H.}~\bibnamefont {Bernien}}, \bibinfo {author} {\bibfnamefont
  {T.}~\bibnamefont {Taminiau}}, \bibinfo {author} {\bibfnamefont
  {D.}~\bibnamefont {Toyli}}, \bibinfo {author} {\bibfnamefont
  {D.}~\bibnamefont {Lidar}}, \bibinfo {author} {\bibfnamefont
  {D.}~\bibnamefont {Awschalom}}, \bibinfo {author} {\bibfnamefont
  {R.}~\bibnamefont {Hanson}}, \ and\ \bibinfo {author} {\bibfnamefont
  {V.}~\bibnamefont {Dobrovitski}},\ }\href {\doibase 10.1038/nature10900}
  {\bibfield  {journal} {\bibinfo  {journal} {Nature}\ }\textbf {\bibinfo
  {volume} {484}},\ \bibinfo {pages} {82} (\bibinfo {year} {2012})}\BibitemShut
  {NoStop}%
\bibitem [{\citenamefont {Johansson}\ \emph {et~al.}(2013)\citenamefont
  {Johansson}, \citenamefont {Nation},\ and\ \citenamefont
  {Nori}}]{johansson2013qutip}%
  \BibitemOpen
  \bibfield  {author} {\bibinfo {author} {\bibfnamefont {J.~R.}\ \bibnamefont
  {Johansson}}, \bibinfo {author} {\bibfnamefont {P.~D.}\ \bibnamefont
  {Nation}}, \ and\ \bibinfo {author} {\bibfnamefont {F.}~\bibnamefont
  {Nori}},\ }\href {\doibase 10.1016/j.cpc.2012.11.019} {\bibfield  {journal}
  {\bibinfo  {journal} {Comput. Phys. Commun.}\ }\textbf {\bibinfo {volume}
  {184}},\ \bibinfo {pages} {1234} (\bibinfo {year} {2013})}\BibitemShut
  {NoStop}%
\bibitem [{\citenamefont {Kolkowitz}\ \emph {et~al.}(2012)\citenamefont
  {Kolkowitz}, \citenamefont {Unterreithmeier}, \citenamefont {Bennett},\ and\
  \citenamefont {Lukin}}]{Kolkowitz_PRL2012}%
  \BibitemOpen
  \bibfield  {author} {\bibinfo {author} {\bibfnamefont {S.}~\bibnamefont
  {Kolkowitz}}, \bibinfo {author} {\bibfnamefont {Q.~P.}\ \bibnamefont
  {Unterreithmeier}}, \bibinfo {author} {\bibfnamefont {S.~D.}\ \bibnamefont
  {Bennett}}, \ and\ \bibinfo {author} {\bibfnamefont {M.~D.}\ \bibnamefont
  {Lukin}},\ }\href {\doibase 10.1103/PhysRevLett.109.137601} {\bibfield
  {journal} {\bibinfo  {journal} {Phys. Rev. Lett.}\ }\textbf {\bibinfo
  {volume} {109}},\ \bibinfo {pages} {137601} (\bibinfo {year}
  {2012})}\BibitemShut {NoStop}%
\bibitem [{\citenamefont {Taminiau}\ \emph {et~al.}(2012)\citenamefont
  {Taminiau}, \citenamefont {Wagenaar}, \citenamefont {van~der Sar},
  \citenamefont {Jelezko}, \citenamefont {Dobrovitski},\ and\ \citenamefont
  {Hanson}}]{Taminiau_PRL2012}%
  \BibitemOpen
  \bibfield  {author} {\bibinfo {author} {\bibfnamefont {T.~H.}\ \bibnamefont
  {Taminiau}}, \bibinfo {author} {\bibfnamefont {J.~J.~T.}\ \bibnamefont
  {Wagenaar}}, \bibinfo {author} {\bibfnamefont {T.}~\bibnamefont {van~der
  Sar}}, \bibinfo {author} {\bibfnamefont {F.}~\bibnamefont {Jelezko}},
  \bibinfo {author} {\bibfnamefont {V.~V.}\ \bibnamefont {Dobrovitski}}, \ and\
  \bibinfo {author} {\bibfnamefont {R.}~\bibnamefont {Hanson}},\ }\href
  {\doibase 10.1103/PhysRevLett.109.137602} {\bibfield  {journal} {\bibinfo
  {journal} {Phys. Rev. Lett.}\ }\textbf {\bibinfo {volume} {109}},\ \bibinfo
  {pages} {137602} (\bibinfo {year} {2012})}\BibitemShut {NoStop}%
\bibitem [{\citenamefont {Zhao}\ \emph {et~al.}(2012)\citenamefont {Zhao},
  \citenamefont {Honert}, \citenamefont {Schmid}, \citenamefont {Klas},
  \citenamefont {Isoya}, \citenamefont {Markham}, \citenamefont {Twitchen},
  \citenamefont {Jelezko}, \citenamefont {Liu}, \citenamefont {Fedder} \emph
  {et~al.}}]{zhao2012sensing}%
  \BibitemOpen
  \bibfield  {author} {\bibinfo {author} {\bibfnamefont {N.}~\bibnamefont
  {Zhao}}, \bibinfo {author} {\bibfnamefont {J.}~\bibnamefont {Honert}},
  \bibinfo {author} {\bibfnamefont {B.}~\bibnamefont {Schmid}}, \bibinfo
  {author} {\bibfnamefont {M.}~\bibnamefont {Klas}}, \bibinfo {author}
  {\bibfnamefont {J.}~\bibnamefont {Isoya}}, \bibinfo {author} {\bibfnamefont
  {M.}~\bibnamefont {Markham}}, \bibinfo {author} {\bibfnamefont
  {D.}~\bibnamefont {Twitchen}}, \bibinfo {author} {\bibfnamefont
  {F.}~\bibnamefont {Jelezko}}, \bibinfo {author} {\bibfnamefont {R.-B.}\
  \bibnamefont {Liu}}, \bibinfo {author} {\bibfnamefont {H.}~\bibnamefont
  {Fedder}},  \emph {et~al.},\ }\href {\doibase 10.1038/nnano.2012.152}
  {\bibfield  {journal} {\bibinfo  {journal} {Nat. Nanotechnol.}\ }\textbf
  {\bibinfo {volume} {7}},\ \bibinfo {pages} {657} (\bibinfo {year}
  {2012})}\BibitemShut {NoStop}%
\bibitem [{\citenamefont {Childress}\ \emph {et~al.}(2006)\citenamefont
  {Childress}, \citenamefont {Gurudev~Dutt}, \citenamefont {Taylor},
  \citenamefont {Zibrov}, \citenamefont {Jelezko}, \citenamefont {Wrachtrup},
  \citenamefont {Hemmer},\ and\ \citenamefont {Lukin}}]{Childress_Science2006}%
  \BibitemOpen
  \bibfield  {author} {\bibinfo {author} {\bibfnamefont {L.}~\bibnamefont
  {Childress}}, \bibinfo {author} {\bibfnamefont {M.~V.}\ \bibnamefont
  {Gurudev~Dutt}}, \bibinfo {author} {\bibfnamefont {J.~M.}\ \bibnamefont
  {Taylor}}, \bibinfo {author} {\bibfnamefont {A.~S.}\ \bibnamefont {Zibrov}},
  \bibinfo {author} {\bibfnamefont {F.}~\bibnamefont {Jelezko}}, \bibinfo
  {author} {\bibfnamefont {J.}~\bibnamefont {Wrachtrup}}, \bibinfo {author}
  {\bibfnamefont {P.~R.}\ \bibnamefont {Hemmer}}, \ and\ \bibinfo {author}
  {\bibfnamefont {M.~D.}\ \bibnamefont {Lukin}},\ }\href {\doibase
  10.1126/science.1131871} {\bibfield  {journal} {\bibinfo  {journal}
  {Science}\ }\textbf {\bibinfo {volume} {314}},\ \bibinfo {pages} {281}
  (\bibinfo {year} {2006})}\BibitemShut {NoStop}%
\bibitem [{\citenamefont {Zhao}\ \emph {et~al.}(2011)\citenamefont {Zhao},
  \citenamefont {Hu}, \citenamefont {Ho}, \citenamefont {Wan},\ and\
  \citenamefont {Liu}}]{zhao2011atomic}%
  \BibitemOpen
  \bibfield  {author} {\bibinfo {author} {\bibfnamefont {N.}~\bibnamefont
  {Zhao}}, \bibinfo {author} {\bibfnamefont {J.-L.}\ \bibnamefont {Hu}},
  \bibinfo {author} {\bibfnamefont {S.-W.}\ \bibnamefont {Ho}}, \bibinfo
  {author} {\bibfnamefont {J.~T.}\ \bibnamefont {Wan}}, \ and\ \bibinfo
  {author} {\bibfnamefont {R.}~\bibnamefont {Liu}},\ }\href {\doibase
  10.1038/nnano.2011.22} {\bibfield  {journal} {\bibinfo  {journal} {Nat.
  Nanotechnol.}\ }\textbf {\bibinfo {volume} {6}},\ \bibinfo {pages} {242}
  (\bibinfo {year} {2011})}\BibitemShut {NoStop}%
\bibitem [{\citenamefont {Sangtawesin}\ \emph {et~al.}(2016)\citenamefont
  {Sangtawesin}, \citenamefont {McLellan}, \citenamefont {Myers}, \citenamefont
  {Jayich}, \citenamefont {Awschalom},\ and\ \citenamefont
  {Petta}}]{sangtawesin2016hyperfine}%
  \BibitemOpen
  \bibfield  {author} {\bibinfo {author} {\bibfnamefont {S.}~\bibnamefont
  {Sangtawesin}}, \bibinfo {author} {\bibfnamefont {C.}~\bibnamefont
  {McLellan}}, \bibinfo {author} {\bibfnamefont {B.}~\bibnamefont {Myers}},
  \bibinfo {author} {\bibfnamefont {A.~B.}\ \bibnamefont {Jayich}}, \bibinfo
  {author} {\bibfnamefont {D.}~\bibnamefont {Awschalom}}, \ and\ \bibinfo
  {author} {\bibfnamefont {J.}~\bibnamefont {Petta}},\ }\href {\doibase
  10.1088/1367-2630/18/8/083016} {\bibfield  {journal} {\bibinfo  {journal}
  {New J. Phys.}\ }\textbf {\bibinfo {volume} {18}},\ \bibinfo {pages} {083016}
  (\bibinfo {year} {2016})}\BibitemShut {NoStop}%
\bibitem [{\citenamefont {Cramer}\ \emph {et~al.}(2016)\citenamefont {Cramer},
  \citenamefont {Kalb}, \citenamefont {Rol}, \citenamefont {Hensen},
  \citenamefont {Blok}, \citenamefont {Markham}, \citenamefont {Twitchen},
  \citenamefont {Hanson},\ and\ \citenamefont
  {Taminiau}}]{Cramer_NatureComm2016}%
  \BibitemOpen
  \bibfield  {author} {\bibinfo {author} {\bibfnamefont {J.}~\bibnamefont
  {Cramer}}, \bibinfo {author} {\bibfnamefont {N.}~\bibnamefont {Kalb}},
  \bibinfo {author} {\bibfnamefont {M.~A.}\ \bibnamefont {Rol}}, \bibinfo
  {author} {\bibfnamefont {B.}~\bibnamefont {Hensen}}, \bibinfo {author}
  {\bibfnamefont {M.~S.}\ \bibnamefont {Blok}}, \bibinfo {author}
  {\bibfnamefont {M.}~\bibnamefont {Markham}}, \bibinfo {author} {\bibfnamefont
  {D.~J.}\ \bibnamefont {Twitchen}}, \bibinfo {author} {\bibfnamefont
  {R.}~\bibnamefont {Hanson}}, \ and\ \bibinfo {author} {\bibfnamefont {T.~H.}\
  \bibnamefont {Taminiau}},\ }\href {\doibase 10.1038/ncomms11526} {\bibfield
  {journal} {\bibinfo  {journal} {Nat. Commun.}\ }\textbf {\bibinfo {volume}
  {7}},\ \bibinfo {pages} {11526} (\bibinfo {year} {2016})}\BibitemShut
  {NoStop}%
\bibitem [{\citenamefont {Kalb}\ \emph {et~al.}(2016)\citenamefont {Kalb},
  \citenamefont {Cramer}, \citenamefont {Twitchen}, \citenamefont {Markham},
  \citenamefont {Hanson},\ and\ \citenamefont
  {Taminiau}}]{Kalb_NatureComm2016}%
  \BibitemOpen
  \bibfield  {author} {\bibinfo {author} {\bibfnamefont {N.}~\bibnamefont
  {Kalb}}, \bibinfo {author} {\bibfnamefont {J.}~\bibnamefont {Cramer}},
  \bibinfo {author} {\bibfnamefont {D.~J.}\ \bibnamefont {Twitchen}}, \bibinfo
  {author} {\bibfnamefont {M.}~\bibnamefont {Markham}}, \bibinfo {author}
  {\bibfnamefont {R.}~\bibnamefont {Hanson}}, \ and\ \bibinfo {author}
  {\bibfnamefont {T.~H.}\ \bibnamefont {Taminiau}},\ }\href {\doibase
  10.1038/ncomms13111} {\bibfield  {journal} {\bibinfo  {journal} {Nat.
  Commun.}\ }\textbf {\bibinfo {volume} {7}},\ \bibinfo {pages} {13111}
  (\bibinfo {year} {2016})}\BibitemShut {NoStop}%
\bibitem [{\citenamefont {Kalb}\ \emph {et~al.}(2017)\citenamefont {Kalb},
  \citenamefont {Reiserer}, \citenamefont {Humphreys}, \citenamefont
  {Bakermans}, \citenamefont {Kamerling}, \citenamefont {Nickerson},
  \citenamefont {Benjamin}, \citenamefont {Twitchen}, \citenamefont {Markham},\
  and\ \citenamefont {Hanson}}]{kalb2017entanglement}%
  \BibitemOpen
  \bibfield  {author} {\bibinfo {author} {\bibfnamefont {N.}~\bibnamefont
  {Kalb}}, \bibinfo {author} {\bibfnamefont {A.~A.}\ \bibnamefont {Reiserer}},
  \bibinfo {author} {\bibfnamefont {P.~C.}\ \bibnamefont {Humphreys}}, \bibinfo
  {author} {\bibfnamefont {J.~J.}\ \bibnamefont {Bakermans}}, \bibinfo {author}
  {\bibfnamefont {S.~J.}\ \bibnamefont {Kamerling}}, \bibinfo {author}
  {\bibfnamefont {N.~H.}\ \bibnamefont {Nickerson}}, \bibinfo {author}
  {\bibfnamefont {S.~C.}\ \bibnamefont {Benjamin}}, \bibinfo {author}
  {\bibfnamefont {D.~J.}\ \bibnamefont {Twitchen}}, \bibinfo {author}
  {\bibfnamefont {M.}~\bibnamefont {Markham}}, \ and\ \bibinfo {author}
  {\bibfnamefont {R.}~\bibnamefont {Hanson}},\ }\href {\doibase
  10.1126/science.aan0070} {\bibfield  {journal} {\bibinfo  {journal}
  {Science}\ }\textbf {\bibinfo {volume} {356}},\ \bibinfo {pages} {928}
  (\bibinfo {year} {2017})}\BibitemShut {NoStop}%
\bibitem [{\citenamefont {Blok}\ \emph {et~al.}(2014)\citenamefont {Blok},
  \citenamefont {Bonato}, \citenamefont {Markham}, \citenamefont {Twitchen},
  \citenamefont {Dobrovitski},\ and\ \citenamefont
  {Hanson}}]{blok2014manipulating}%
  \BibitemOpen
  \bibfield  {author} {\bibinfo {author} {\bibfnamefont {M.}~\bibnamefont
  {Blok}}, \bibinfo {author} {\bibfnamefont {C.}~\bibnamefont {Bonato}},
  \bibinfo {author} {\bibfnamefont {M.}~\bibnamefont {Markham}}, \bibinfo
  {author} {\bibfnamefont {D.}~\bibnamefont {Twitchen}}, \bibinfo {author}
  {\bibfnamefont {V.}~\bibnamefont {Dobrovitski}}, \ and\ \bibinfo {author}
  {\bibfnamefont {R.}~\bibnamefont {Hanson}},\ }\href {\doibase
  10.1038/nphys2881} {\bibfield  {journal} {\bibinfo  {journal} {Nat. Phys.}\
  }\textbf {\bibinfo {volume} {10}},\ \bibinfo {pages} {189} (\bibinfo {year}
  {2014})}\BibitemShut {NoStop}%
\bibitem [{\citenamefont {Meurer}\ \emph {et~al.}(2017)\citenamefont {Meurer},
  \citenamefont {Smith}, \citenamefont {Paprocki}, \citenamefont
  {\v{C}ert\'{i}k}, \citenamefont {Kirpichev}, \citenamefont {Rocklin},
  \citenamefont {Kumar}, \citenamefont {Ivanov}, \citenamefont {Moore},
  \citenamefont {Singh}, \citenamefont {Rathnayake}, \citenamefont {Vig},
  \citenamefont {Granger}, \citenamefont {Muller}, \citenamefont {Bonazzi},
  \citenamefont {Gupta}, \citenamefont {Vats}, \citenamefont {Johansson},
  \citenamefont {Pedregosa}, \citenamefont {Curry}, \citenamefont {Terrel},
  \citenamefont {Rou\v{c}ka}, \citenamefont {Saboo}, \citenamefont {Fernando},
  \citenamefont {Kulal}, \citenamefont {Cimrman},\ and\ \citenamefont
  {Scopatz}}]{sympy}%
  \BibitemOpen
  \bibfield  {author} {\bibinfo {author} {\bibfnamefont {A.}~\bibnamefont
  {Meurer}}, \bibinfo {author} {\bibfnamefont {C.~P.}\ \bibnamefont {Smith}},
  \bibinfo {author} {\bibfnamefont {M.}~\bibnamefont {Paprocki}}, \bibinfo
  {author} {\bibfnamefont {O.}~\bibnamefont {\v{C}ert\'{i}k}}, \bibinfo
  {author} {\bibfnamefont {S.~B.}\ \bibnamefont {Kirpichev}}, \bibinfo {author}
  {\bibfnamefont {M.}~\bibnamefont {Rocklin}}, \bibinfo {author} {\bibfnamefont
  {A.}~\bibnamefont {Kumar}}, \bibinfo {author} {\bibfnamefont
  {S.}~\bibnamefont {Ivanov}}, \bibinfo {author} {\bibfnamefont {J.~K.}\
  \bibnamefont {Moore}}, \bibinfo {author} {\bibfnamefont {S.}~\bibnamefont
  {Singh}}, \bibinfo {author} {\bibfnamefont {T.}~\bibnamefont {Rathnayake}},
  \bibinfo {author} {\bibfnamefont {S.}~\bibnamefont {Vig}}, \bibinfo {author}
  {\bibfnamefont {B.~E.}\ \bibnamefont {Granger}}, \bibinfo {author}
  {\bibfnamefont {R.~P.}\ \bibnamefont {Muller}}, \bibinfo {author}
  {\bibfnamefont {F.}~\bibnamefont {Bonazzi}}, \bibinfo {author} {\bibfnamefont
  {H.}~\bibnamefont {Gupta}}, \bibinfo {author} {\bibfnamefont
  {S.}~\bibnamefont {Vats}}, \bibinfo {author} {\bibfnamefont {F.}~\bibnamefont
  {Johansson}}, \bibinfo {author} {\bibfnamefont {F.}~\bibnamefont
  {Pedregosa}}, \bibinfo {author} {\bibfnamefont {M.~J.}\ \bibnamefont
  {Curry}}, \bibinfo {author} {\bibfnamefont {A.~R.}\ \bibnamefont {Terrel}},
  \bibinfo {author} {\bibfnamefont {v.}~\bibnamefont {Rou\v{c}ka}}, \bibinfo
  {author} {\bibfnamefont {A.}~\bibnamefont {Saboo}}, \bibinfo {author}
  {\bibfnamefont {I.}~\bibnamefont {Fernando}}, \bibinfo {author}
  {\bibfnamefont {S.}~\bibnamefont {Kulal}}, \bibinfo {author} {\bibfnamefont
  {R.}~\bibnamefont {Cimrman}}, \ and\ \bibinfo {author} {\bibfnamefont
  {A.}~\bibnamefont {Scopatz}},\ }\href {\doibase 10.7717/peerj-cs.103}
  {\bibfield  {journal} {\bibinfo  {journal} {PeerJ Comput. Sci.}\ }\textbf
  {\bibinfo {volume} {3}},\ \bibinfo {pages} {e103} (\bibinfo {year}
  {2017})}\BibitemShut {NoStop}%
\bibitem [{\citenamefont {Zopes}\ \emph {et~al.}(2018)\citenamefont {Zopes},
  \citenamefont {Herb}, \citenamefont {Cujia},\ and\ \citenamefont
  {Degen}}]{zopes2018three}%
  \BibitemOpen
  \bibfield  {author} {\bibinfo {author} {\bibfnamefont {J.}~\bibnamefont
  {Zopes}}, \bibinfo {author} {\bibfnamefont {K.}~\bibnamefont {Herb}},
  \bibinfo {author} {\bibfnamefont {K.}~\bibnamefont {Cujia}}, \ and\ \bibinfo
  {author} {\bibfnamefont {C.}~\bibnamefont {Degen}},\ }\href {\doibase
  10.1103/PhysRevLett.121.170801} {\bibfield  {journal} {\bibinfo  {journal}
  {Phys. Rev. Lett.}\ }\textbf {\bibinfo {volume} {121}},\ \bibinfo {pages}
  {170801} (\bibinfo {year} {2018})}\BibitemShut {NoStop}%
\bibitem [{\citenamefont {Sasaki}\ \emph {et~al.}(2018)\citenamefont {Sasaki},
  \citenamefont {Itoh},\ and\ \citenamefont {Abe}}]{sasaki2018determination}%
  \BibitemOpen
  \bibfield  {author} {\bibinfo {author} {\bibfnamefont {K.}~\bibnamefont
  {Sasaki}}, \bibinfo {author} {\bibfnamefont {K.~M.}\ \bibnamefont {Itoh}}, \
  and\ \bibinfo {author} {\bibfnamefont {E.}~\bibnamefont {Abe}},\ }\href
  {\doibase 10.1103/PhysRevB.98.121405} {\bibfield  {journal} {\bibinfo
  {journal} {Phys. Rev. B}\ }\textbf {\bibinfo {volume} {98}},\ \bibinfo
  {pages} {121405} (\bibinfo {year} {2018})}\BibitemShut {NoStop}%
\bibitem [{\citenamefont {Laraoui}\ \emph {et~al.}(2015)\citenamefont
  {Laraoui}, \citenamefont {Pagliero},\ and\ \citenamefont
  {Meriles}}]{laraoui2015imaging}%
  \BibitemOpen
  \bibfield  {author} {\bibinfo {author} {\bibfnamefont {A.}~\bibnamefont
  {Laraoui}}, \bibinfo {author} {\bibfnamefont {D.}~\bibnamefont {Pagliero}}, \
  and\ \bibinfo {author} {\bibfnamefont {C.~A.}\ \bibnamefont {Meriles}},\
  }\href {\doibase 10.1103/PhysRevB.91.205410} {\bibfield  {journal} {\bibinfo
  {journal} {Phys. Rev. B}\ }\textbf {\bibinfo {volume} {91}},\ \bibinfo
  {pages} {205410} (\bibinfo {year} {2015})}\BibitemShut {NoStop}%
\bibitem [{\citenamefont {Nielsen}\ and\ \citenamefont
  {Chuang}(2002)}]{nielsen2002quantum}%
  \BibitemOpen
  \bibfield  {author} {\bibinfo {author} {\bibfnamefont {M.~A.}\ \bibnamefont
  {Nielsen}}\ and\ \bibinfo {author} {\bibfnamefont {I.}~\bibnamefont
  {Chuang}},\ }\href {\doibase 10.1017/CBO9780511976667} {\enquote {\bibinfo
  {title} {Quantum computation and quantum information},}\ } (\bibinfo {year}
  {2002})\BibitemShut {NoStop}%
\end{thebibliography}%

\end{document}


\preprint{}

\title{Supplemental Material: A 10-qubit solid-state spin register with quantum memory up to one minute}

\date{\today}
\maketitle

\section{Experimental Details}

\subsection{Sample}

All experiments are performed on a high-purity, chemical-vapour-deposition homoexpitaxially grown diamond (type IIa) with natural abundance of carbon isotopes (1.1\% $^{13}$C). The diamond was grown and cleaved (along the $\langle 111 \rangle$ crystal
axis) by Element Six. We work with a naturally occurring NV center which was selected based on the absence of strongly coupled ($>$500 kHz hyperfine coupling) $^{13}$C spins, but without any other criteria on the spin environment. Microwave and RF fields are applied via a lithographically defined gold stripline. A solid-immersion lens enhances photoluminescence collection efficiency \cite{hadden2010strongly,robledo2011high}, and an aluminium-oxide layer is grown by atomic-layer-deposition to serve as an anti-reflection coating \cite{pfaff2013demonstration,yeung2012anti}.

\subsection{Confocal microscope and NV operations}

The sample is held in a home-built confocal microscope based upon a closed cycle cryostat (Montana Cryostation, 3.7 K). We measure long NV electron spin relaxation and spin coherence times ($T_{1}$ $>$ 1 h, $T^{DD}_{2}>$ 1 s) \cite{abobeih2018one}, and we achieve fast spin operations (Rabi frequency: 14 MHz) and readout/initialization ($\sim$ 10 \textmu s \cite{robledo2011high}). We perform spin-selective optical readout of the NV-center to measure the spin state in a single shot, with fidelities of 0.905(2) for the bright state ($m_{s} = 0, := \ket{0}$) and 0.986(2) for the dark state ($m_{s} = -1, := \ket{1}$). The $E_{x}$ and $E_y$ optical transitions are split by $4.0\,$GHz due to lateral strain. All measurements are corrected for the finite readout fidelity in order to provide a best estimate for the actual state.

\subsection{Magnetic field stabilization}

An external magnetic field is applied along the NV-axis, with magnitude $\sim$ 403 G. A PI loop (Team-Wavelength WEC3293) is used to stabilize the temperature of the permanent neodymium magnet, and thus the field. The temperature stability is measured to be 1.13 mK across a 24 hour period (one standard deviation), corresponding to a magnetic field stability of $\sim$ 450 \textmu G, or a 0.5 Hz frequency shift for the $^{13}$C spins, which is negligible compared to the intrinsic linewidth (1/$\sqrt{2}\pi T^{*}_{2}$) of $\sim$ 20 Hz. The field alignment is calibrated to within 0.1 degrees using a thermal echo sequence \cite{Abobeih_arXiv2019}. 

\subsection{Microwave and RF setup}\label{sec:mwRF}

Microwave amplifier (AR 25S1G6) noise is suppressed by a fast microwave switch (TriQuint TGS2355-SM) with a suppression ratio of 40 dB. Video leakage noise generated by the switch is filtered with a high pass filter. To obtain effective MW pulses independent of the $^{14}$N nuclear spin state, all microwave operations are performed using Hermite envelopes \cite{vandersypen2005nmr,warren1984effects}. During dynamical decoupling, pulse errors are mitigated by implementation of the XY8 scheme \cite{gullion1990new}. In the initial two-qubit DDRF gate experiments (Fig. 4 in the main text), we use an RF amplifier (Analog Devices ADA4870). To reduce ringing of the RF signal due to sub-period switching in the presence of an AC-coupling from a high-pass filter (3rd order Butterworth, 52 kHz, home-built), we use erf pulse shapes, with envelope function

\begin{equation}
    f(t) = 1 - \frac{1}{2}\erf \left(\frac{2(\Delta t - t + t_0)}{\Delta t}\right) - \frac{1}{2}\erf \left(\frac{2(\Delta t + t - t_\text{pulse})}{\Delta t}\right),
\end{equation}
where $\Delta t$ is the risetime, $t_0$ is the start time of the pulse, $t_\text{pulse}$ is the pulse length and $\erf(x)$ is the error function. We also use erf pulses for the RF spin echo pulses (see section \ref{sec:echoes}). In the multi-qubit experiments, the RF signal is generated directly by an AWG (Tektronix 5014C, 4.5V pk-pk) to avoid heating of the chip and to allow shorter RF pulses without ringing from filters in the RF electronics. In all cases, we ensure that the RF pulses are comprised of an integer number of periods of the RF waveform, i.e. we ensure that $\omega t_\text{pulse} = 2\pi n$ for integer $n$, where $t_\text{pulse}$ is the RF pulse length. This ensures that any phase picked up on the electron spin due to the RF pulse is cancelled. We set $n$ to maximize the length of $t_\text{pulse}$ while ensuring that $t_\text{pulse} < \tau$ for the first and last RF pulse in the DDRF sequence. For the remainder of the pulses, we use the value $2n$.

\subsection{Optical setup}
Laser pulses are generated by direct current modulation (515 nm laser for charge state control, Cobolt MLD, on/off ratio of $>$135 dB) or by acoustic optical modulators (637 nm lasers for spin pumping and single-shot readout, Toptica DL Pro and New Focus TLB-6704-P). By placing two modulators in series (Gooch and Housego Fibre Q), an on/off ratio of $>$100 dB is obtained for the 637 nm lasers \cite{abobeih2018one}. The laser frequencies are stabilized to within 2 MHz using a wavemeter (HF-ANGSTROM WS/U-10U). 

\section{10 Qubit Register Characterization}

\subsection{Qubit frequencies}

\begin{table}[h!]
  \centering
  \begin{tabular}{c|c|c|c|c|c}
  			& $\omega_{0} / 2\pi$ [Hz]& $\omega_{-1} / 2\pi$ [Hz] & $\omega_{+1} / 2\pi$ [Hz] & $A_\parallel /2\pi$ [kHz] & $A_\perp/2\pi$ [kHz] \\ \hline \hline
   C1	    &	431961(1)       & 	218828(1)   & 645123(1)  &   213.154(1)  &	3.0(4)	\\
   C2	    &	431956(1)       & 	469020(1)   & 396542(1)  &   -36.308(1)  &	26.62(4) 	\\
   C3 	    &	431958(1)       &   413477(1)   & 454427(1)  &   20.569(1) 	& 	41.51(3) 	\\
   C4	    &   431951(1)       &   447234(1)   & 424752(3)  &   -11.346(2)	&   59.21(3) 	\\
   C5	    &   431962(1)       &   408317(1)   & 457035(1)  &   24.399(1) 	&	24.81(4) 	\\
   C6	    &	431962(1)       &   480625(1)   & 383480(40) &   -48.58(2) 	&	9(2)	\\
   C7      &	431956(1)       &   451802(1)   & 412175(5)  &   -19.815(3) &	5.3(5) 	\\
   C8      &	431973(1)       &   414407(1)   & 449687(2)  &   17.643(1) 	&	8.6(2)	\\
   $^{14}$N	    &	5069110(1)      &   2884865(1)  & 7263440(1) &   2189.288(1)   &	- 	\\
   \end{tabular}
  \caption{\textbf{Precession frequencies and hyperfine couplings for the nuclear spins.} $\omega_{0}$, $\omega_{-1}$, and $\omega_{+1}$ are the measured nuclear precession frequencies for the $m_s = 0, -1$ and $+1$ electron spin projections respectively, obtained from least-squares fits of Ramsey signals. $A_\parallel$ and $A_\perp$ are the hyperfine interaction components parallel and perpendicular to the applied magnetic field, obtained using the approximate relations in Eq. \ref{eq:Hyperfine}. For the $^{14}$N spin the frequencies given are for the $m_I = 0 \leftrightarrow m_I = -1$ transition, and the parallel hyperfine component is taken as $(\omega_{+1} - \omega_{-1})/2$. Note that we use the definition $\omega_1 = \omega_{-1}$ in the main text and other sections of the Supplemental Material.}
  \label{Tab:Spins}
\end{table}

In table \ref{Tab:Spins}, we give the measured precession frequencies for each nuclear spin, determined from Ramsey spectroscopy with the electron stored in the $m_{s} = 0,-1$ and $+1$ spin projections. For the $^{13}$C spins, under both the secular approximation and the assumption of a perfectly aligned field, the hyperfine parameters can then be obtained from the relations

\begin{equation}
\begin{split}
A_{\parallel} &= \frac{\omega_{+1}^{2} - \omega_{-1}^{2}}{4\omega_{0}} \\
A_{\bot} &= \sqrt{\frac{\omega_{+1}^{2}+\omega_{-1}^{2} -2\omega_{0}^{2}-2A_{\parallel}^{2}}{2}}.
\label{eq:Hyperfine}
\end{split}
\end{equation}
Note that due to the high accuracy of the Ramsey measurements, small deviations from Eqs. \ref{eq:Hyperfine} (due to non-secular Hamiltonian terms and magnetic field misalignment) are likely to be larger than the uncertainties in the parameters extracted by propagating the measurement errors. Therefore, the quoted values are approximate. 

\subsection{Coherence measurements}

\begin{table}[ht!]
  \centering
  \begin{tabular}{c|c|c|c|c}
  			& $T^{*}_{2}$ ($m_{s} = -1$) [ms] & $T^{*}_{2}$ ($m_{s} = 0$) [ms] & $T_{2}$  ($m_{s} = -1$) [s] & $T^{\alpha=256}_{2}$ ($m_{s} = -1$) [s] \\ \hline \hline
   C1	    & 12.0(6)   & 10.0(3)    & 0.77(4) &  25(4)      \\
   C2	    & 9.2(8)    & 9.1(5)     & 0.53(6) &  6.8(8) \ [$\alpha = 256$]     \\
            &           &            &         &  10.6(7) [$\alpha = 400$]          \\
   C3 	    & 11.9(5)   & 12.3(3)    & 0.68(3) &  7.4(8)      \\
   C4	    & 5.7(2)    & 5.3(4)     & 0.53(6) &  4.1(4)         \\
   C5	    & 15.6(8)   & 17.2(4)    & 0.62(3) &  12.9(4)    \\
   C6	    & 3.7(2)    & 3.6(2)     & 0.59(2) &  13(2)    \\
   C7       & 4.1(6)    & 4.6(7)     & 0.52(3) &  4(1)      \\
   C8       & 7.6(4)    & 7.6(3)     & 0.26(3) &  1.2(2) [$\alpha = 8$]   \\
   $^{14}$N	& 23.2(7)   & 25.1(7)    & 2.3(2)  &  63(2) \\
   \end{tabular}

  \caption{\textbf{Coherence times for the nuclear spins.} $T^{*}_{2}$ is obtained from a least-squares fit of Ramsey signals, with evolution for the $m_{s} = -1$ and $m_s = 0$ electron spin projections. $T_{2}$ is obtained from a least-squares fit of a spin echo experiment, with free evolution measured for the $m_{s} = -1$ electron spin projection. $T^{\alpha=256}_{2}$ is obtained from a least-squares fit of a dynamical decoupling experiment with $\alpha = 256$ pulses, with the electron spin in the $m_{s} = -1$ spin projection.}
  \label{Tab:SpinsCoh}
\end{table}

\begin{figure*}[!ht]
    \centering
    \includegraphics[width=0.5\textwidth]{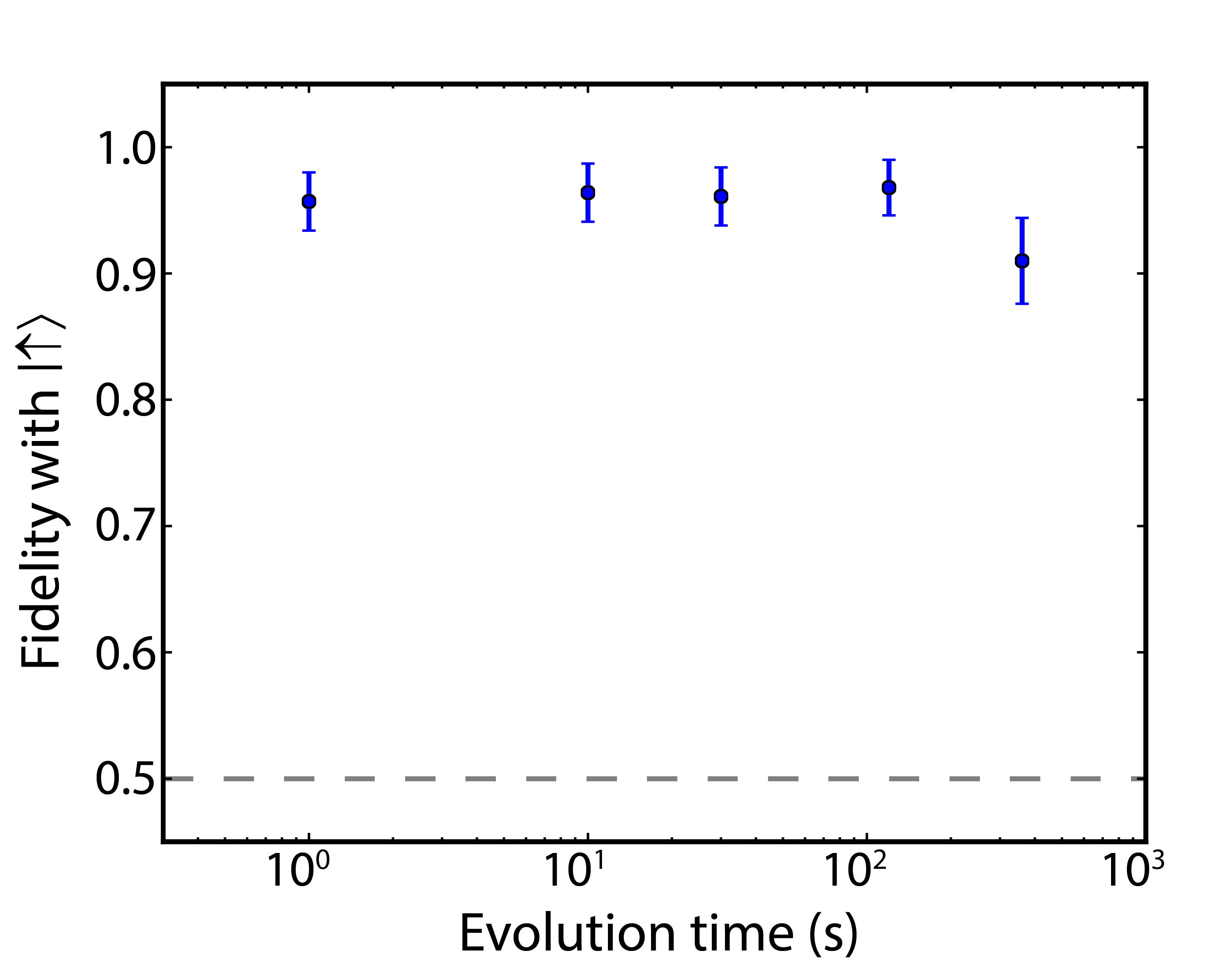}
    \caption{Longitudinal relaxation ($T_{1}$) measurement for nuclear spin C5. We prepare the nuclear spin in the $\ket{\uparrow}$ eigenstate, and the electron spin in the $m_{s} = -1$ projection. On timescales up to 6 minutes, only a minor decay is observed.} 
    \label{fig:T1}
\end{figure*}

Table \ref{Tab:SpinsCoh} contains the measured coherence properties of each spin. $T^{*}_{2}$ times are measured with the electron in the $m_{s} = -1$ and $m_s = 0$ spin projections, while $T_{2}$ measurements are performed with a single spin echo pulse for the $m_s = -1$ electron spin projection. A significantly lower $T_{2}$ time is observed for the $m_{s} = 0$ spin projection ($\sim$ 15-90 ms). This effect is attributed to the frozen core created by the NV center, whereby the hyperfine interaction suppresses magnetic field noise due to flip-flop interactions \cite{khutsishvili1962spin}. A similar effect is observed for the longitudinal relaxation times of the $^{13}$C spins. Due to the times required to acquire data, we do not perform full $T_{1}$ characterizations for all nuclear spins with $m_s = -1$. However, a verification measurement is performed for a single nuclear spin, with the electron spin prepared in the $m_{s} = -1$ spin projection. We measure the decay from the $\ket{\uparrow}$ state for time periods up to 6 minutes, as shown in Fig. \ref{fig:T1}. On these timescales, only a small decay is observed. Dynamical decoupling measurements are performed with $\alpha=256$ pulses with the electron again in the $m_s = -1$ projection. For spins C6 and C8, the best signal was achieved with all pulses performed around the $\hat{x}$-axis. For the other spins, $XY8$ sequences were used \cite{gullion1990new}. For C8, we were unable to measure the coherence time for 256 pulses due to strong pulse errors caused by spectral overlap with other spins. Optimized pulse sequences for such situations will be investigated further in the future.

\subsection{Gate parameters}

\begin{table}[ht!]
  \centering
  \begin{tabular}{c|c|c|c|c}
  			& N & $\tau$ (\textmu s)& Gate duration (\textmu s) & RF $\pi$-pulse duration (\textmu s) \\ \hline \hline
   C1$^{*}$ &	8           & 	 39.356    & 629    & 469       \\
   C1       &   48          &   16.204     & 1556   & 1330      \\
   C2	    &	44          & 	 7.218     & 635    & 1096	    \\
   C3 	    &	22          &   11.250     & 495    & 929	    \\
   C4	    &   20          &   16.480     & 659    & 734	    \\
   C5	    &   32          &   6.540      & 419    & 1606	    \\
   C6	    &	90          &   4.932      & 888    & 1173	    \\
   C7       &	64          &   18.522     & 2371   & 1024      \\
   C8       &	48          &   23.152     & 2222   & 1592      \\
   $^{14}$N	&	12          &   16.204     & 389    & 278       \\
   \end{tabular}

  \caption{\textbf{Gate parameters for the nuclear spins.} $N$ and $\tau$ parameterise a dynamical decoupling gate with $N$ electron $\pi$-pulses, separated by 2$\tau$ (see Fig. 2(a) of the main text). Two sets of parameters are given for C1; the starred entry corresponds to the case with use of an RF amplifier (see text for details).}
  \label{Tab:SpinsGate}
\end{table}

The electron-nuclear gate parameters for each spin are given in table \ref{Tab:SpinsGate}. Spins C1, C7, C8 and $^{14}$N are all controlled using the DDRF gate scheme, while the remaining spins are controlled using previously developed dynamical decoupling based gates \cite{taminiau2014universal}. Two parameter regimes are given for the electron-nuclear gate on spin C1. To produce the results shown in Fig. 4(b) of the main text, we used a RF amplifier to increase the nuclear spin Rabi frequency. However, operating with this power caused issues during experiments involving multiple nuclear spins. This is due to two effects; firstly, a loss of readout photons was observed due to heating of the sample. For the final readout, this could be mitigated by a wait duration of $\sim 1$ ms, as the longitudinal relaxation time of the electron spin is long ($>1$ hour). Secondly, the use of an amplifier required additional high-pass filters due to the presence of low-frequency noise (see section \ref{sec:mwRF}). The AC-coupling induced by these filters led to ringing effects over significant timescales ($\sim 10-100$ \textmu s) which can induce unwanted phases on the electron spin. We expect that these issues can be mitigated in future experiments by using a free-space RF antenna to prevent heating of the sample and an improved RF delivery system to reduce ringing. For spins C7 and C8, longer gate durations are used to avoid crosstalk with additional spectrally close nuclear spins ($\sim$ 1 kHz) \cite{Abobeih_arXiv2019}.

\section{Two qubit gates: theory}\label{sec:DDRFtheory}

In this section we derive equations for the evolution of the NV-$^{13}$C system under the dynamical decoupling with RF (DDRF) gate sequence (main text Fig. 2). Here, we consider the case of a single $^{13}$C nuclear spin interacting with the electron spin of an NV center through the hyperfine interaction, under the presence of a radio frequency driving field. We first consider an approximate idealized case for which the evolution operator can be calculated analytically. We then consider the generalized case, where time-independent simulation of the system is enabled by the application of a piecewise treatment which accounts for the shift of nuclear spin quantization axis due to a change of the electron spin state. We set $\hbar = 1$ for all following derivations.

\subsection{Hamiltonian}
In the interaction picture with respect to the electron energy splitting and neglecting non-secular terms, the Hamiltonian of the NV-$^{13}$C system is given by
\begin{equation}\label{eq:Hgeneral}
H = \omega_{L} I_{z} + A_{\parallel} S_{z} I_{z} + A_{\bot} S_{z} I_{x} + 2\Omega\cos(\omega t + \phi)I_x
\end{equation}
where the final term describes the interaction of the nuclear spin with a radio frequency (RF) driving field polarized along $x$ with frequency $\omega$, phase $\phi$ and Rabi frequency $\Omega$. Here, $S_{\alpha}$ and $I_{\alpha}$ are the spin-1 and spin-1/2 operators of the electron and nuclear spins respectively. $\omega_L = \gamma B_z$ is the nuclear spin Larmor frequency, where $\gamma$ is the gyromagnetic ratio of the $^{13}$C spin and $B_z$ is the external magnetic field strength along the NV axis. $A_\parallel$ ($A_\perp$) is the component of the electron-nuclear hyperfine interaction parallel (perpendicular) to the NV axis. 

\subsection{Basic derivation}

Here we derive the dynamics under a simple approximate model that gives the idealized evolution of the DDRF gate. Setting $A_\perp = 0$ and restricting to the $m_s =\{0,-1\}$ subspace of the electron spin, we can write Eq. \ref{eq:Hgeneral} as

\begin{equation}
    H = \ket{0}\bra{0}\otimes \omega_L I_z + \ket{1}\bra{1}\otimes \omega_{1}I_z + 2\Omega\cos(\omega t + \phi)I_x,
\end{equation}
where $\ket{0}$ ($\ket{1}$) indicates the electron $m_s = 0$ ($m_s = -1$) spin projection and $\omega_{1} = \omega_{L} - A_{\parallel}$ is the nuclear spin precession frequency when the electron is in the state $\ket{1}$. We can now move to the rotating frame at frequency $\omega$ and make the rotating wave approximation (RWA), giving

\begin{equation}\begin{split}
    H' &= R_0(t)(H - \omega I_z)R_0(t)^\dagger \\
    &= \ket{0}\bra{0}\otimes (\omega_L-\omega) I_z + \ket{1}\bra{1}\otimes (\omega_{1}-\omega)I_z + \Omega(\cos(\phi)I_x + \sin(\phi)I_y),
\end{split}\end{equation}
where $R_0(t) = e^{i\omega t I_z}$. Setting the RF frequency resonant with the nuclear spin when the electron is in the state $\ket{1}$ ($\omega = \omega_{1}$) and neglecting driving of the nuclear spin when the electron is in the state $\ket{0}$ (i.e. assuming $\Omega \ll (\omega_L - \omega_{1})$), we find \cite{van2012decoherence}

\begin{equation}\label{eq:Hsimple}
    H' = \ket{0}\bra{0} \otimes (\omega_L - \omega_{1})I_z + \ket{1}\bra{1}\otimes \Omega(\cos(\phi)I_x + \sin(\phi)I_y).
\end{equation}
This is equal to the Hamiltonian given in Eq. 2 of the main text. Writing equation \ref{eq:Hsimple} in the form $H' = \ket{0}\bra{0}\otimes H_0 + \ket{1}\bra{1}\otimes H_{1}$, we can consider the evolution of the nuclear spin for the two electron eigenstates separately, giving the unitary operators

\begin{equation}\begin{split}
    U_0(t) &= e^{-i H_0 t} = \left(
    \begin{array}{cc} 
    e^{-i(\omega_L - \omega_{1}) t/2} & 0 \\
    0 & e^{i(\omega_L - \omega_{1}) t/2} \end{array} \right), \\
    U_{1}(t,\phi) &= e^{-i H_{1} t} = \left(
    \begin{array}{cc} 
    \cos(\Omega t/2) & -(i\cos(\phi) + \sin(\phi))\sin(\Omega t/2) \\
    -(i\cos(\phi) - \sin(\phi))\sin(\Omega t/2) & \cos(\Omega t/2) \end{array} \right).
\end{split}\end{equation}

We now construct evolution operators for the dynamical decoupling sequence $(\tau- \pi - 2\tau- \pi - \tau)^{N/2}$. For simplicity, decoupling ($\pi$) pulses on the electron are treated to be perfect and instantaneous. We can consider separately the cases in which the electron starts in either $\ket{0}$ or $\ket{1}$, giving the total evolution operators $V_0$ and $V_{1}$ respectively, defined as
\begin{equation}\label{eq:totalevo}\begin{split}
    V_0 &= U_0(\tau) \cdot U_{1}(2\tau,\phi_{K-1}) \cdot U_0(2\tau) \cdots U_0(2\tau) \cdot U_{1}(2\tau,\phi_2) \cdot U_0(\tau) \\
    V_{1} &= U_{1}(\tau,\phi_{K}) \cdot U_0(2\tau) \cdot U_{1}(2\tau,\phi_{K-2}) \cdots U_{1}(2\tau,\phi_3) \cdot U_0(2\tau) \cdot U_{1}(\tau,\phi_1),
\end{split}\end{equation}
where $\phi_k$ are the phases of each RF pulse, $k = 1,...,K$ and $K = N + 1$. The total evolution of the two-spin system is then given by
\begin{equation}\label{eq:Vtotal}
    V = \ket{0}\bra{0} V_{0} + \ket{1}\bra{1} V_{1}.
\end{equation}
As can be seen from Eq. \ref{eq:totalevo}, the even $k$ RF pulses only affect the dynamics of $V_0$, while the odd $k$ pulses only affect the dynamics of $V_{1}$. Therefore, we can design the sequences individually in order to achieve the desired conditional rotation of the nuclear spin. In order to ensure the rotations due to each RF pulse add up constructively, the phases $\phi_k$ should be adapted to account for the periods of free precession, which build up in integer multiples of $\phi_\tau = (\omega_L-\omega_1)\tau$. For the case of $V_0$, the required phases should be updated according to the sequence $\phi_\tau, 3\phi_\tau, 5\phi_\tau,\dots$, while for the case of $V_{1}$ the sequence is $0, 2\phi_\tau, 4\phi_\tau,\dots$. This would yield an unconditional rotation of the nuclear spin. To achieve a conditional rotation, we can additionally add a relative $\pi$ phase shift to each phase in the $V_1$ sequence (odd $k$). Furthermore, we can add a phase $\varphi$ to all RF pulses which can be used to set the rotation axis for the full gate. Combining these elements, the required RF pulse phases for a conditional rotation (two-qubit gate) are summarized as $\phi_k = \varphi + \phi_k'$, where

\begin{equation}\label{eq:phicondition}
    \phi_k' = \left\{ 
    \begin{array}{ll} 
    (k-1)\phi_\tau + \pi & \quad k\text{ odd} \\
    (k-1)\phi_\tau & \quad k\text{ even},
    \end{array}
    \right.
\end{equation}
which is equal to Eq. 3 in the main text. We can then substitute the phases defined by Eq. \ref{eq:phicondition} into the sequence given in Eq. \ref{eq:totalevo}, which gives 

\begin{equation}\begin{split}
    V_0 &= \left(
    \begin{array}{cc} 
    e^{-i N (\omega_L - \omega_{1}) \tau/2}\cos(N\Omega\tau/2) & -ie^{-i N (\omega_L - \omega_{1}) \tau/2}e^{-i\varphi}\sin(N\Omega \tau/2) \\
    -ie^{i N (\omega_L - \omega_{1}) \tau/2}e^{i\varphi}\sin(N\Omega \tau/2) & e^{i N (\omega_L - \omega_{1}) \tau/2}\cos(N\Omega\tau/2) \end{array} \right) \\
    &= R_z(N(\omega_L - \omega_{1})\tau)\cdot R_\varphi(N\Omega\tau), \\
    V_1 &= \left(
    \begin{array}{cc} 
    e^{-i N (\omega_L - \omega_{1}) \tau/2}\cos(N\Omega\tau/2) & ie^{-i N (\omega_L - \omega_{1}) \tau/2}e^{-i\varphi}\sin(N\Omega \tau/2) \\
    ie^{i N (\omega_L - \omega_{1}) \tau/2}e^{i\varphi}\sin(N\Omega \tau/2) & e^{i N (\omega_L - \omega_{1}) \tau/2}\cos(N\Omega\tau/2) \end{array} \right) \\
    &= R_z(N(\omega_L - \omega_{1})\tau)\cdot R_\varphi(-N\Omega\tau),
\end{split}\end{equation}
where $R_z(\theta) = e^{-i\theta I_z}$ and $R_\varphi(\theta) = e^{-i\theta (\cos(\varphi)I_x + \sin(\varphi)I_y)}$. From Eq. \ref{eq:Vtotal}, the total evolution operator is therefore described by $V = V_z \cdot V_\text{CROT}$, where $V_z$ is an unconditional rotation of the nuclear spin around $z$, given by

\begin{equation}
    V_z = \mathbb{1}\otimes R_z(N(\omega_L - \omega_{1})\tau),
\end{equation}
and $V_\text{CROT}$ is a  controlled rotation of the nuclear spin with tuneable rotation angle (set by $N$, $\Omega$ and $\tau$) and rotation axis (set by $\varphi$), given by

\begin{equation}
V_\text{CROT} = \ket{0}\bra{0}\otimes R_\varphi(N\Omega\tau) + \ket{1}\bra{1}\otimes R_\varphi(-N\Omega\tau),
\end{equation}
which is equal to Eq. 4 in the main text. Setting $N\Omega \tau = \pi/2$, a maximally entangling two-qubit operation is achieved. With $\varphi = 0$, this operation is related to a controlled-not (CNOT) gate by the local rotations: \begin{equation}
    V_\text{CROT} = (R_z(\pi/2)\otimes \mathbb{1})\cdot (\mathbb{1}\otimes R_x(\pi/2))\cdot \text{CNOT}
\end{equation}
where 
\begin{equation}
   \text{CNOT} = \ket{0}\bra{0}\otimes \mathbb{1} + \ket{1}\bra{1}\otimes X.
\end{equation}

\subsection{Generalized case}\label{sec:generalized}

We now consider the dynamics under the more general case where $A_\perp \neq 0$, and where we do not neglect driving of the nuclear spin when the electron is in the state $\ket{0}$ (i.e. we do not assume that $\Omega \ll (\omega_L - \omega_{1})$). In this case, when the electron is in the state $\ket{1}$, the nuclear spin precesses at frequency $\omega_{1} = \sqrt{A_{\perp}^{2}+(\omega_{L}-A_{\parallel})^{2}}$, with quantization axis $A_{\perp} \hat{x} + (\omega_{L}-A_{\parallel}) \hat{z}$ at an angle $\beta$ from $\hat{z}$, defined by $\cos(\beta) = (\omega_L - A_\parallel)/\omega_{1}$. As the nuclear spin quantization axis is now dependent on the electron spin state, we rewrite
Eq. \ref{eq:Hgeneral} as

\begin{equation}\begin{split}\label{H_tilted}
H &= \ket{0}\bra{0}H_{0} + \ket{1}\bra{1}H_{1} \\
H_{0} &= \omega_{L}I_{z} + 2\Omega\cos(\omega t+\phi)I_{x} \\
H_{1} &= \omega_{1}\widetilde{I}_{z} + 2\widetilde{\Omega}_x\cos(\omega t+\phi)\widetilde{I}_{x} + 2\widetilde{\Omega}_z\cos(\omega t+\phi)\widetilde{I}_{z},
\end{split}\end{equation}
where $\widetilde{I}_z = R_y(\beta)I_zR_y(\beta)^\dagger = \cos(\beta)I_z + \sin(\beta)I_x$, $\widetilde{I}_x = R_y(\beta)I_xR_y(\beta)^\dagger = \cos(\beta)I_x - \sin(\beta)I_z$, $\widetilde{\Omega}_x = \Omega\cos(\beta)$, $\widetilde{\Omega}_z = \Omega\sin(\beta)$ and $R_y(\theta) = e^{-i\theta I_y}$. We can define two different rotating frames depending on the electron spin state; $R_0(t) = e^{-i\omega t I_z}$ and $R_{1}(t) = e^{-i\omega t \widetilde{I}_z}$. After making the RWA, the interaction picture Hamiltonian terms become

\begin{equation}\label{eq:Hint}\begin{split}
    H_{0}' &= R_{0}(t)(H_{0}-\omega I_{z})R_{0}(t)^{\dagger} \\
    &= (\omega_{L}-\omega)I_{z} + \Omega(\cos(\phi)I_{x}+\sin(\phi)I_{y}), \\
    H_{1}' &= R_{1}(t)(H_{1}-\omega \widetilde{I}_{z})R_{1}(t)^{\dagger} \\
    &= (\omega_{1}-\omega)\widetilde{I}_{z} + \widetilde{\Omega}_x(\cos(\phi)\widetilde{I}_{x}+\sin(\phi)\widetilde{I}_{y}) \\
    &= (\omega_{1}-\omega)(\cos(\beta)I_{z}+\sin(\beta)I_{x}) + \Omega\cos(\beta)(\cos(\phi)(\cos(\beta)I_{x}-\sin(\beta)I_{z})+\sin(\phi)I_{y}).
\end{split}\end{equation}
The Hamiltonian contains additional terms compared to the idealized case with $A_\perp = 0$ (Eq. \ref{eq:Hsimple}). We would therefore like to analyze the effect of these terms on the gate dynamics. Since we have used a different rotating frame for the two electron states, the evolution through the dynamical decoupling sequence should be calculated in a piecewise manner, where we change rotating frame between each electron spin flip. The generalized evolution operators for the two initial electron states from Eq. \ref{eq:totalevo} are then given by

\begin{equation}\label{eq:totalevogeneral}\begin{split}
    V_0 &= U_0(\tau) \cdot R_{0}(N'\tau) \cdot R_{1}(N'\tau)^\dagger \cdot U_{1}(2\tau,\phi_{K-1}) \cdot R_{1}((N'-2)\tau) \cdot R_{0}((N'-2)\tau)^\dagger \cdot U_0(2\tau) \cdots \\
    &\cdots U_0(2\tau)\cdot R_{0}(3\tau) \cdot R_{1}(3\tau)^\dagger \cdot U_{1}(2\tau,\phi_2) \cdot R_{1}(\tau) \cdot R_{0}(\tau)^\dagger \cdot U_0(\tau) \\
    V_{1} &= U_{1}(\tau,\phi_{K}) \cdot R_{1}(N'\tau) \cdot R_{0}(N'\tau)^\dagger \cdot U_0(2\tau) \cdot R_{0}((N'-2)\tau) \cdot R_{1}((N'-2)\tau)^\dagger \cdot U_{1}(2\tau,\phi_{K-2}) \cdots \\ &\cdots U_{1}(2\tau,\phi_3) \cdot R_{1}(3\tau) \cdot R_{0}(3\tau)^\dagger \cdot U_0(2\tau) \cdot R_{0}(\tau) \cdot R_{1}(\tau)^\dagger \cdot U_{1}(\tau,\phi_1),
\end{split}\end{equation}
where $N' = 2N-1$. 

\begin{figure*}[!t]
    \centering
    \includegraphics[width=0.4\textwidth]{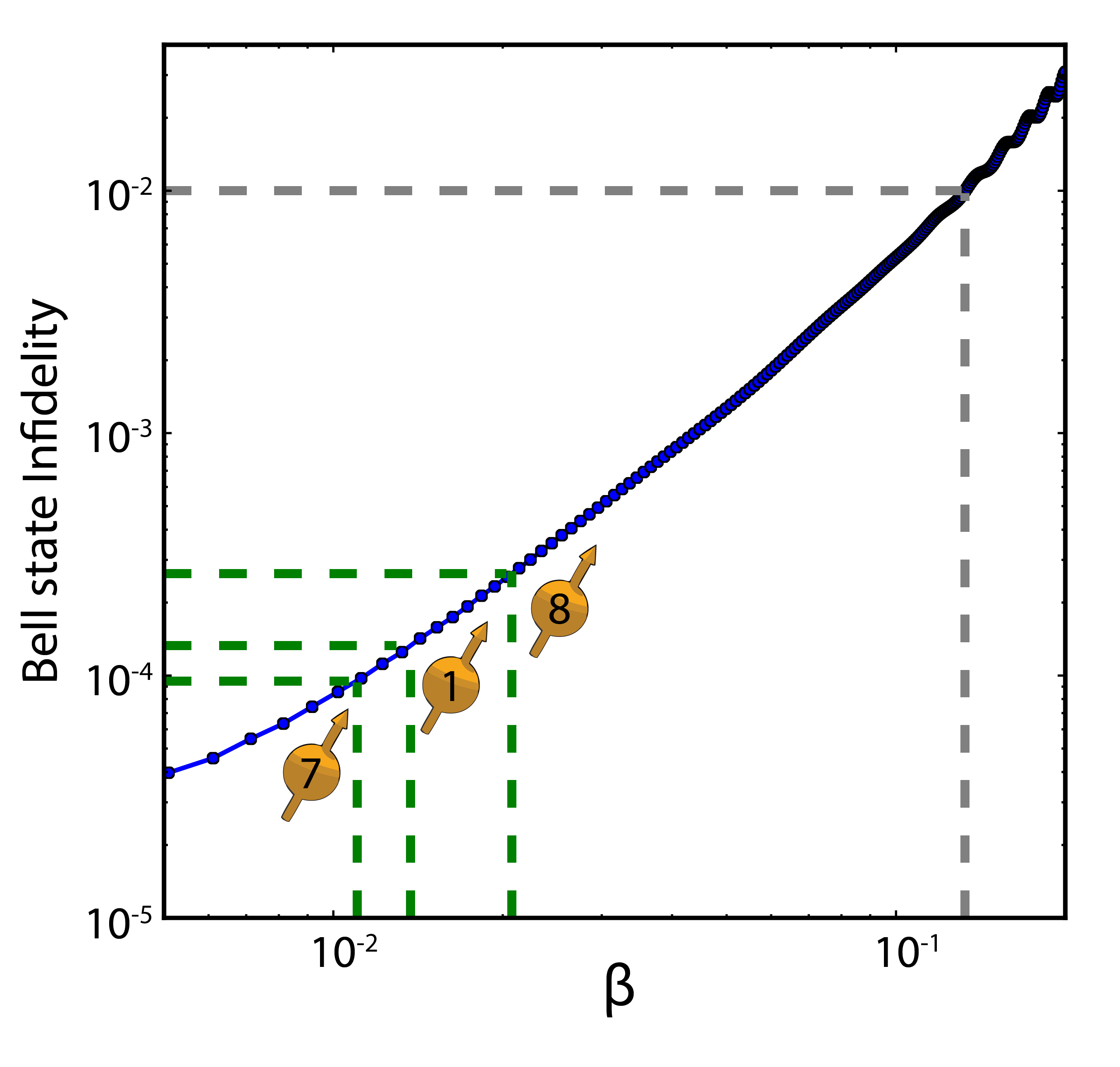}
    \caption{Theoretical infidelity of a Bell state generated by the DDRF gate as a function of the angle $\beta$ between the nuclear spin quantization axes for the electron spin projections $m_{s}=0$, $m_{s} = -1$. The dashed green lines indicate the values of $\beta$ for the 3 $^{13}$C spins for which we use DDRF gates. The dashed gray line indicates the $\beta$ required for a gate infidelity of $1\%$. } 
    \label{fig:Aperp}
\end{figure*}

We calculate the evolution numerically following the treatment derived above, making use of the QuTip Python package \cite{johansson2013qutip}. To understand the effect of the additional terms from a non-zero $A_\perp$, we consider a test spin with a parallel hyperfine component $A_{\parallel}/2\pi$ = 50 kHz. As a measure of fidelity we apply the DDRF gate to create a Bell state (starting from an ideal initialized state) and take the fidelity with the ideal Bell state. In Fig. \ref{fig:Aperp}, we plot the simulated Bell state infidelity against the relative strength of the perpendicular hyperfine component $A_{\bot}$, quantified in terms of $\beta$. For reference, we mark the values of $\beta$ for the 3 $^{13}$C spins for which we use DDRF gates in the experiments described in the main text. In all cases, the simulated infidelity from this component is $<$ 3$\times$10$^{-4}$. We also mark the value of $\beta$ for which the infidelity reaches 1\% ($\sim$ 0.13). This corresponds to $A_{\bot}/2\pi \sim\,$60 kHz for the $^{13}$C Larmor frequency of $\omega_L/2\pi \sim 432\,$kHz in our experiments, approximately equal to the largest $A_\perp$ observed in our register. It is important to note that this effect can be heavily suppressed by moving to larger magnetic fields. For example, at a field strength of $2000\,$G, a $\beta$ value of 0.13 at $403\,$G would be reduced to 0.03, with a corresponding contribution to the infidelity less than 5$\times$10$^{-4}$. Alternatively, one could also consider modifications to the gate design which account for the effect of non-zero $A_\perp$.
 
\section{Two-qubit gates: experiment}

\subsection{DDRF Spectroscopy}

\begin{figure}[]
    \centering
    \includegraphics[width=0.85\textwidth]{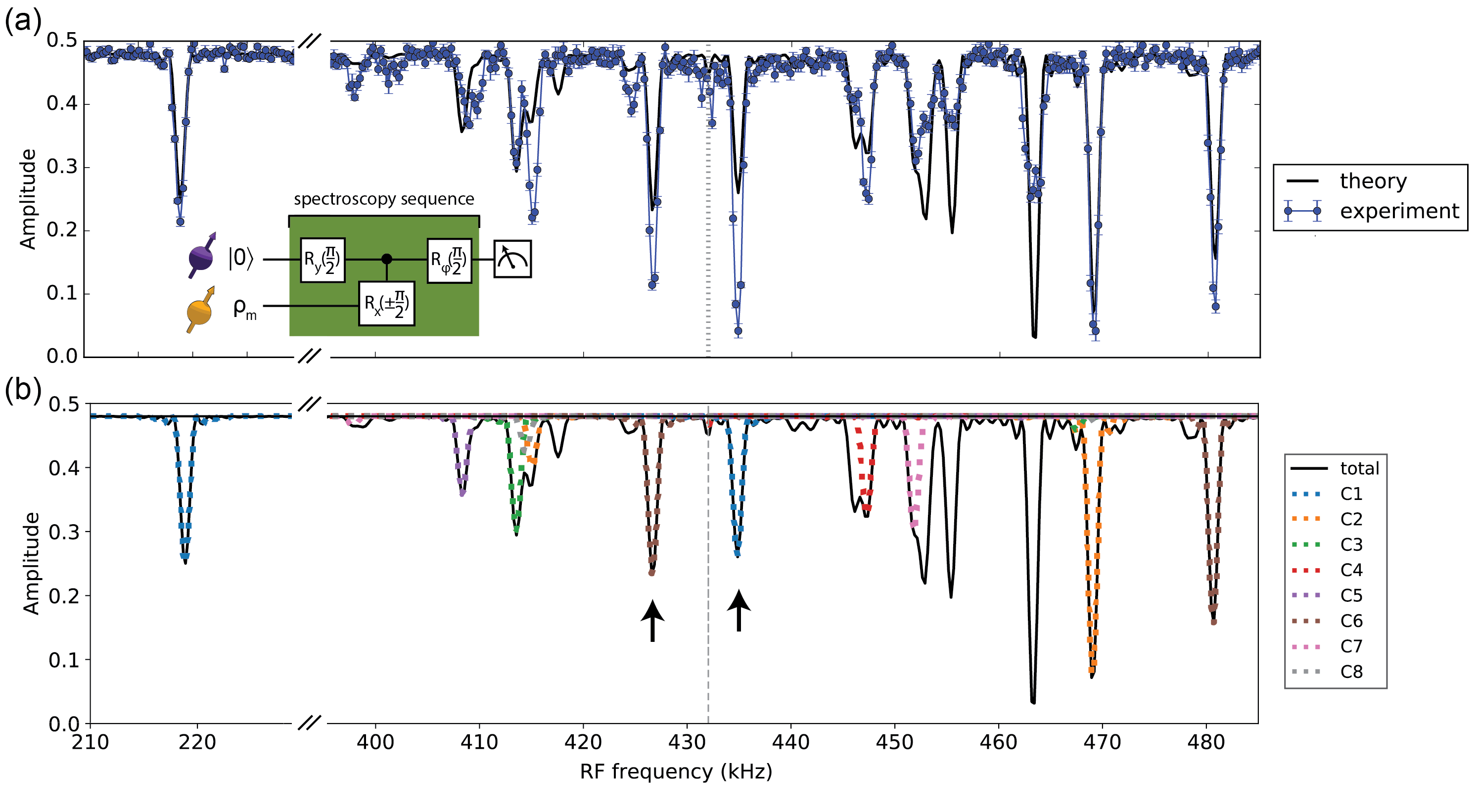}
    \caption{Spectroscopy of the nuclear spin environment using the DDRF interaction. (a) Measured spectrum (blue points) using the sequence described in the text and illustrated in the inset. Each data point corresponds to the fitted amplitude A of the function $f(\varphi) = a + A \cos(\varphi + \varphi_{0})$, where $\varphi$ is swept from $0$ to $360$ deg and $\varphi_{0}$ accounts for deterministic phase shifts induced on the electron by the RF field. By fitting the amplitude, we distinguish these phase shifts from loss of coherence due to entangling operations. The black line is the result of a numerical simulation incorporating all 27 known spins surrounding the NV center \cite{Abobeih_arXiv2019}. (b) Individual numerical simulations for each of the 8 $^{13}$C spins used in this work. The black arrows indicate two prominent dips that are caused by off-resonant interactions with $^{13}$C spins from the DDRF interaction (see text), corresponding to $m=4$ for C1 (blue dashed line) and $m=-1$ for C6 (brown dashed line).} 
    \label{fig:DDRFspectroscopy}
\end{figure}

A common approach to characterize the nuclear spin environment of an NV center is to perform dynamical decoupling spectroscopy, where the spacing between a sequence of $\pi$-pulses on the electron spin is varied \cite{Kolkowitz_PRL2012, Taminiau_PRL2012, zhao2012sensing}. The DDRF interaction presented in this work can be used in a similar way, with the additional benefit that spins with small perpendicular hyperfine components can also be detected. The sequence for the spectroscopy experiment is shown in the inset of Fig. \ref{fig:DDRFspectroscopy}. First a $\pi/2$ pulse rotates the electron spin to $\ket{+}$, after which the DDRF interaction is applied with fixed $N$ and $\tau$. Finally a second $\pi/2$ pulse is applied with varying phase $\varphi$. By fitting the resulting oscillation, we can distinguish between deterministic electron phase shifts caused by the RF field and loss of electron coherence due to interaction with the nuclear spin environment (see Fig. 3(b,c) of the main text). We can repeat this procedure while varying the frequency of the RF pulse, which will result in a dip in the amplitude if the RF frequency is resonant with one or more nuclear spins. 

An example spectrum is shown in Fig. \ref{fig:DDRFspectroscopy} (for the NV center considered in the main text). We also show a theory curve from a numerical simulation of the sequence based on the identified nuclear spins. In addition to the 8 $^{13}$C spins used in this work, an additional 19 $^{13}$C spins in the local environment of this NV center were identified in parallel work \cite{Abobeih_arXiv2019}. Therefore, we plot the expected signal due to all 27 known spins. The Rabi frequencies of each spin are observed to vary (see table \ref{Tab:SpinsGate}), which has not been investigated fully, but could be caused by frequency dependencies of the RF transmission as well as by the non-secular interaction terms in the electron-nuclear hyperfine coupling Hamiltonian \cite{Childress_Science2006, zhao2011atomic,sangtawesin2016hyperfine,Abobeih_arXiv2019}. Since the parameters associated with these non-secular terms were not measured, we do not take them into account directly in the model, but rather input the measured Rabi frequencies for each spin individually. 

As described in section \ref{sec:DDRFtheory}, the DDRF interaction realizes an entangling operation with a nuclear spin if the RF frequency matches the precession frequency when the electron is in the $m_s = -1$ spin projection, i.e. when $\omega = \omega_{1}$. This is due to the constructive buildup of small rotations from each RF pulse within the dynamical decoupling sequence, which is ensured by setting the phase of each pulse according to Eq. \ref{eq:phicondition}. However, constructive buildup is also possible at other RF frequencies if the condition $\phi_k'(\omega) - \phi_k'(\omega_{1}) = 0 \,\text{mod}\, 2\pi$ is satisfied, where $\phi_k'(\omega)$ is the generalized version of Eq. \ref{eq:phicondition} where $\omega_1 \rightarrow \omega$. Solving for $\omega$, we find resonance conditions

\begin{equation}
    \omega_m = \omega_{1} + \frac{2\pi m}{\tau},
\end{equation}
for integer $m$. Examples of such off-resonant interactions are indicated by the arrows in Fig. \ref{fig:DDRFspectroscopy}(b). The strengths of the off-resonant interactions are set by the Rabi frequency in combination with the detuning of the RF field from both $\omega_{1}$ and $\omega_L$. For the parameters used in Fig. \ref{fig:DDRFspectroscopy}, the off-resonant interactions are separated by integer multiples of $2\pi\times 54\,$kHz. The two highlighted peaks therefore correspond to $m=4$; $\omega_m/2\pi = 434.828\,$kHz for C1 and $m=-1$; $\omega_m/2\pi = 426.625\,$kHz for C6. Additional unconditional rotations of the nuclear spins can also occur when $\omega_p = \omega_1 + (2p + 1)\pi/\tau$ for integer $p$, for which $\phi_k'(\omega_p) = (k-1)\phi_\tau + \pi$ for all $k$. These rotations do not entangle the electron and nuclear spins and therefore do not show up in the spectroscopy measurements, but can cause additional unwanted crosstalk. In order to avoid such unwanted crosstalk in the 10-qubit register, we ensure that any conditional or unconditional off-resonant interactions do not overlap with the resonances of any targeted spins.

\subsection{Sources of infidelity for the DDRF gate}

As shown in section \ref{sec:generalized}, the perpendicular component of the electron-nuclear hyperfine interaction contributes $< 10^{-3}$ to the infidelity of the DDRF gate for the spins used in this work. We now consider two additional sources of infidelity in our numerical model: crosstalk between spins and nuclear spin dephasing. To model crosstalk, we consider a system composed of the NV electron spin, the target nuclear spin, and any additional nuclear spins within a 5 kHz spectral range of the target spin that were identified in Abobeih et al. \cite{Abobeih_arXiv2019}. If there are no spins within 5 kHz, we include only the nearest spin. To account for nuclear spin dephasing, we use Monte Carlo methods. At the start of each simulation, we randomly draw a detuning of the angular frequency from a Gaussian distribution with $\sigma$ = $(\sqrt{2}/T_{2}^{*})$ and shift the nuclear spin frequencies $\omega_{L}$ and $\omega_{1}$ accordingly. All simulations are performed with 500 samples. The gate parameters, $T^{*}_{2}$ values, and hyperfine couplings used to simulate each spin are the same as those measured experimentally (tables \ref{Tab:Spins}, \ref{Tab:SpinsCoh} and \ref{Tab:SpinsGate}). We assume a magnetic field of $403.55\,$G, aligned along the NV axis.

\begin{table}
  \centering
  \begin{tabular}{c|c|c}
  		Spin	& $\mathcal{F}_\text{Bell}$ (measured) & $\mathcal{F}_\text{Bell}$ (simulated) \\ \hline \hline
   C1*      &   0.972(8)    & 0.998 \\
   C1	    &   0.93(1)     & 0.990 \\
   C7       &   0.85(2)     & 0.797 \\
   C8       &   0.81(2)     & 0.892 \\
   $^{14}$N	&   0.93(1)     & 0.99996 \\

   \end{tabular}
  \caption{\textbf{Measured and simulated electron-nuclear spin Bell state fidelities for the nuclear spins controlled by DDRF gates.} For C1, the spectrally closest spin is C5. For C7, there are three spins within $5\,$kHz; C4 and two unused spins with hyperfine components $\{ A_\parallel/2\pi, A_\perp/2\pi \} = \{ -20.7,12\}\,$kHz and $\{ -23.2, 13\} \,$kHz \cite{Abobeih_arXiv2019}. For C8, there are two spins within $5\,$kHz; C3 and one unused spin with hyperfine components $\{ A_\parallel/2\pi, A_\perp/2\pi \} = \{ 14.5,10\}\,$kHz. *Measured with use of RF amplifier, see section \ref{sec:mwRF}.}
  \label{Tab:SpinsSim}
\end{table}

We calculate the Bell state fidelity from a simulated experiment following the sequence shown in the Fig. 4(a) of the main text. In the simulation, the electron is initialized in the state $\ket{0}$, while the nuclear spins are initially in a mixed state. We then apply the `initialize' and `entangle' blocks of the sequence shown in Fig. 4(a) of the main text to the target nuclear spin, where each two-qubit gate is simulated using the unitary operator as described in Eq. \ref{eq:totalevogeneral}. We simulate the Bell state fidelities for both gate regimes on the nuclear spin C1, along with the gates used for spins C7, C8, and $^{14}$N. For simplicity, the $^{14}$N spin is treated analogously to the $^{13}$C spins, with appropriately modified Larmor frequency and hyperfine interaction strength. Additionally, we assume perfect initialization for the $^{14}$N spin. In the simulations, we allow for optimization of the Rabi frequency $\Omega$ and of the nuclear spin readout basis (the azimuthal angle) to maximize the fidelity, following the calibration procedure used in the experiments. In table \ref{Tab:SpinsSim} we compare the simulated and measured values. For the spins C1 and $^{14}$N, the simulated Bell state fidelities are quite high compared to the measured values, indicating that nuclear spin dephasing and crosstalk are not the dominant infidelity mechanisms in these cases. Conversely, for C7 and C8, the simulation predicts a large contribution from these two effects.

For the case shown in Fig. 4(b) of the main text (C1* in table \ref{Tab:SpinsSim}), we independently measure the loss of coherence of the electron spin during the DDRF gate by applying the same dynamical decoupling sequence without any RF pulses. We perform the experiment shown in the inset of Fig. 3(a) with $\varphi = 90$ deg, after which the electron spin should ideally end up in the state $\ket{0}$. We perform this experiment for two cases: one where the RF source is connected and one where it is disconnected. We interleave these two cases in blocks of 20000 repetitions (10 blocks in total) to ensure that any measured difference is not due to drifts in the experimental setup. For the case where the RF source is disconnected, we measure $P(0) = 0.999(2)$ and for the case where it is connected, we measure $P(0) = 0.992(2)$. This indicates that loss of coherence on the electron spin during the DDRF gate due to noise from the RF source is a significant source of infidelity in this case. Careful design of a low noise RF delivery system could therefore lead to significant improvements in the gate fidelities.

\section{Multi-qubit Entanglement Experiments}

\subsection{Qubit Initialization}\label{sec:init}

\begin{table}[h!]
  \centering
  \begin{tabular}{c|c}
  		Spin	& $\mathcal{F}_\text{init}$\\ \hline \hline
  e        &	0.998(2)    \\
  C1*      &   0.983(1)    \\
  C1	    &   0.965(5)     \\
  C2	    &	0.985(5)   \\
  C3	    &	0.970(5)    \\
  C4	    &	0.965(5)    \\
  C5       &   0.980(5)     \\
  C6 	    &	0.985(5)     \\
  C7	    &	0.86(1)     \\
  C8	    &	0.83(1)      \\
  $^{14}$N &   0.997(11)    \\
  \end{tabular}
  \caption{\textbf{Measured initialization fidelities for the spin register.}}
  \label{Tab:initialization}
\end{table}

Initialization is performed at the start of all experiments. Initialization fidelities for all spins are given in table \ref{Tab:initialization}. The electron spin can be prepared in the $m_{s} = 0$ spin projection with high fidelity (0.998(2)) by resonant optical excitation \cite{robledo2011high}. 

The $^{13}$C nuclear spins are sequentially initialized by a partial swap sequence which maps the $\ket{0}$ state of the electron spin onto the target nuclear spin \cite{taminiau2014universal}. We assume symmetric state preparation and measurement errors (see section \ref{sec:ROcorrection}). The spread of values is due to the different two-qubit gate fidelity for each nuclear spin \cite{Cramer_NatureComm2016}. 

\begin{figure}[!ht]
    \centering
    \includegraphics[width=0.8\textwidth]{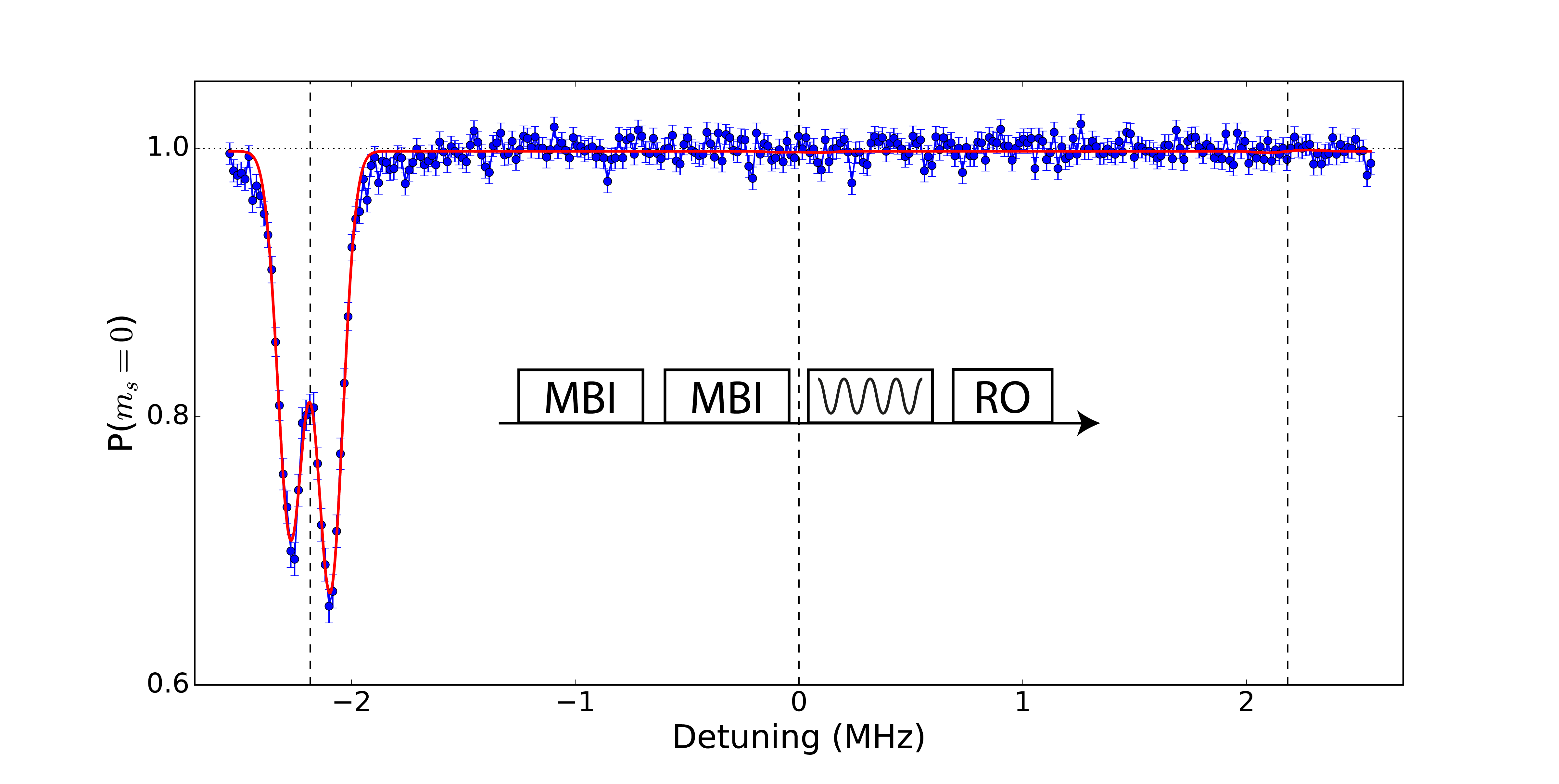}
    \caption{Electron spin resonance experiment from which the $^{14}$N spin initialization may be inferred. We repeat a measurement based initialization sequence twice in order to suppress initialization errors from erroneous photon detection events. We then perform a highly selective microwave pulse (duration: $10\,$\textmu s), followed by optical readout of the NV electron spin. We sweep the frequency of the microwave pulse; the detuning shown is relative to the $m_{s}= 0 \leftrightarrow m_{s}= -1$ transition frequency (1.74667(1) GHz). We fit six Lorentzian lines, accounting for a stronger coupled $^{13}$C nuclear spin (C1) along with the $^{14}$N splitting.} 
    \label{fig:N_init}
\end{figure}

The $^{14}$N spin is initialized by a measurement-based initialization (MBI) at the start of the sequence \cite{robledo2011high}. Unlike previous implementations of this scheme, where the electron spin was prepared in a mixture of the $m_{s} = \pm 1$ projections, we initialize into the $m_{s} = 0$ state, after which a microwave $\pi$-pulse prepares the electron in the $m_{s} = -1$ spin projection with high probability ($>$99\%). In this way, we immediately double the success rate of the procedure. We also perform the MBI sequence twice, thereby suppressing $^{14}$N initialization errors due to imperfect readout. Fig. \ref{fig:N_init} shows an example electron spin resonance experiment following nitrogen initialization in $m_I = -1$. The transition is split into six lines due to hyperfine coupling to the $^{14}$N ($\sim 2.18$ MHz) and a $^{13}$C spin (C1; $\sim 213$ kHz). We fit six Lorentzian lines to determine the relative height of the dips, which correspond to the populations in each of the nitrogen spin states. The fitted amplitudes reveal populations of $p_{-1}$ = 0.997(11), $p_{0}$ = 0.003(6), and $p_{+1}$ = 0.000(6). We thus improve on the initialization fidelity for this nuclear spin by an order of magnitude relative to previous work (previously measured $\sim 0.96$ for this NV \cite{Kalb_NatureComm2016}). After initialization of the spin-1 $^{14}$N, we work in a two-level subspace $m_{I} = \{-1,0\}$, and perform the remainder of the operations analogously to the $^{13}$C spins, including two-qubit gates using the DDRF scheme. 

\subsection{Measurement of electron-nuclear entangled states}

To characterize electron-nuclear entangled states, we generalize the method described in Kalb et al. \cite{kalb2017entanglement} to multiple spins and arbitrary electron states. We first measure the electron spin in a chosen basis, before measuring the relevant multi-qubit expectation values for the nuclear spins. The latter step is performed by parity measurements using the electron as an ancilla. In order to minimize dephasing on the nuclear spins during the first electron measurement, we only continue and perform the nuclear spin measurements in the case that we receive a photon detection event within the first $60\,$\textmu s ($10\,$\textmu s for the experiment shown in Fig. 4 of the main text, $5\,$\textmu s for the electron-nitrogen experiment in Fig. 6 of the main text), and dynamically stop the measurement on receipt of the photon. This has multiple advantages. First, we simultaneously minimize unwanted spin-flips after heralding the electron state, and dephasing of the nuclear spins due to the optical cycling process. Second, we enhance the projectivity of our measurement. Compared to single-shot readout (average fidelity 0.945(2)), we now project into the electron spin state $\ket{0}$ with a fidelity of 0.992 \cite{Cramer_NatureComm2016}. To appropriately measure the entanglement correlations for both the $M_i = +1$ and $M_i = -1$ electron spin measurement outcomes, we perform the sequence twice. Here $M_i$ is the outcome of an electron measurement in the $i = \{X,Y,Z\}$ basis. In the first sequence (sequence $A$), we measure the electron after the appropriate basis rotation, while in the second sequence (sequence $B$), we perform an additional $\pi$-pulse prior to the electron readout (see Fig. 4(a) of the main text). The electron measurement probabilities $p(M_i = \pm 1)$ are then reconstructed from the probability of photon detection across each pair of measurements, using the relations

\begin{equation}
\begin{split}
p(M_i = +1) &= \frac{p_i^A(n>0)}{p_i^A(n>0) + p_i^B(n>0)}, \\
p(M_i = -1) &= \frac{p_i^B(n>0)}{p_i^A(n>0) + p_i^B(n>0)}, \\
\end{split}
\end{equation}
where $p_i^A(n>0)$ is the probability to detect $>0$ photons during the measurement in sequence $A$ (no $\pi$-pulse) and $p_i^B(n>0)$ is the probability to detect $>0$ photons during the measurement in sequence $B$ (with $\pi$-pulse). We can then calculate the electron-nuclear expectation values as
\begin{equation}\begin{split}
\langle I_e \otimes O_n \rangle &=  p(M_Z = +1) \langle O_n \rangle_{M_Z = +1}
+ p(M_Z = -1) \langle O_n \rangle_{M_Z = -1}, \\
\langle X_e \otimes O_n \rangle &=  p(M_X = +1) \langle O_n \rangle_{M_X = +1}
- p(M_X = -1) \langle O_n \rangle_{M_X = -1}, \\
\langle Y_e \otimes O_n \rangle &=  p(M_Y = +1) \langle O_n \rangle_{M_Y = +1}
- p(M_Y = -1) \langle O_n \rangle_{M_Y = -1}, \\
\langle Z_e \otimes O_n \rangle &=  p(M_Z = +1) \langle O_n \rangle_{M_Z = +1}
- p(M_Z = -1) \langle O_n \rangle_{M_Z = -1}, \\
\end{split}
\end{equation}
where $O_n$ is the nuclear spin Pauli operator and $\langle O_n \rangle_{M_i = \pm 1}$ is the expectation value of $O_n$ given that the electron measurement gave the outcome $+1$ (sequence $A$) or $-1$ (sequence $B$). While this measurement procedure minimizes dephasing for the $^{13}$C spins, the large excited state hyperfine coupling between the electron and the $^{14}$N can lead to significant dephasing within a few optical cycles \cite{blok2014manipulating}. Consequently, for the multi-qubit experiments shown in Fig. 7 of the main text, we protect the nitrogen spin state by performing a basis rotation that maps the desired measurement basis to the $Z$-basis prior to the electron measurement.

\subsection{Nuclear spin readout correction}\label{sec:ROcorrection}

In order to provide best estimates for the state fidelities, we correct the results for infidelities in the readout sequence. We base the correction on methods developed in a previous work \cite{Cramer_NatureComm2016}. We first calculate a single qubit readout fidelity for each nuclear spin based on a symmetric initialization and readout scheme. That is, we prepare a nuclear spin into the eigenstate $\ket{\uparrow}$, and measure in the $Z$-basis. Assuming the process is symmetric, for each nuclear spin we can write

\begin{equation}
    \langle Z_{j} \rangle = C^{2}_{\text{Qj}},
\end{equation}
where $\langle Z_{j} \rangle$ is the measured expectation value in the $Z$-basis for spin j, and $C_{Qj}$ is the error associated with the initialization and readout processes. For the nitrogen spin, we use an independently measured value of 0.997(11) for the initialization fidelity (see Fig. \ref{fig:N_init}), and can directly extract the correction factor through the relationship

\begin{equation}
    \langle Z_{^{14}\text{N}} \rangle = C_{^{14}\text{N,init}}C_{^{14}\text{N,RO}}.
\end{equation}
To correct a multi-qubit readout, we now prepare the corresponding multi-qubit state, for example: $\ket{\uparrow \uparrow \uparrow}$. Measurement of the expectation value $\langle Z_{1}Z_{2}Z_{3}\rangle$ allows one to calculate an appropriate readout correction, following the relation
\begin{equation}
    \langle Z_{1}Z_{2}Z_{3} \rangle = C_{\text{Q1,init}}C_{\text{Q2,init}}C_{\text{Q3,init}} C_{\text{Q1,Q2,Q3}},
\end{equation}
where $C_{\text{Q1,Q2,Q3}}$ accounts for the error associated with the 3-qubit readout (for example, due to decoherence of the electron spin and crosstalk to other nuclear spins).  

To calculate the factor $C_{\text{Q1,Q2,Q3}}$, we require the initialization fidelities, $C_{\text{Q1,init}},C_{\text{Q2,init}},C_{\text{Q3,init}}$, which may now differ from the values measured in the single qubit experiments ($C_{\text{Qj}}$) due to crosstalk during the multi-qubit initialization procedure. To characterize these values, we also measure the expectation values $\langle Z_{1}I_{2}I_{3}\rangle$, $\langle I_{1}Z_{2}I_{3}\rangle$ and $\langle I_{1}I_{2}Z_{3}\rangle$. Taking the previously measured single qubit readout fidelities, and following the relations
\begin{equation}
\begin{split}
    \langle Z_{1}I_{2}I_{3} \rangle = C_{\text{Q1,init}}C_{\text{Q1}} \\
    \langle I_{1}Z_{2}I_{3} \rangle = C_{\text{Q2,init}}C_{\text{Q2}} \\
    \langle I_{1}I_{2}Z_{3} \rangle = C_{\text{Q3,init}}C_{\text{Q3}},
\end{split}
\end{equation}
we retrieve the initialization fidelities, and thus arrive at a value for $C_{\text{Q1,Q2,Q3}}$. Similar analysis enables characterization of multi-qubit parity readout for all combinations of spins used in the entanglement experiments.

\subsection{Spin echoes}\label{sec:echoes}

In order to protect nuclear spin coherences across the multi-qubit entanglement and measurement sequences, composed of up to 14 two-qubit gates and 7 single-qubit gates on nuclear spins, we integrate RF spin echo pulses into the measurement sequence. These pulses are inserted in two positions; after the entanglement step, and after the basis rotations required for measurement of the appropriate multi-qubit Pauli operators. In this way, we extend typical nuclear spin coherence times of order 10 ms to 0.2 - 0.8 s (see table \ref{Tab:SpinsCoh}). The exact sequencing of the echo pulses, along with the required single and two-qubit gates, is tailored to the specific measurement basis.

\begin{figure}[!t]
    \centering
    \includegraphics[width=0.8\textwidth]{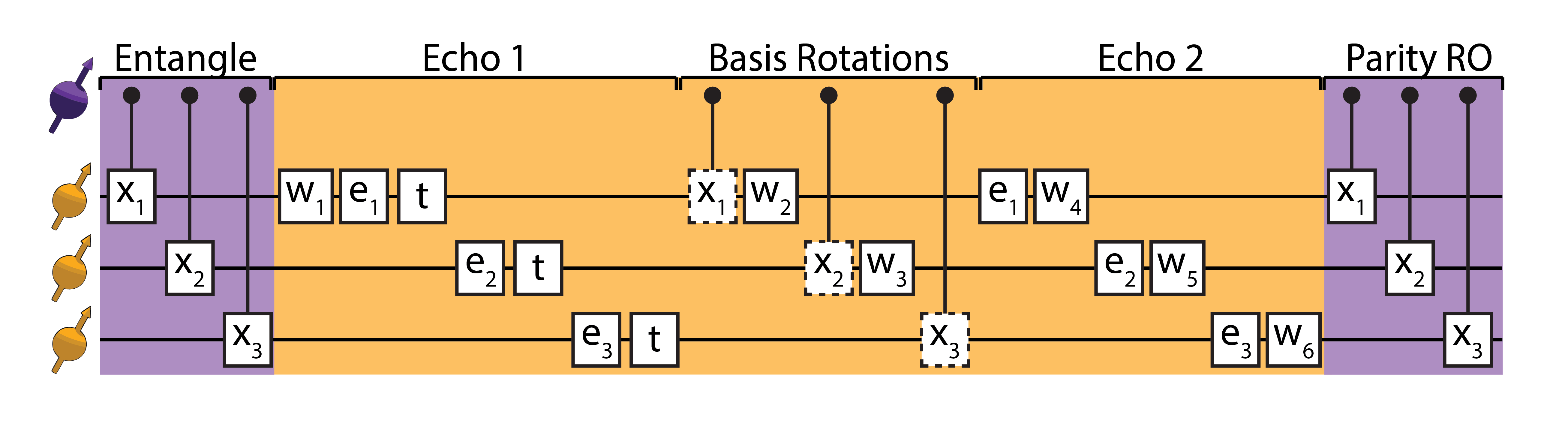}
    \caption{Example of a nuclear spin measurement sequence interspersed with spin echo pulses for the case of 3 nuclear spins. Purple (orange) boxes correspond to periods in which the electron spin state is in superposition (the $m_{s} = -1$ spin projection). Boxes $x_{1}$, $x_{2}$, $x_{3}$ correspond to gate operations on nuclear spins 1, 2, and 3 respectively, with their individual durations. Dashed boxes indicate that - in the case a basis rotation is not required on a given spin - idling is performed for an equal duration to the operation time. Similarly, boxes $e_{1}$, $e_{2}$, and $e_{3}$ correspond to spin echo pulses on those spins, also with unique lengths. Boxes $w_{1}-w_{6}$ are unique wait durations which must be derived from a set of simultaneous equations (Eqs. \ref{eq:echo1} and \ref{eq:echo2}). Electron pulses, repumping operations and optical readout are omitted for simplicity and can be considered part of the wait boxes. t is a spacing parameter for the first round of echoes, which can be increased in the case that a solution is not found for those equations due to the relative lengths of the gate and echo durations for each nuclear spin.} 
    \label{fig:Echo}
\end{figure}

The timings for the echo pulses are calculated from the point of the final microwave pulse of the entanglement sequence. We identify the nuclear spins to be read out, the time since the last operation acting upon them, and their respective gate and echo durations. A specific challenge for our sequence is to time the refocussing points of the second round of echoes such that no additional waiting time is added during the parity readout. In this way, we minimize the duration for which the electron spin state is in superposition and thus sensitive to dephasing noise. Instead, the electron spin is in the $m_{s} = -1$ projection during idling times, and only suffers depolarization due to longitudinal relaxation ($T_{1}$ timescales $>$1 hour). We identify a general solution, for which an example for 3 nuclear spins is given in Fig. \ref{fig:Echo}.

Considering Fig. \ref{fig:Echo}, we wish to balance the times between successive operations on a given nuclear spin and the spin echo pulse which separates them. For example, to balance the delays such that the echo pulse $e_{1}$ refocusses the nuclear spin between the first and second $x_{1}$ operations, we must solve the equation (see Fig. \ref{fig:Echo} for timing definitions)
\begin{equation}\label{eq:echo1}
    x_{2} + x_{3} + w_{1} = 3t + e_{2} + e_{3}
\end{equation}
For the remainder of the echoes, we must simultaneously solve the following relations alongside Eq. \ref{eq:echo1}:
\begin{equation}\label{eq:echo2}
\begin{split}
    x_{3}+ w_{1} + e_{1} = t + e_{3} + x_{1} + w_{2} \\
    w_{1} + e_{1} + e_{2} + t = x_{1} + w_{2} + x_{2} + w_{3} \\
    w_{2} + x_{2} + w_{3} + x_{3} = w_{4} + e_{2} + w_{5} +e_{3} + w_{6} \\
    w_{3} + x_{3} + e_{1} + w_{4} = w_{5} + e_{3} + w_{6}+ x_{1} \\
    e_{1} + w_{4} + e_{2} + w_{5} = w_{6} + x_{1} + x_{2}\\
\end{split}
\end{equation}
The identification of a solution ensures that all nuclear spins are refocussed at the point of each operation acting upon them. The example given here is readily scaled for an increasing number of nuclear spins. In experiment, we calculate the specific delays for each measurement basis by translating this sequence structure into a matrix equation which is algorithmically generated and solved using the Sympy Python package \cite{sympy}. The output of this algorithm is directly fed into the control software to program the experimental sequences.

Echo pulses are implemented by RF driving at the unique precession frequency of each nuclear spin (in the $m_{s} = -1$ electron spin state). An error-function envelope with a rise time of $7.5\,$\textmu s is used to mitigate ringing of the RF signal due to sharp switching (see section \ref{sec:mwRF}). $\pi$-pulse durations for the echo pulses are given in table \ref{Tab:SpinsGate}. To estimate the fidelity of the echo pulses, we numerically simulate the effect of a 1 ms, square-envelope RF pulse with Rabi frequency $\Omega/2\pi$ = 500 Hz following the numerical model described in section \ref{sec:DDRFtheory}. The RF pulse is set to be resonant with a test spin whose hyperfine interaction components are chosen to be $A_{\parallel}/2\pi$ = 50 kHz and $A_{\perp}/2\pi$ = 20 kHz and dephasing time $T^{*}_{2}$ = 10 ms. Assuming perfect electron spin initialization in the $m_{s} = -1$ projection, and perfect nuclear spin initialization in each of the 6 cardinal states, we find the average fidelity with the ideal states to be 0.998. 

\begin{figure}[t]
    \centering
    \includegraphics[width=0.8\textwidth]{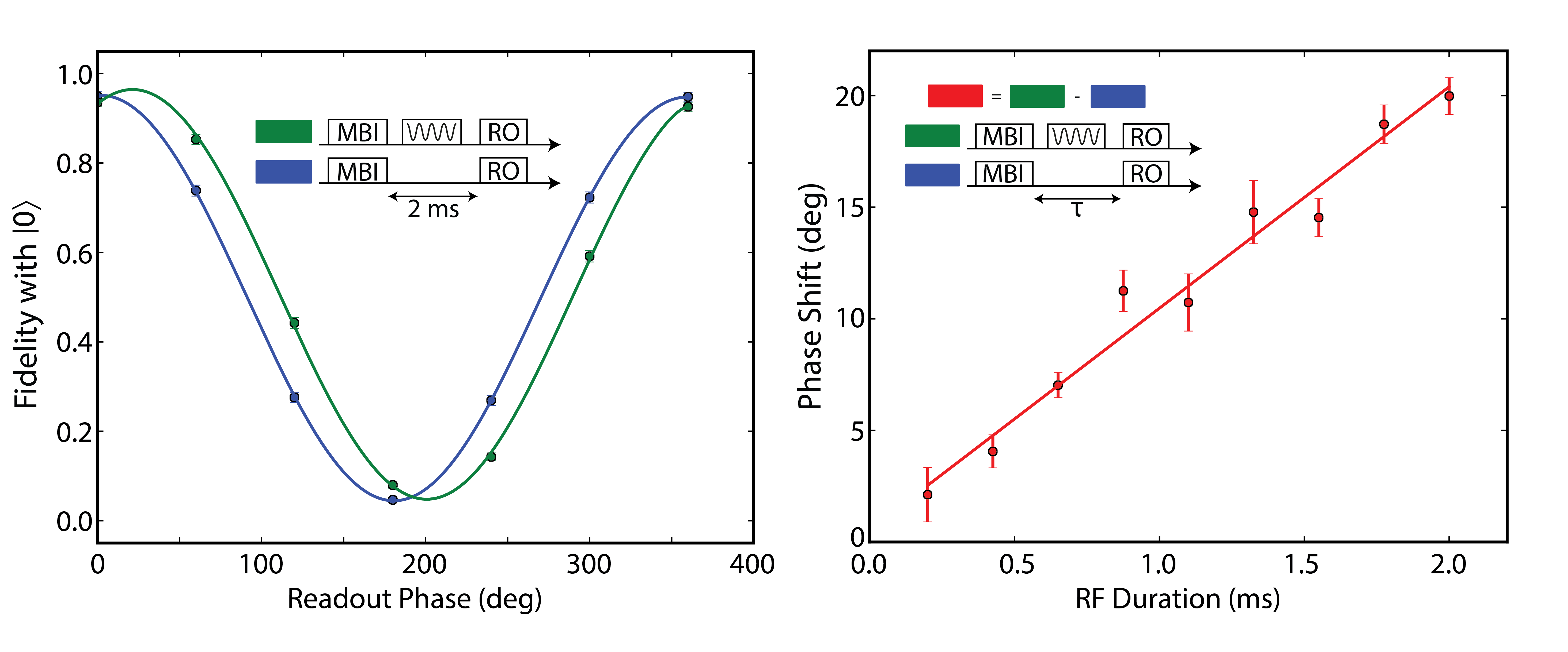}
    \caption{Characterization of the AC-Stark effect. Experiments are performed on the two $^{13}$C spins C3 and C5, detuned by 5160(2) Hz. (a) Measurement of the phase shift induced on spin C3 by a 2 ms RF pulse applied on spin C5.  We first initialise spin C3 in the $X$-basis. We then either apply the RF pulse (green) or allow the qubit to idle for the same duration (blue). Finally, we sweep the phase of the nuclear spin readout (in the $\hat{x}$-$\hat{y}$ plane). We fit both curves to the function $f(\phi) = 0.5 + A \cos(\phi + \phi_{0})$, and extract the phase shift $\delta\phi_{0}$. (b) Plot of the fitted phase shifts $\delta\phi_{0}$, against the duration of the applied RF pulse, $t_\text{pulse}$. From a linear fit $f(t_\text{pulse}) = a + A t_\text{pulse}$, we extract a value of 28(2) Hz for the AC-Stark shift, in agreement with the predicted value of 28(1) Hz.} 
    \label{fig:AC_Stark}
\end{figure}

We note that the measured Rabi frequencies associated with each spin are not equal, an effect which we attribute to frequency dependencies of the RF transmission, along with the non-secular interaction terms in the electron-nuclear hyperfine coupling Hamiltonian \cite{Childress_Science2006,zhao2012sensing,sangtawesin2016hyperfine,Abobeih_arXiv2019}. For nuclear spins which are initialized and read out using standard dynamical decoupling sequences, we calibrate the phase of the RF pulse to match the azimuthal angle of the hyperfine interaction for that spin \cite{zopes2018three,sasaki2018determination,laraoui2015imaging}. 

We also account for the AC-Stark shift imposed on the other spins by the echo pulses. As an example, consider two nuclear spins A and B. We prepare spin A in the $X$-basis, apply the echo pulse on spin B, and then measure the phase of spin A. We then repeat this sequence, but do not apply any RF power. The difference in the measured phases of spin A reveals the shift due to the AC-Stark effect. The AC-Stark shift is approximately given by 
\begin{equation}
\Delta{\omega_{-1}} = \frac{1}{2}\frac{\Omega^{2}}{\omega-\omega_{-1}},   
\label{eq:AC_Stark}
\end{equation}
where $\Omega$ is the Rabi frequency, $\omega$ is the RF frequency, and $\omega_{-1}$ is the $m_{s}=-1$ nuclear spin precession frequency. Fig. \ref{fig:AC_Stark} shows a measurement of this frequency shift for spins C3 and C5, which have close spectral proximity (detuning of $5160(2)\,$Hz). We measure a Rabi frequency of $538(12)\,$Hz for spin C3. By sweeping the duration of the RF pulse on spin C5, we can extract a phase shift of 9.9(6) degrees/ms of applied RF, corresponding to a frequency shift of $28(2)\,$Hz.  A calculation using Eq. \ref{eq:AC_Stark} for the known RF frequency and Rabi frequency predicts a value of $28(1)\,$Hz, in agreement with the measurement. We perform a similar calibration for all spins in the register. 

\subsection{Bell state measurements}

\begin{table}[h!]
  \centering
  \begin{tabular}{c||cccccccccc}
  			& e       & $^{14}$N& C1      & C2      & C3      & C4      & C5      & C6      & C7      & C8 \\ \hline \hline
   e	    & -       & 0.93(1) & 0.93(1) & 0.97(1) & 0.94(1) & 0.93(1) & 0.97(1) & 0.93(1) & 0.85(1) & 0.81(2) \\
   $^{14}$N & 0.93(1) & -       & 0.84(3) & 0.91(2) & 0.87(2) & 0.83(2) & 0.89(2) & 0.81(3) & 0.72(4) & 0.63(3) \\
   C1	    & 0.93(1) & 0.84(3) & -       & 0.83(2) & 0.84(2) & 0.80(2) & 0.79(2) & 0.83(2) & 0.75(4) & 0.82(5) \\
   C2	    & 0.97(1) & 0.91(2) & 0.83(2) & -       & 0.92(2) & 0.84(2) & 0.94(2) & 0.92(2) & 0.76(3) & 0.73(3) \\
   C3 	    & 0.94(1) & 0.87(2) & 0.84(2) & 0.92(2) & -       & 0.95(3) & 0.87(2) & 0.79(2) & 0.69(3) & 0.86(7) \\
   C4	    & 0.93(1) & 0.83(2) & 0.80(2) & 0.84(2) & 0.95(3) & -       & 0.84(2) & 0.89(2) & 0.77(4) & 0.68(4) \\
   C5	    & 0.97(1) & 0.89(2) & 0.79(2) & 0.94(2) & 0.87(2) & 0.84(2) & -       & 0.79(2) & 0.75(3) & 0.74(6) \\
   C6	    & 0.93(1) & 0.81(3) & 0.83(2) & 0.92(2) & 0.79(2) & 0.89(2) & 0.79(2) & -       & 0.69(3) & 0.74(5) \\
   C7       & 0.85(2) & 0.72(4) & 0.75(4) & 0.76(3) & 0.69(3) & 0.77(4) & 0.75(3) & 0.69(3) & -       & 0.65(6) \\
   C8       & 0.81(2) & 0.63(3) & 0.82(5) & 0.73(3) & 0.86(7) & 0.68(4) & 0.74(6) & 0.74(5) & 0.65(6) & - \\  

   \end{tabular}
  \caption{\textbf{Bell state fidelities for all pairs of spins.} Data as presented in Fig. 6 of the main text. The target state is ($\ket{0+} + \ket{1-})/\sqrt{2}$ for electron-nuclear Bell states, and ($\ket{++} + \ket{--})/\sqrt{2}$ for the nuclear-nuclear Bell states. The fidelities are calculated based on measurements of the Pauli operators with non-zero expectation values for those states. That is, $\mathcal{F} = ( 1+ \langle XZ\rangle + \langle YY \rangle + \langle ZX \rangle)/4$ for the electron-nuclear Bell states, and $\mathcal{F} = ( 1+ \langle XX\rangle - \langle YY \rangle + \langle ZZ \rangle)/4$ for nuclear-nuclear Bell states. Statistical errors (one standard deviation) are given in parentheses.}
  \label{Tab:EntanglementGrid}
\end{table}

In table \ref{Tab:EntanglementGrid}, we present the numerical values for the Bell state matrix presented in Fig. 6 of the main text. The data utilizes a single set of gate parameters, rather than separately optimizing the parameters to avoid crosstalk between each pair of qubits.

\subsection{Theoretical predictions for multi-qubit state fidelities}

We use a simple depolarizing noise model to estimate the gate fidelities and predict the scaling of the GHZ state fidelity $\mathcal{F}_\text{GHZ} = \bra{\text{GHZ}_N}\rho\ket{\text{GHZ}_N}$ with the number of qubits added, where $\ket{\text{GHZ}_N} = (\ket{0}\otimes \ket{+}^{\otimes (N-1)} + \ket{1}\otimes \ket{-}^{\otimes (N-1)})/\sqrt{2}$ is the state ideally created by the application of the entanglement sequence (Fig. 7(a) of the main text) to the initial state $\ket{0}\otimes\ket{\uparrow}^{\otimes (N-1)}$. We use two measured quantities: the initialization fidelity $\mathcal{F}_{\text{init},j} = \bra{\uparrow_j}\rho_{\text{init},j}\ket{\uparrow_j} = (1 + C_{Qj,\text{init}})/2$ of each nuclear spin $j$ (table \ref{Tab:initialization}), and the Bell state fidelity $\mathcal{F}_{\text{Bell},j}$ after an entangling sequence between the electron and nuclear spin $j$ (table \ref{Tab:SpinsFidelities}). For the entanglement sequence, we assume perfect single-qubit rotations on the electron spin, while each two-qubit gate is modelled by a noisy operation \cite{nielsen2002quantum}

\begin{table}
  \centering
  \begin{tabular}{c|c|c}
  		Spin	& $\mathcal{F}_\text{Bell}$ (measured) & $\mathcal{F}_\text{gate}$ (extracted) \\ \hline \hline
   C1*          &   0.972(8)    & 0.991(9) \\
   C1	        &   0.93(1)     & 0.97(1) \\
   C2	        & 	0.97(1)     & 0.99(1) \\
   C3	        &   0.94(1)     & 0.97(1) \\
   C4	        &   0.93(1)     & 0.97(1) \\
   C5           &   0.97(1)     & 0.99(1) \\
   C6 	        &   0.93(1)     & 0.95(1) \\
   C7	        &   0.85(2)     & 0.99(3)$^*$ \\
   C8	        &   0.81(2)     & 0.98(3)$^*$ \\
   $^{14}$N     &   0.93(1)     & 0.94(1)$^\dagger$ \\
   \end{tabular}
  \caption{\textbf{Measured Bell state fidelities between the electron and each nuclear spin, as well as the extracted two-qubit electron-nuclear gate fidelities.} The extracted two-qubit gate fidelities are calculated with Eq. \ref{eq:gatefid}, using initialization fidelities taken from table \ref{Tab:initialization}. $^*$The measured initialization and Bell state fidelities are similar for these spins, suggesting significant effects that are not taken into account in this model. $^\dagger$The optical projective measurement on the electron spin is expected to have a greater effect on the nitrogen spin due to the strong hyperfine coupling when the electron is in the excited state. This is not taken into account in the model, and therefore the gate fidelity is expected to be higher than calculated here.}
  \label{Tab:SpinsFidelities}
\end{table}

\begin{equation}\label{eq:depolarizing}
    \mathcal{E}_{j}(\rho) = (1-p_j)U_j\rho U_j^\dagger + \frac{p_j}{16}\sum_{\alpha, \beta}\alpha_e \beta_j \rho \alpha_e \beta_j,
\end{equation}
where $\alpha,\beta \in \{I,X,Y,Z\}$ are single qubit Pauli operators, $p_j$ is the error probability and $U_j$ is the ideal unitary operation given by

\begin{equation}\label{eq:CROT}
    U_j = \ket{0}\bra{0}_e \otimes R_y(\pi/2)_j + \ket{1}\bra{1}_e \otimes R_y(-\pi/2)_j.
\end{equation}
Applying the two-qubit entanglement sequence (Fig. 7(a) of the main text) using Eq. \ref{eq:depolarizing} for the two-qubit gate, we can find an analytical solution for the error probability in terms of the measured initialization and Bell state fidelities, given by

\begin{equation}
    p_j = 1-\frac{1 - 4\mathcal{F}_{\text{Bell},j}}{1- 4\mathcal{F}_{\text{init},e}\mathcal{F}_{\text{init},j}}.
\end{equation}
The two-qubit gate fidelity can then be estimated from the error probability using the relation \cite{nielsen2002quantum}

\begin{table}
  \centering
  \begin{tabular}{c|c|c|c|c|c|c}
  		Number of & Spin & $\mathcal{F}_\text{init}$ & $\mathcal{F}_\text{init}$ & $\mathcal{F}_\text{GHZ}$  & $\mathcal{F}_\text{GHZ}$ & $\mathcal{W}_\text{GHZ}$  \\
  		qubits & added & [predicted] & [measured] & [predicted] & [measured] & violation ($\sigma$) \\ \hline \hline
   2        &   C5        & 0.978   & 0.970(11)  &   0.972    & 0.966(9)  &   51    \\
   3	    &   C2        & 0.963   & 0.968(10)  &   0.947    & 0.949(9)  &   49   \\
   4 	    &   C6        & 0.949   & 0.917(8)   &   0.889    & 0.827(7)  &   46 	\\
   5	    &   $^{14}$N  & 0.946   & 0.905(7)   &   0.836    & 0.813(6)  &   52 	\\
   6	    &   C1        & 0.913   & 0.885(8)   &   0.782    & 0.695(7)  &   27    \\
   7	    &   C3        & 0.886   & 0.862(6)   &   0.740    & 0.615(5)  &   23 	\\
   8        &   C4        & 0.855   & 0.682(6)   &   0.692    & 0.365(5)  &   -  
   \end{tabular}
  \caption{\textbf{Predicted and measured initialization and GHZ state fidelities for different numbers of qubits.} Predicted fidelities for the initial states are calculated as the fidelity of $\rho_\text{init}$ (Eq. \ref{eq:initstate}) with the ideal initial state $\ket{0}\otimes\ket{\uparrow}^{\otimes (N-1)}$, while the measured values are calculated from measurements of the non-zero expectation values for this state (see Fig. \ref{fig:FullROC}). Predicted fidelities for the GHZ states are calculated by applying the entanglement sequence shown in Fig. 7(a) of the main text to the initial state $\rho_\text{init}$, where each two-qubit gate is modelled as the noisy operation given in Eq. \ref{eq:depolarizing}. Also tabulated are the experimental violations (in statistical standard deviations) of the witness $\mathcal{W}_\text{GHZ} = \mathds{1} - 2\ket{\text{GHZ}_{N}}\bra{\text{GHZ}_{N}}$.}
  \label{table:GHZ}
\end{table}

\begin{equation}\label{eq:gatefid}\begin{split}
    \mathcal{F}_{\text{gate},j} &= \min_{\ket{\psi_j}} \left[\bra{\psi_j}U_j^\dagger \mathcal{E}_j(\ket{\psi_j}\bra{\psi_j}) U_j \ket{\psi_j}  \right] \\
    &= 1-\frac{3 p_j}{4},
\end{split}\end{equation}
where the minimization is over all possible pure electron-nuclear two-qubit states $\ket{\psi_j}$. Calculated values of $\mathcal{F}_{\text{gate},j}$ are shown in table \ref{Tab:SpinsFidelities}. We can then use $\mathcal{F}_{\text{init},j}$ and $p_j$ to predict $F_\text{GHZ}$ for a larger number of qubits by applying the multi-qubit entanglement sequence shown in Fig. 7(a) of the main text to the initial state

\begin{equation}\begin{split}\label{eq:initstate}
    \rho_\text{init} = &\left(\mathcal{F}_{\text{init},e}\ket{0}\bra{0}_e + (1 - \mathcal{F}_{\text{init},e})\ket{1}\bra{1}_e\right) \\
    &\bigotimes_{j=1}^{N-1} \left(\mathcal{F}_{\text{init},j}\ket{\uparrow}\bra{\uparrow}_j + (1 - \mathcal{F}_{\text{init},j})\ket{\downarrow}\bra{\downarrow}_j\right).
\end{split}\end{equation}
The resulting values are shown in table \ref{table:GHZ} and Fig. 7(d) of the main text. Note that this model does not capture the effects of correlated noise and crosstalk between spins, which provides a possible explanation for the deviation of the measured fidelities from the theoretically predicted values as the number of qubits is increased.

\subsection{Additional data for the $N$-qubit GHZ state experiments}

\begin{figure*}[!ht]
    \centering
    \includegraphics[width=0.9\textwidth]{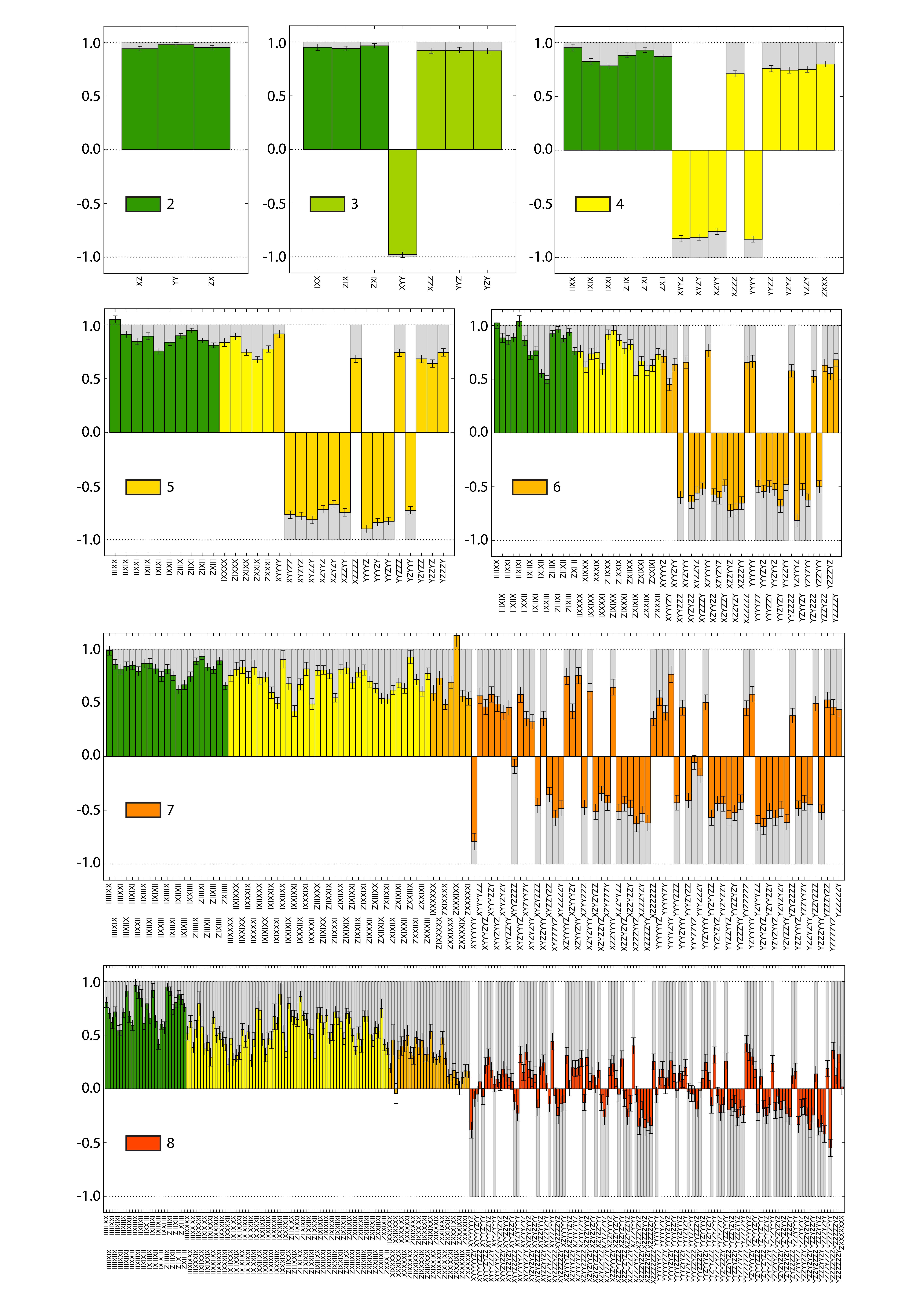}
    \caption{Bar plots showing the non-zero expectation values for 2-8 qubit GHZ states, used to calculate the GHZ state fidelity plotted in Fig. 7(d) of the main text. The colors of the bars indicate the number of qubits for which the measurement basis is not identity, shown in the insets. Gray bars show the ideal expectation values.} 
    \label{fig:FullTomo}
\end{figure*}

\begin{figure*}[!ht]
    \centering
    \includegraphics[width=0.9\textwidth]{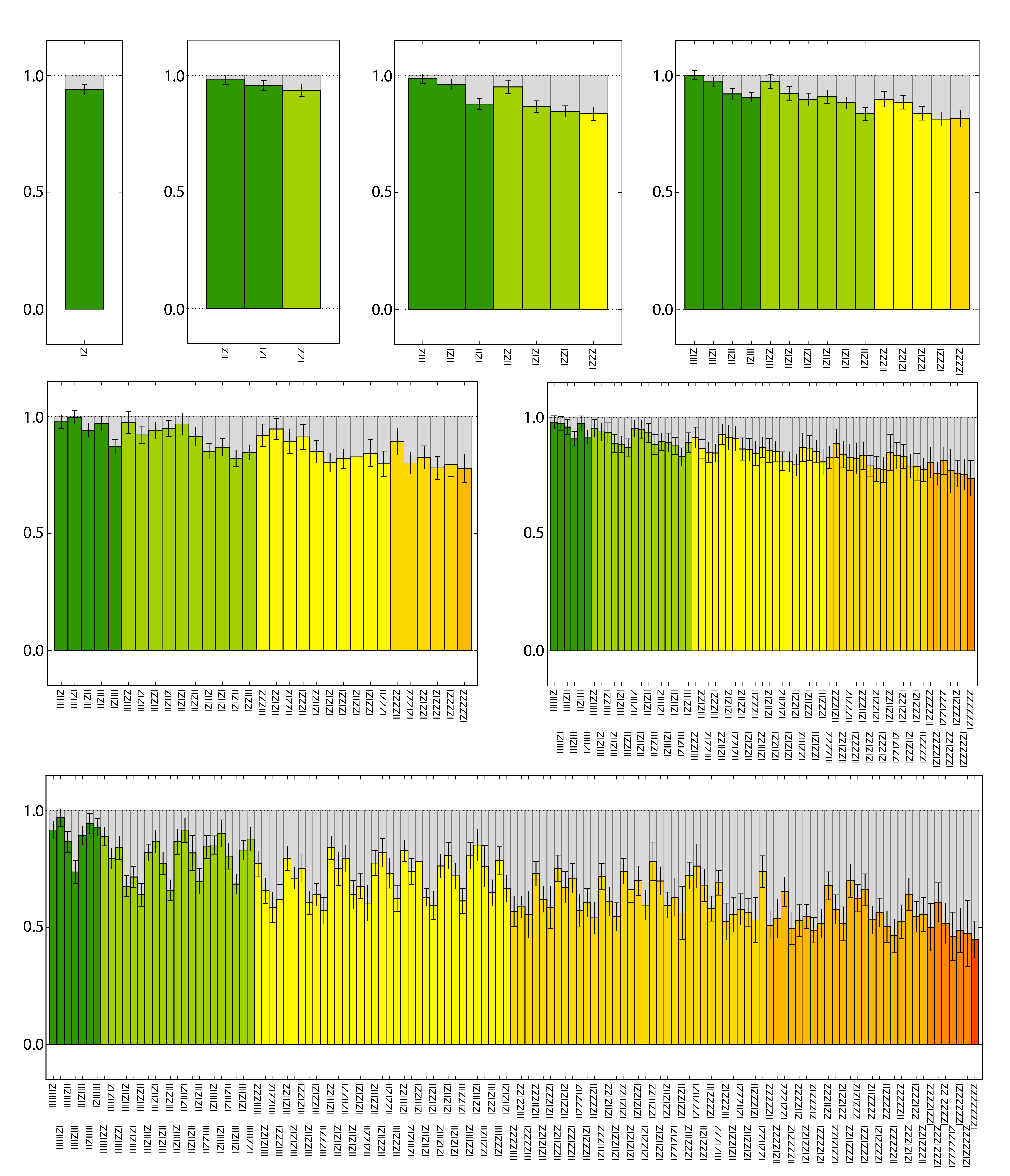}
    \caption{Bar plots showing the non-zero expectation values for 1-7 nuclear spin qubits after initializing in the state $\ket{0}\otimes\ket{\uparrow}^{\otimes (N-1)}$. These measurements are used to correct for readout errors (see section \ref{sec:ROcorrection}). The colors of the bars indicate the number of qubits for which the measurement basis is not identity, shown in the insets of Fig. \ref{fig:FullTomo}. Gray bars show the ideal expectation values.} 
    \label{fig:FullROC}
\end{figure*}

Figs. 7(b,c) of the main text show two example bar plots of the non-zero expectation values for 5 and 7 qubit GHZ states. Fig. \ref{fig:FullTomo} shows the bar plots for 2-8 qubit GHZ states, from which the fidelities shown in table \ref{table:GHZ} and plotted in Fig. 7(d) of the main text are calculated. Fig. \ref{fig:FullROC} shows the bar plots of the non-zero expectation values after initializing 1-7 nuclear spin qubits, which are used for correct to readout errors (see section \ref{sec:ROcorrection}).

\bibliography{bib_supp}